\documentclass[pra,aps,psfig,epsf,twocolumn,amsmath,amssymb,groupedaddress]{revtex4}

\newcommand{\vx}{{\vec{x}}}
\newcommand{\rv}{{\bf r}}
\newcommand{\xv}{{\bf x}}
\newcommand{\kv}{{\bf k}}
\newcommand{\kh}{{\hat{\bf k}}}
\newcommand{\Qh}{{\hat{\bf q}}}
\newcommand{\eps}{{\varepsilon}}

\newcommand{\vv}{{\bf v}}
\newcommand{\qv}{{\bf q}}
\newcommand{\qtv}{\tilde{\bf q}}
\newcommand{\qt}{\tilde{q}}
\newcommand{\Qv}{{\bf q}}
\newcommand{\jv}{{\bf j}}

\newcommand{\Nv}{\vec{N}}
\newcommand{\cNv}{{\vec{\mathcal N}}}
\newcommand{\nh}{{\hat n}}

\newcommand{\oh}{{\frac{1}{2}}}

\newcommand{\cH}{{\mathcal H}}
\newcommand{\cL}{{\mathcal L}}
\def\rf#1{(\ref{#1})}
\def\rfs#1{Eq.~\rf{#1}}

\newcommand{\as}{a_s}

\newcommand{\phdag}{{\phantom{\dagger}}}
\newcommand{\phstar}{{\phantom{*}}}
\newcommand{\vol}{{V}}
\newcommand{\kf}{k_{\rm F}}

\newcommand{\tphi}{\tilde{\phi}}
\newcommand{\ttheta}{\tilde{\theta}}

\newcommand{\ef}{\epsilon_{\rm F}}

\newcommand{\BdG}{Bogoliubov--de~Gennes}

\newcommand{\bk}{{\bf k}}
\newcommand{\bq}{{\bf q}}
\newcommand{\bQ}{{\bf q}}
\newcommand{\bp}{{\bf p}}

\newcommand{\grad}{{\bm{\nabla}}}
\newcommand{\zh}{\hat{z}}
\newcommand{\xh}{\hat{x}}

\newcommand{\ch}{\hat{c}}
\newcommand{\alphah}{\hat{\alpha}}

\newcommand{\Psih}{\hat{\Psi}}
\newcommand{\psib}{\bar{\psi}}

\newcommand{\Deltab}{\bar{\Delta}}

\newcommand{\Vt}{\tilde{V}}
\newcommand{\Gt}{\tilde{G}}
\newcommand{\muh}{\hat{\mu}}
\newcommand{\hh}{\hat{h}}

\newcommand{\br}{{\bf r}}

\newcommand{\be}{\begin{equation}}
\newcommand{\ee}{\end{equation}}
\newcommand{\bea}{\begin{eqnarray}}
\newcommand{\eea}{\end{eqnarray}}
\newcommand{\bse}{\begin{subequations}}
\newcommand{\ese}{\end{subequations}}

\newcommand{\fermiint}{g}

\newcommand{\Z}{{\mathbb{Z}}}
\input{epsf}
\usepackage{bm}
\usepackage{epsfig}

\begin{document}
\title{Fluctuations and phase transitions in Larkin-Ovchinnikov liquid
  crystal states of population-imbalanced resonant Fermi gas}
\author{Leo Radzihovsky}
\affiliation{Department of Physics, University of Colorado, 
Boulder, CO 80309}

\date{\today}

\begin{abstract}
  Motivated by a realization of imbalanced Feshbach-resonant atomic
  Fermi gases, we formulate a low-energy theory of the Fulde-Ferrell
  and the Larkin-Ovchinnikov (LO) states and use it to analyze
  fluctuations, stability, and phase transitions in these enigmatic
  finite momentum-paired superfluids.  Focusing on the unidirectional
  LO pair-density wave state, that spontaneously breaks the continuous
  rotational and translational symmetries, we show that it is
  characterized by two Goldstone modes, corresponding to a superfluid
  phase and a smectic phonon. Because of the liquid-crystalline
  ``softness'' of the latter, at finite temperature the 3d state is
  characterized by a {\em vanishing} LO order parameter, quasi-Bragg
  peaks in the structure and momentum distribution functions, and a
  ``charge''-4, paired Cooper-pairs, off-diagonal-long-range order,
  with a superfluid-stiffness anisotropy that diverges near a
  transition into a nonsuperfluid state. In addition to conventional
  integer vortices and dislocations the LO superfluid smectic exhibits
  composite half-integer vortex-dislocation defects. A proliferation
  of defects leads to a rich variety of descendant states, such as the
  ``charge''-4 superfluid and Fermi-liquid nematics and topologically
  ordered nonsuperfluid states, that generically intervene between the
  LO state and the conventional superfluid and the polarized
  Fermi-liquid at low and high imbalance, respectively. The fermionic
  sector of the LO gapless superconductor is also quite unique,
  exhibiting a Fermi surface of Bogoliubov quasiparticles associated
  with the Andreev band of states, localized on the array of the LO
  domain-walls.
\end{abstract}
\pacs{}

\maketitle


\section{Introduction}
\label{intro}
\subsection{Background}
\subsubsection{Imbalanced resonant atomic gases}
Experimental progress in trapping, cooling and coherently manipulating
Feshbach-resonant atomic gases opened an unprecedented opportunity to
study degenerate strongly interacting quantum many-body systems in a
broad range of previously unexplored regimes
\cite{BlochReview,KetterleZwierleinReview,GRaop,SRaop,GiorginiRMP,RSreview}.
These include paired fermionic superfluids (SF) (with $s$-wave SF now
readily realized
\cite{Regal2004prl,Zwierlein2004prl,Kinast2004prl,Bartenstein2004prl,Bourdel2004prl},
and $p$-wave SF under extensive current
study\cite{ZhangPRApwave,GaeblerPRLpwave,BotelhoSdeMeloPwave,
  GRApwave,ChengYipPRLpwave,GRaop}), the associated
Bardeen-Cooper-Schrieffer (BCS) to Bose-Einstein condensation (BEC)
crossover\cite{Eagles,Leggett,NSR,SdeMelo,Timmermans01,Holland,Ohashi,
  AGRswave,Stajic,GRaop}, Bose-Fermi mixtures\cite{OlsenBFmixture},
bosonic molecular
superfluids\cite{Cornish2000prl,RPWprl,RomansPRL,RPWaop}, and many
other states and regimes\cite{Reviews} under both equilibrium and
nonequilibrium conditions\cite{BLprl,AGRswave,AltmanVishwanathPRL}.
In addition to the tunability of the Feshbach-resonant interaction
strength (and its effective sign), temperature, and many types of
external perturbations, a species number imbalance in e.g., a two
atomic hyperfine-states mixture turned out to be an extremely fruitful
experimental
knob\cite{Zwierlein06Science,Partridge06Science,Shin2006prl,Navon2009prl}.

A nonzero species imbalance frustrates conventional BCS pairing of a
two-species Fermi gas\cite{Combescot01,Liu03,Bedaque03,Caldas04} and
the associated BCS-BEC
crossover\cite{CarlsonReddy05,Cohen05,SRprl,Pao06,Son06}, driving
quantum phase transitions out of a paired superfluid to a variety of
interesting possible ground states and thermodynamic
phases\cite{Castorina05,Sedrakian05nematic,SRprl,SRcomment,Son06,
  Bulgac06pwavePRL,Dukelsky}. This rekindled considerable theoretical
activity in the context of species-imbalanced resonant Fermi
gases\cite{Mizushima,YangFFLOdetect,YangSachdev,Pieri,Torma,
  Yi,Chevy,He,DeSilva,HaqueStoof,SachdevYang,LiuHu,Chien06prl,
  Gubbels06prl,YiDuan,PaoYip,SRaop,Martikainen,Parish07nature,
  BulgacFFLO,Parish1dLO,Sheehy}. The identification of the number
species imbalance with the magnetization of an electronic system, and
the chemical potential difference with an effective Zeeman energy,
connects these recent atomic gases studies with a large body of
research on solid state electronic superconductors under a Zeeman
field\cite{Clogston,Sarma,FF,LO}, as well as extensively studied
realizations in nuclear and particle
physics\cite{Alford,Bowers,Casalbuoni,Combescot}. The obvious
advantage of the current atomic system is the aforementioned
tunability, disorder-free ``samples'', and absence of the orbital part
of the magnetic field, that always accompanies a solid-state charged
superconductor in a magnetic field\cite{footnoteSRaop}. In these
neutral paired superfluids the orbital field effects can be
independently controlled by a rotation of the atomic
cloud\cite{DuineMacDonaldPRA}.

\begin{figure}[tbp]
\epsfxsize=9cm 
\centerline{\epsfbox{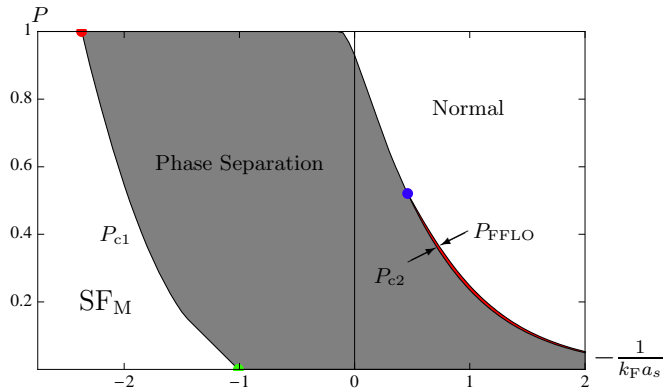}} \vskip-.25cm
\caption{A mean-field zero-temperature phase diagram from
  Refs.\onlinecite{SRprl,SRaop} of an imbalanced Fermi gas, as a
  function of the inverse scattering length and normalized species
  imbalance $P=(N_\uparrow-N_\downarrow)/N\equiv\Delta N/N$, showing
  the magnetized (imbalanced) superfluid ($SF_{\rm M}$), the FFLO
  state (approximated as the simplest FF state, confined to a narrow
  red sliver bounded by $P_{\rm FFLO}$ and $P_{{\rm c}2}$) and the
  imbalanced normal Fermi liquid.}
\label{SRphasediagram}
\end{figure}

As illustrated in Fig.\ref{SRphasediagram}, among many interesting
features, such as the gapless imbalanced superfluid
($SF_M$)\cite{SRprl,SRcomment,Pao06,Son06}, ubiquitous phase
separation\cite{Bedaque03,SRprl,SRcomment,SRaop}, tricritical
point\cite{Parish07nature,Sheehy}, etc., observed
experimentally\cite{Zwierlein06Science,Partridge06Science,Shin2006prl,
  Navon2009prl} and studied extensively
theoretically\cite{SRaop,RSreview}, the interaction--imbalance BEC-BCS
phase diagram is also
predicted\cite{Mizushima,SRprl,SRaop,Son06,BulgacFFLO,Parish1dLO} to
exhibit the enigmatic Fulde-Ferrell-Larkin-Ovchinnikov state
(FFLO)\cite{FF,LO}. First predicted in the context of solid-state
superconductors over 45 years ago\cite{FF,LO}, the FFLO states has so
far eluded a definitive observation, though some promising solid
state\cite{evidenceFFLO} and quasi-1d atomic\cite{Hulet1dLO} candidate
systems have recently been realized.

At its most generic level the FFLO state\cite{commentClassFFLO} is a
fermionic superfluid, paired at a finite center of mass
momentum. Generically such a state spontaneously ``breaks'' gauge and
translational symmetry, i.e., it is a periodically-paired superfluid
(superconductor), akin to a
supersolid\cite{Andreev69,Chester70,Leggett70,KimChan}, and thus can
appropriately be called a pair-density wave
(PDW)\cite{commentNotSS,ZhangPDW}.  Microscopically, it is driven by
Fermi surface mismatch\cite{FF,LO} due to an imposed pairing species
number (and/or mass\cite{WuPaoYip06mass}) imbalance.  As a compromise
between the superfluid pairing and an imposed imbalance, at
intermediate values of the latter, the superconducting order parameter
condenses at a set of finite center-of-mass momenta determined by the
details of the Fermi surface mismatch and interactions, thereby
self-consistently leading to FFLO pairing between these imbalanced
fermionic species. At sufficiently large imbalance, $\Delta N_{c2}$ or
equivalently at the upper-critical Zeeman field $h_{c2}$
(corresponding to the chemical potential difference $\Delta\mu_{c2}$
of the two pairing species) no compromise is possible, and instead a
transition to the normal state takes place.

\subsubsection{Mean-field energetic stability of FFLO states}

The original predictions by Fulde and Ferrell (FF), and by Larkin and
Ovchinnikov (LO) were followed by extensive studies of the FFLO
states\cite{Alford,Bowers,Casalbuoni,Combescot}, though exclusively
confined to their energetics in the BCS limit. Stimulated by the
aforementioned potential realization in imbalanced resonant Fermi
gases, the recent revival of the subject extended the analysis to the
full range of the BCS-BEC
crossover\cite{Mizushima,SRprl,SRaop,Parish07nature,BulgacFFLO,LevinFFLO}. As
illustrated in Fig.\ref{SRphasediagram}, the key observation is that,
despite of strong interactions, within a single $\qv_0$ BdG treatment
the conventional FFLO state (those originally proposed by FF and LO)
remains quite fragile, confined to a narrow sliver of polarization in
the BCS regime\cite{SRprl,SRaop}.

\begin{figure}[tbp]
\vskip0.25cm 
\epsfig{file=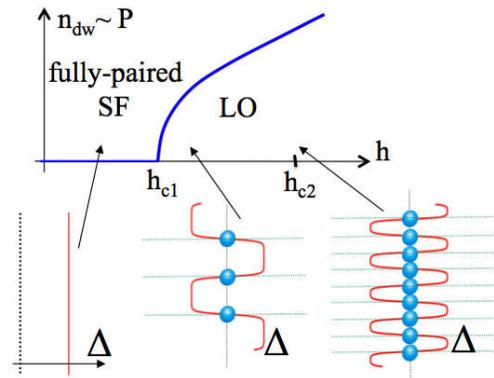,width=7cm,angle=0}
\caption{An illustration of a continuous commensurate-incommensurate
  (CI) transition at $h_{c1}$ from a fully-gapped (balanced)
  paired-superfluid to an imbalanced Larkin-Ovchinnikov
  superfluid. The excess majority atoms are localized on the domain
  walls in (zeros of) the LO order parameter, whose number $n_{dw}(h)$
  is then proportional to the imbalance $P(h)$ and grows continuously
  with the chemical potential difference (Zeeman energy), $h-h_{c1}$.}
\label{fig:CItransition}
\end{figure}

However, motivated by earlier studies of the Bogoliubov-de Gennes
(BdG) equation\cite{MachidaNakanishiLO,BurkhardtRainerLO,MatsuoLO},
combined with finding of a negative domain-wall energy in an otherwise
fully-paired singlet BCS superfluid in Zeeman
field\cite{MatsuoLO,YoshidaYipLO}, recent studies have quite
convincingly argued, that a more generic pair-density wave state (that
includes a larger set of collinear wavevectors) may be significantly
more stable. Analogously to a strongly type-II superconductor, that
undergoes a continuous transition into a vortex state at a
lower-critical orbital field
$H_{c1}$\cite{deGennes,Tinkham,Schrieffer} (that is significantly
below the thermodynamic field, $H_c$) in the current system, a
Zeeman-field $h$ (the chemical potential imbalance) can drive a
nucleation of domain-walls in the superfluid order parameter above the
lower-critical Zeeman field $h_{c1}$ that is below the bulk mean-field
value $h_c$. Thereby, the excess of the majority fermionic atoms
(polarization) in an imbalanced system can be {\em continuously }
accommodated by the sub-gap states localized on the self-consistently
induced domain-walls, through a continuous commensurate-incommensurate
(CI) Pokrovsky-Talapov (PT) type of transition\cite{PokrovskyTalapov}
from a fully paired s-wave superfluid to a Larkin-Ovchinnikov like
periodic state of domain
walls\cite{commentCIcontinuous,Brazovskii}. This picture, illustrated
in Fig.\ref{fig:CItransition} resembles the soliton mechanism for
doping of polyacetylene\cite{SuSchriefferHeeger}. The $\pm\Delta$
domain-wall description, that is explicitly realized in one-dimension
(1d) through exact BdG\cite{MachidaNakanishiLO} and Bethe
ansatz\cite{OrsoBA,HuBA} solutions and via
bosonization\cite{YangLL,ZhaoLiuLL}, is complementary, but not
qualitatively distinct from the more familiar single cosine form of
the LO state\cite{LO}.

The latter, ``soft'' pair-density wave is a more appropriate
description near the $h_{c2}$ transition into the normal state, where
the pair-order parameter is naturally small, and a Landau expansion in
the leading harmonic, $\Delta_{q}$ is expected to be valid. On the
other hand, clearly, far below $h_{c2}$ (e.g., near $h_{c1}$
transition from the fully-paired uniform singlet BCS state) the
periodic soliton state is a quantitatively better description as it
more accurately captures the strong pairing at $\Delta$, confining the
``normal'' regions that accommodate the imposed fermion imbalance to
the narrow gapless domain-walls between $\pm\Delta$.  Upon increasing
$h$ above $h_{c1}$ the density of domain-walls grows, eventually
overlapping at $h_{c2}$, thereby interpolating between the two
limiting descriptions of the LO state. This picture is explicitly
realized in the exact 1d
solutions\cite{MachidaNakanishiLO,OrsoBA,HuBA,YangLL} and emerges from
the numerical BdG studies\cite{BurkhardtRainerLO,MatsuoLO}.
Furthermore, the lack of a single $\qv$ nesting for $d > 1$, to which
the fragility of the LO state is usually attributed is irrelevant when
the LO order parameter exhibits a broad spectrum of $\qv$ (set by
$1/\xi_0$), as in its soliton form above $h_{c1}$.

The LO state can be equivalently thought of as a periodically ordered
{\em micro}-phase separation between the normal and paired states,
that naturally replaces the {\em macro}-phase
separation\cite{Combescot,Bedaque03} ubiquitously found in the BCS-BEC
detuning-imbalance phase diagram\cite{SRprl,SRcomment,SRaop}.

It is clear that the standard (even multi-wavevector) Landau
treatments valid near $h_{c2}$ and analytical BdG analysis of a single
FF plane-wave state (exhibiting no amplitude nodes)\cite{SRaop} fail
to capture above quantitatively important ingredients. These
treatments are therefore not necessarily quantitatively trustworthy in
their prediction of the energetic range of stability (location of
$h_{c1}$) of the LO state (they are however reliable for the
prediction of $h_{c2}$), and in our view need to be reexamined.

\subsubsection{Fluctuation in the FFLO states}

The microscopic question of the energetic stability of FFLO states is
certainly an extremely important one and has dominated most of the
research on the subject to date.  However, given the extensive 45-year
history of the topic, it is astounding that the equally basic
complementary question of the nature of Goldstone modes description
and their fluctuations within the FFLO states received so little
attention,\cite{Shimahara98,SamokhinFFLO} until our study of the
problem, reported in a recent Letter\cite{RVprl}.  From the general
symmetry principles the FFLO states' low-energy phenomenology is
expected to be significantly richer than that of a homogeneous fully
gapped superconductor, whose low-energy phenomenological
(Ginzburg-Landau and xy-model) description long predated the
microscopic theory by BCS\cite{deGennes,Tinkham,Schrieffer}. Namely,
in addition to a local superfluid phase degree of freedom, the
low-energy modes include the phonons (a single scalar one in the case
of a uniaxial LO state) of the periodic superconducting structure, as
well as gapless polarized (single-species) fermionic atoms confined to
a fluctuating periodic array of two-dimensional domain
walls\cite{RVprl,SamokhinFFLO}.

In the isotropic realization (e.g., in cold atoms in an isotropic
trap) of interest to us, the FFLO states {\em spontaneously} break a
{\em continuous} rotational (in addition the translational) symmetry
akin to smectic liquid crystals, in contrast to their solid state
density-wave analogs. Consequently, as was originally anticipated by
Shimahara\cite{Shimahara98}, and was demonstrated in our recent
work\cite{RVprl}, to be explored in greater detail below, their
Goldstone modes are qualitatively ``softer'', and therefore exhibit
far stronger fluctuations. These can either completely destabilize the
(otherwise energetically stable) FFLO state, or can qualitatively
modify its mean-field form and properties.  This general picture
therefore reveals that the complexity of the FFLO state beyond its
mean-field cartoon requires the understanding of subtle interplay of
superfluidity, liquid-crystallization, and anisotropic Fermi surface
physics.

\begin{figure}[bth]
\vspace{5.5cm}
\centering
\setlength{\unitlength}{1mm}
\begin{picture}(40,5)(0,0)
\put(-25,-37){\begin{picture}(0,0)(0,0)
\includegraphics{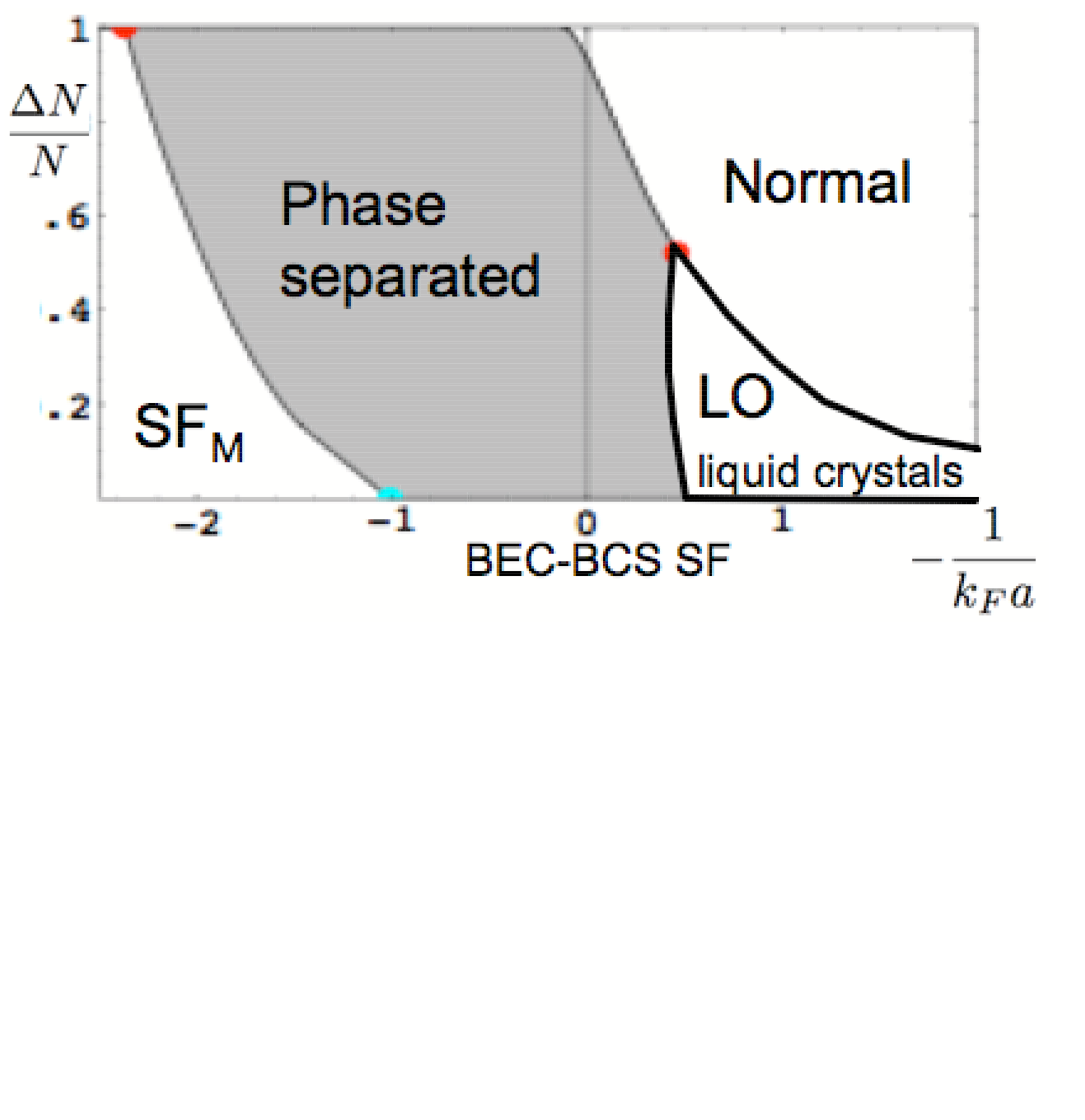}
\end{picture}}
\end{picture}
\vspace{-.5cm}
\caption{A proposed $\Delta N/N$ vs. $1/(\kf\as)$ phase diagram for an
  imbalanced resonant Fermi gas, showing the more stable LO liquid
  crystal phases (discussed in the text and illustrated in detail
  Fig.\ref{fig:phasediagramLO3d}) replacing a portion of the
  phase-separated regime.}
\label{fig:phasediagramLOmft}
\end{figure}

With the above motivation in mind, in a recent Letter\cite{RVprl}
above questions were formulated and carefully explored. Namely, as
illustrated in a schematic phase diagram in
Fig.\ref{fig:phasediagramLOmft}, supported by the aforementioned
studies\cite{MachidaNakanishiLO,BurkhardtRainerLO} and the exact 1d
solutions\cite{MachidaNakanishiLO,YangLL,ZhaoLiuLL}, we assumed that a
striped (unidirectional) FFLO state (that we refer to as LO),
characterized by a collinear set of wavevectors is {\em energetically}
(microscopically) stable over an experimentally accessible portion of
the detuning-imbalance phase diagram. We then formulated the model for
the low-energy Goldstone modes fluctuations and fermionic excitations
in the LO state and used it to study the stability of the state to
quantum and thermal fluctuations, as well as explored the
fluctuation-driven phenomenology, topological defects, novel quantum
liquid crystal phases, and the associated phase
transitions\cite{RVprl}. In the present manuscript we present a
significantly more extensive description of these findings and the
details of the associated calculations.


\subsubsection{Relation to other systems and studies}
Although physically quite distinct, some of our motivation and
findings\cite{RVprl} are closely related to studies in solid state
realizations, such as the putative FFLO states in heavy-fermion
(CeCoIn$_5$) and organic ($\kappa$-(BEDT-TTF)$_2$Cu(NCS)$_2$)
superconductors\cite{Agterberg08nature,Agterberg08prl}, the striped
(and spiral) states in high-T$_c$ superconductors and nickelates and
more generally strongly correlated doped Mott
insulators\cite{KFE98nature,FK99prb,OKF01prb,Berg09prb,Berg09nature,Sachdev03aop},
spiral states in helimagnets\cite{BakJensen80,BKRprb06,Green09prl}, 
partially-filled high Landau level 2d electron
gas\cite{Fogler96,MacdonaldFisher00prb,RD02prl}, charge-density waves
in anisotropic metals\cite{Gruner88rmp}, and others. Among a number of
physical properties special to resonant atomic Fermi gases, one key
qualitatively distinguishing feature in our study is that the FF and
LO states {\em spontaneously} breaks {\em continuous translational and
  orientational} symmetries of an isotropically trapped Fermi gas. As
first emphasized in Refs.\onlinecite{Shimahara98,RVprl} this latter
property is responsible for the qualitatively enhanced fluctuations,
that lead to the predicted universal power-law correlations
\rf{results:Sq},\rf{results:nk} persisting throughout the 3d
quasi-long-range ordered FFLO phase, rather than just at a critical
point. Thus, such FF and LO states (and its descendents) are
qualitatively distinct from their mean-field form\cite{FF,LO}, akin to
a distinction between, e.g., a 2d xy model and a long-range ordered 3d
ferromagnet.  In contrast, the aforementioned states in the solid
state systems, by their very definition only break {\em discrete}
point-group crystal symmetries, and therefore exhibit quite tame
Gaussian fluctuations inside the phase, with these periodic states
only quantitatively distinct from their mean-field form.

However, as we will see, despite enhanced fluctuations the LO state
studied here does lead to a number of novel features recently
discussed in the literature, such as the fractional vortex-dislocation
defects, ``charge''-4 superfluidity, Andreev-like mid-gap states, and
many others\cite{Agterberg08nature,Berg09prb}.

\subsubsection{Outline}
The rest of the paper is organized as follows. We conclude the
Introduction with the summary of our main results. Then, in
Sec.\ref{sec:microscopics} we use a microscopic description of the
imbalanced resonant atomic Fermi gas to derive in the
weakly-interacting BCS limit the corresponding Landau theory (along
with microscopic expressions for its parameters) valid near $h_{c2}$
transition to the normal state. In Sec.\ref{sec:GM} we discuss
symmetries of two generic class of FFLO states, construct
corresponding order parameters and derive from the Landau theory the
corresponding models for their Goldstone modes. As we argue, these
models provide universal descriptions of low-energy fluctuations in
the FF and LO class of states, beyond the limit of validity of their
microscopic BCS-limit derivation. In Sec.\ref{sec:LOhc1} we complement
this derivation (valid near $h_{c2}$) by an analysis of an array of
fluctuating domain walls, valid near $h_{c1}$. In
Sec.\ref{sec:GMfluctuations} we use these low-energy universal models
to analyze quantum and thermal fluctuations in the FF and LO class of
states, paying particular attention to elastic nonlinearities that (as
we show) are qualitatively important at a finite temperature.  Then in
Sec.\ref{sec:defects} we classify and discuss energetics of
topological defects in the LO state. In Sec.\ref{sec:transitions} we
use this information to uncover a variety of exotic ``daughter''
liquid-crystal phases that emerge as a result of unbinding different
combinations of topological defects, and discussion corresponding
phase transitions.  In Sec.\ref{sec:fermions} we extend the model to
include the coupling of the Goldstone modes to gapless fermionic
atoms, and comment on their effects. In Sec.\ref{sec:LDA} we use local
density approximation (LDA) to put our results in the context of
trapped atomic gases and discuss experimental probes of our
predictions. We conclude in
Secs.\ref{sec:experiments},\ref{sec:open},\ref{sec:summary} with
discussions of experimental probes, open questions and a summary of
our study. In appendices A and B we provide technical details for the
microscopic derivation of the Ginzburg-Landau expansion near $h_{c2}$
and finite-size scaling analysis of smectic fluctuations,
respectively.

\subsection{Results}
\label{sec:results}

Before turning to the derivation and analysis of the model, we
summarize the key predictions of our work, previously reported in
Ref.\onlinecite{RVprl}. Because our predictions\cite{commentIsotropic}
are based on general symmetry principles, supported by detailed
microscopic weak-coupling calculations, they are generic and robust to
variation in microscopic details.  At a very general level, we
demonstrate that (in contrast to the conventional uniform superfluid
and FF states) a unidirectional (striped) LO superfluid exhibits {\it
  two\/} Goldstone modes, $\phi$ and $\theta$. These correspond to two
coupled smectics phonons, or equivalently the superfluid phase $\phi$
and a nonlinearly-coupled smectic phonon mode $u=-\theta/q_0$, with
$\qv_0$ the wavevector characterizing the LO state.

Through robust symmetry arguments, complemented by an explicit
microscopic derivation (valid in the BCS regime, near the LO to normal
state transition at $h_{c2}$), we show that the low-energy universal
(classical) LO Hamiltonian governing fluctuations of these Goldstone
modes is given by
\begin{eqnarray}
  \cH^{GM}_{LO} &=&\frac{K}{2}(\nabla^2 u)^2 + 
  \frac{B}{2}\bigg(\partial_\parallel u - \frac{1}{2}(\nabla u)^2\bigg)^2
  +\frac{\rho_s^i}{2}(\nabla_i\phi)^2\nonumber\\
\end{eqnarray}
This Hamiltonian form, familiar from studies of conventional smectic
liquid crystals, encodes the underlying rotational invariance through
the vanishing of the $(\nabla_{\perp} u)^2$ modulus and the specific
form of the nonlinear elastic terms in
$u$\cite{deGennesProst,ChaikinLubensky,GP,GW}. In above
$\parallel,\perp$ refer to axes that are parallel and perpendicular to
the LO ordering wavevector $\qv_0$, respectively. In the weak-coupling
BCS limit, we derive explicit expressions for the above
Goldstone-modes moduli, $K$, $B$, $\rho_s^{i}$ ($i=\parallel,\perp$),
given by
\bse
\begin{eqnarray}
K&\approx&\frac{0.8 n\Delta_{BCS}^2}{\epsilon_F q_0^2}\ln(h/h_{c2}),\\
B&\approx&q_0^2\rho_s^\parallel
\approx\frac{3.3n\Delta_{BCS}^2}{\epsilon_F}\ln(h/h_{c2}),\\
\rho_s^\perp 
&\approx&\frac{0.8 n\Delta_{BCS}^2}{\epsilon_F q_0^2}\ln^2(h/h_{c2}),
\end{eqnarray}
\label{intro:KBrhos}
\ese
also given in Eqs.\rf{KB},\rf{rhos_pp} of the main text. Among these
predictions, we find that the LO state is a highly anisotropic
superfluid, with the ratio of superfluid stiffnesses given by
\begin{equation}
\rho_s^\perp/\rho_s^\parallel=
\frac{3}{4}\left(\Delta_{q_0}/\Delta_{\rm BCS}\right)^2
\approx\frac{1}{4}\ln(h_{c2}/h)\ll 1,
\label{result:ratio}
\end{equation}
vanishing on the approach to the upper-critical Zeeman field
$h\rightarrow h_{c2}^-$, that marks the mean-field transition to the
normal state at which the LO order parameter $\Delta_{q_0}$ vanishes.
$\Delta_{BCS}$ is the zero-field BCS order parameter.
 
We find the FF state to be even more exotic. In contrast to other
homogeneous superfluids (described by an xy-model), its single
superfluid Goldstone mode $\phi$ is described by above smectic
Hamiltonian, with an identically vanishing transverse superfluid
stiffness, $\rho_{s,FF}^\perp=0$, a reflection of the rotational
invariance of the spontaneous current to an energy-equivalent ground
state. Thus we show that a resonant imbalanced Fermi gas, confined to
an isotropic trap gives a natural realization of a quantum
(superfluid) smectic liquid crystal.

As a consequence of the spontaneous breaking of {\em continuous}
spatial symmetries (contrasting with its solid state realizations,
where only {\em discrete} point-group (lattice) symmetries are
spontaneously
broken)\cite{Agterberg08nature,KFE98nature,Berg09prb,Berg09nature,Sachdev03aop,
  BakJensen80,Green09prl,Fogler96,MacdonaldFisher00prb} the Goldstone
mode excitations in the FF and LO states are of {\em qualitatively}
lower energy.  As a result, the fluctuations in such superfluid
smectic states are qualitatively stronger. Specifically, we find that
while they are stable to quantum fluctuations, in 3d the LO and FF
long-range orders are marginally unstable at any nonzero $T$, with the
LO order parameter (with Dirichlet boundary conditions; for a more
general case, see Sec.\ref{sec:experiments} of the main text)
\begin{eqnarray}
  \langle\Delta_{LO}(\rv)\rangle_R
  &=&\langle2\Delta_{q_0}e^{i\phi(\rv)}\cos\big(\qv_0\cdot\rv +
  \theta(\rv)\big)\rangle_R,\nonumber\\
  &\sim&\frac{1}{R^\eta}\cos\qv_0\cdot\rv\ \ \longrightarrow\ \ 0,
\label{LOvanish}
\end{eqnarray}
vanishing in the thermodynamic limit (a large cloud with atom number
$N$ and cloud size $R\rightarrow\infty$), suppressed to zero by
thermal phonon $u=-\theta/q_0$ fluctuations, and is therefore strictly
speaking homogeneous on long scales. The resulting superfluid state is
thus an ``algebraic topological'' phase with no long-ranged
translational order. Namely, beyond mean-field theory it is instead
characterized by power-law order-parameter correlations, distinguished
from the spatially short-ranged disordered phase by confined
topological defects (bound dislocations), not by a nonzero LO order
parameter. It is therefore a 3d analog of the more familiar
quasi-long-range ordered superfluid film and a 2d easy-plane
ferromagnet.

There are a number of interesting consequences of this finding. For
example, as illustrated in Fig.\ref{fig:Sq}, we predict that in
3d\cite{commentSq} the static structure function $S(\qv)$ in the LO
state exhibits universal anisotropic {\em quasi}-Bragg peaks (akin to
a conventional smectic liquid crystal\cite{Caille,SqSmExp}), with
$n$-th order peak given by
\begin{eqnarray}
  S(q_\parallel,\qv_\perp=0)\sim\frac{1}{|q_\parallel - 2n q_0|^{2-4n^2\eta}},
\label{results:Sq}
\end{eqnarray}
rather than the true $\delta$-function Bragg peaks of e.g., a
long-range-ordered pair-density wave in a crystalline environment. In
above the anomalous Caill\'e exponent\cite{Caille} is given by
\begin{eqnarray}
\eta = \frac{q_0^2 T}{8\pi\sqrt{B K}}.
\label{results:eta}
\end{eqnarray}

We similarly find that the momentum distribution function of pairs displays a
power-law form around the reciprocal lattice momenta set by of $q_0$,
\begin{eqnarray}
  n_\kv &\sim& \frac{1}{|k_z - n q_0|^{2-n^2\eta}}
\label{results:nk}
\end{eqnarray}
as illustrated in Fig.\rf{fig:nk}. This power-law is a reflection of a
striking pair-condensate depletion to zero by the divergent finite $T$
LO Goldstone-mode fluctuations even in 3d, akin to the
Landau-Peierls\cite{Landau1dsolid,Peierls1dsolid} behavior of films of
a conventional superfluid and 2d
crystals\cite{MerminWagner,Hohenberg,Coleman,KT}. Such static
correlations in the LO state can be computed asymptotically exactly,
as was first done for a conventional smectic liquid
crystal\cite{Caille,GP,GW}.  This fluctuation-driven 3d power-law
phenomenology is a unique feature of a unidirectional ({\em collinear}
wavevectors) FFLO state. It is not exhibited by crystalline FFLO
phases with multiple {\em non-collinear} ordering
wavevectors\cite{Bowers,Alford}, that, in contrast are characterized
by the long-range positional order and a nonzero pair-condensate, that
is stable to thermal fluctuations.

%

As with treatments of the LO state, where long-range order is
assumed\cite{Agterberg08nature,Berg09prb}, in this algebraic LO phase
we also find an unusual topological excitation -- a half vortex bound
to a half dislocation -- in addition to integer vortices and
dislocations.  These are illustrated in
Figs.\ref{fig:vortex},\ref{fig:dislocation}, and
\ref{fig:vortexdislocation}.

In 2d, at nonzero $T$ the LO state is even more strongly disordered,
at intermediate scales characterized by universal power-law phonon
correlations and concomitant short-range positional order with
Lorentzian structure function peaks, controlled by a nontrivial
exactly calculable fixed point\cite{GW}. Asymptotically, however, at
arbitrary low temperature the 2d LO state is unstable to proliferation
of dislocations\cite{TonerNelsonSm}.  The state that results from such
dislocated superfluid smectic is either a ``charge''-4 (paired Cooper
pairs)\cite{commentCharge} nematic superfluid\cite{RVprl,Berg09nature}
or a nematic (possibly ``fractionalized''b) Fermi
liquid\cite{OKF01prb,RD02prl}, latter qualitatively the same as the
deformed Fermi surface state \cite{Sedrakian05nematic}.

\begin{widetext}

\begin{figure}[bth]
\centering
\setlength{\unitlength}{1mm}
\begin{picture}(100,110)(0,0)
\put(-30,-45){\begin{picture}(110,130)(0,0)
\includegraphics{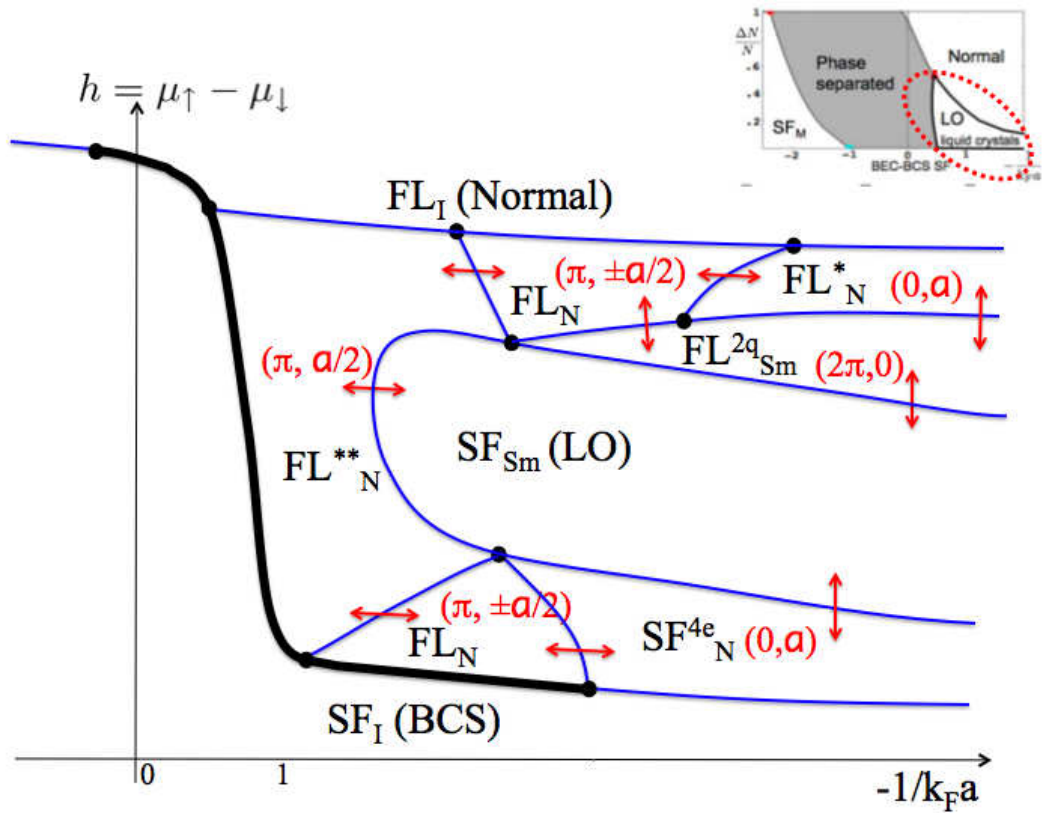}
\end{picture}}
\end{picture}
\caption{A schematic imbalance-chemical potential (Zeeman energy),
  $h=\mu_\uparrow-\mu_\downarrow$ vs detuning (interaction strength),
  $-1/k_Fa$ phase diagram, illustrating the 3d LO smectic phase
  ($SF_{Sm}$) and its descendant (described in the text), driven by a
  proliferation of various combinations of topological defects.  The
  inset shows the global imbalance-interaction BCS-BEC phase diagram,
  illustrating the location of these putative phases.}
\label{fig:phasediagramLO3d}
\end{figure}
\end{widetext}

More generally, while analyzing defects-driven continuous transitions
out of the LO state, we uncover a rich array of descendent states,
that generically must intervene between the LO superfluid and a
fully-paired conventional (isotropic and homogeneous) superfluid and a
conventional Fermi liquid.  If indeed the 3d LO state is energetically
stable, as argued above, we expect these novel states to appear in the
region collectively denoted ``LO liquid crystals'' of the
detuning-polarization phase diagram of
Fig.\ref{fig:phasediagramLOmft}. They include a nonsuperfluid smectic
($FL_{Sm}^{2q}$, driven by an unbinding of integer $2\pi$-vortices),
and a superfluid ($SF_{N}^4$, driven by a proliferation of integer
$a$-dislocations) and a nonsuperfluid ($FL_{N}$, driven by an
unbinding of both vortices and dislocations) nematics, and the
corresponding isotropic states, when disclinations also condense. In
addition, we predict a variety of topologically-ordered isotropic and
nematic ``fractionalized'' Fermi-liquid states ($FL_N^{*}$,
$FL_N^{**}$, $FL_I^{*}$, and others), that are distinguished from
their more conventional fully-disordered forms by {\em gapped} (bound)
half-integer defects. These phases are summarized by a schematic phase
diagram illustrated in Fig.\ref{fig:phasediagramLO3d}. We now turn to
the derivation of these predictions.

\section{BCS theory of imbalanced resonant Fermi gas}
\label{sec:microscopics}

\subsection{Model}
We begin with the one-channel model of two-species, resonantly
interacting Fermi gas, appropriate for the experimentally relevant
broad Feshbach resonance\cite{SRaop,GiorginiRMP,RSreview}. In the
grand-canonical ensemble, it is described by a Hamiltonian
\begin{eqnarray}
  H = \sum_{\bk,\sigma}(\epsilon_k - \mu_\sigma) \ch_{\bk\sigma}^\dagger  
  \ch_{\bk\sigma}^\phdag 
  + \frac{\fermiint}{V}\sum_{\bk\bq\bp} 
  \ch_{\bk\uparrow}^\dagger \ch_{\bp\downarrow}^\dagger 
  \ch_{\bk+\bq\downarrow}^\phdag\ch_{\bp-\bq\uparrow}^\phdag,
\label{eq:Hfermi}
\end{eqnarray}
with the two atomic species (labeling the hyperfine states $\sigma =
\uparrow,\downarrow$) open-channel fermions created by the
anticommuting operator $\ch_{\bk\sigma}^\dagger$
\be
\label{eq:ac}
\{\ch_{\bk\sigma}^\phdag,\ch_{\bk'\sigma'}^\dagger\} 
= \delta_{\sigma,\sigma'} \delta_{\bk,\bk'},
\ee
with the single-particle energy $\epsilon_k =\hbar^2 k^2/2m$, mass
$m$, and system's volume $\vol$ (henceforth set to $1$).

In above we introduced two distinct chemical potentials,
$\mu_\sigma=(\mu_\uparrow,\mu_\downarrow)$ to impose numbers of two
separately conserved atomic species, $N_\sigma =
(N_\uparrow,N_\downarrow)$, or equivalently the total fermion number
$N=N_\uparrow + N_\downarrow$ and the atom species imbalance $\Delta N
= N_\uparrow - N_\downarrow$.  Equivalently, the two species chemical
potentials $\mu_\uparrow = \mu + h$ and $\mu_\downarrow = \mu - h$ are
related to the total-number chemical potential $\mu$ and the
species-imbalance chemical potential $h$, latter corresponding to the
pseudo Zeeman energy.  The imbalanced resonant Fermi gas
thermodynamics as a function of $N,\Delta N, T, \as$, i.e., the
extension of the BEC-BCS crossover to a finite imbalance can be
computed by a variety of theoretical techniques, including quantum
Monte Carlo~\cite{CarlsonReddy05}, mean-field
theory~\cite{SRprl,SRcomment,Pao06,SRaop,Parish07nature}, the
large-$N_f$ (fermion flavor)~\cite{Nikolic07largeN,Veillette07largeN}
and $\epsilon$-expansions~\cite{Nishida06eps}.

The attractive interaction is parameterized by a short-range s-wave
pseudopotential with a strength $\fermiint<0$.  Through a standard
T-matrix scattering calculation\cite{LandauLifshitzQM,seeGRaop}, that
gives the two-atom s-wave scattering amplitude, $f_s(k) =
-\frac{m}{4\pi\hbar^2}T_k$, the pseudo-potential parameter $\fermiint$
can be related to the experimentally determined, magnetic field
dependent~\cite{RegalJinRF03} scattering length
\be 
\frac{m}{4\pi\hbar^2\as} = \frac{1}{\fermiint}+ \frac{1}{\vol}\sum_\bk
 \frac{1}{2\epsilon_k},
\label{eq:scatt}
\ee
where the ultraviolet-divergent second term is regularized by a
microscopic momentum cutoff scale $\Lambda\sim 1/d$ set by the range
of the potential, $d$. This gives (with $\hbar=1$ hereafter)
\bse
\begin{eqnarray}
\label{eq:deltalength} 
\as(\fermiint)& =& \left({4\pi \over m \fermiint} + {2 \Lambda
    \over  \pi} \right)^{-1} \equiv \frac{m}{4\pi}\fermiint_R,\\
&=&\frac{m}{4\pi}\frac{\fermiint}{1+\fermiint/\fermiint_c},
\end{eqnarray}
\label{as}
\ese
where $\fermiint_R$ is the effective screened coupling and
$\fermiint_c={2\pi^2 \over\Lambda m}$ is the critical coupling $\fermiint$
(set by the zero-point energy at scale $d$) at which a molecular bound
state appears and the scattering length diverges.  The above relation
allows a definition of the model and therefore a reexpression of
physical observables in terms of the experimentally defined (UV-cutoff
independent) scattering length $\as$.

To treat the many-body problem, \rf{eq:Hfermi}, we utilize the
standard mean-field analysis\cite{SRprl,SRaop}
(quantitatively valid deep in the BCS regime, $k_F\as\ll 1$, but
expected to be qualitatively valid throughout) by first assuming an
expectation value
\begin{eqnarray}
\label{eq:Delta_q}
\fermiint\langle \ch_\downarrow(\br) \ch_\uparrow(\br) \rangle
&=&\Delta(\br),\nonumber\\
&=&\sum_{\bq}\Delta_{\bq} {\rm e}^{i{\bq}\cdot\br}, 
\end{eqnarray}
corresponding to pair condensation at a superposition of finite
momenta $\bq$, with the set of $\Delta_\bq$ and $\bq $ to be
self-consistently determined.  With this mean-field assumption, $H$,
in Eq.~(\ref{eq:Hfermi}), reduces to the standard BCS mean-field form:
\bea
H &=& -\sum_\bq\frac{|\Delta_\bq|^2}{\fermiint}+ 
\sum_{\bk} (\epsilon_k - \mu_\sigma) \ch_{\bk \sigma}^\dagger  
\ch_{\bk\sigma}^\phdag,\\
&+& \sum_{\bq,\bk}\Big(
\Delta_\bq^*\ch_{\bk+\frac{\bq}{2}\downarrow}^{\phdag}
\ch_{-\bk+\frac{\bq}{2}\uparrow}^{\phdag}+ 
\ch_{-\bk+\frac{\bq}{2}\uparrow}^{\dagger}
\ch_{\bk+\frac{\bq}{2}\downarrow}^{\dagger}\Delta_\bq
\Big),\nonumber
\eea
that can equivalently be obtained using a Hubbard-Stratonovich
transformation and a saddle-point approximation on the coherent state
path-integral formulation of the problem.

Although the resulting mean-field Hamiltonian is quadratic in the
fermionic operators, its diagonalization for a generic $\Delta(\br)$
(an arbitrary set of Fourier components $\Delta_\bq$) is only possible
through a numerical self-consistent solution of the \BdG\
equations\cite{MachidaNakanishiLO,BurkhardtRainerLO}. 

Analytical progress is however possible through two complementary
approaches. One is to specialize to a single Fourier component,
$\Delta_\bq$ FF state\cite{FF}, that is self-consistently determined
through the $\bq$-dependent gap equation, \rf{eq:Delta_q}, equivalent
to the ground-state energy minimization. This approach allows for a
computation of the ground-state energy that is fully nonlinear in
$\Delta_\bq$. However, as emphasized in the Introduction with regard
to learning about a more generic amplitude-modulated FFLO state this
simplifying specialization to a single Fourier component is only
harmless near the normal-to-FFLO state transition at the mean-field
upper-critical field $h_{c2}$, where $\Delta_\bq$ is small and physics
is well-approximated by the lowest dominant harmonic.

An alternative analytical approach is through the Ginzburg-Landau
expansion, that allows a treatment of a general form of the
superconducting order parameter, \rf{eq:Delta_q}, but is again limited
to the vicinity of $h_{c2}$, where $\Delta(\br)$ is small, permitting
Taylor expansion of the ground-state energy in terms of it\cite{LO}.

\subsection{\BdG\ analysis of  the Fulde-Ferrell state}
\label{sec:BdGff}
\subsubsection{Ground state solution}

Specializing to the Fulde-Ferrell (single harmonic, $\Delta_\bq$)
state simplifies the mean-field Hamiltonian, \rf{eq:Hfermi}, allowing
for a straightforward diagonalization\cite{FF,SRprl,SRaop} of the the
associated \BdG\ equations. A standard analysis\cite{SRaop} gives the
(fermionic) Bogoliubov quasi-particle operators, $\alphah_{\bk\sigma}$
\bse\label{eq:alphasFF}
\bea
\alphah_{\bk\uparrow} 
&=& u_k  \ch_{-\bk+\frac{\bQ}{2}\uparrow}^{\phdag} + 
v_k  \ch_{\bk+\frac{\bQ}{2}\downarrow}^{\dagger},
\\
\alphah_{\bk\downarrow}^\dagger 
&=& -v_k^*  \ch_{-\bk+\frac{\bQ}{2}\uparrow}^{\phdag} +
u_k^*  \ch_{\bk+\frac{\bQ}{2}\downarrow}^{\dagger},
\eea
\ese
with the orthonormal coherence factors~\cite{deGennes,Tinkham,Schrieffer}
\bse
\label{eq:coherencefactors}
\bea
u_k &=& \frac{1}{\sqrt{2}} \sqrt{1+ \frac{\varepsilon_k}{E_k}},
\\
v_k &=& \frac{1}{\sqrt{2}} \sqrt{1- \frac{\varepsilon_k}{E_k}},
\eea
\label{uvFF}
\ese
ensuring the canonical anticommutation relations, $\{\alphah_{\bk
  \sigma}^\phdag ,\alphah_{\bk \sigma'}^\dagger\} =
\delta_{\sigma,\sigma'}$. In terms of these quasi-particles, the
Hamiltonian reduces to a diagonal quadratic form
\begin{eqnarray}
  H_{FF} &=& \sum_{\bk\sigma} \Big(  E_{\bk \sigma}\Theta(E_{\bk \sigma}) 
  \alphah_{\bk \sigma}^\dagger \alphah_{\bk \sigma}^\phdag 
  - E_{\bk \sigma}\Theta(-E_{\bk \sigma}) 
  \alphah_{\bk \sigma}^\phdag  \alphah_{\bk \sigma}^\dagger
  \Big)\nonumber\\
&& + E^{FF}_{GS}(\Delta_\bq),
  \label{eq:HalphaFF}
\end{eqnarray}
with the ground state energy given by
\begin{eqnarray}
E^{FF}_{GS}&=& \sum_\bk \big(
\eps_{k} - E_{k}
+ E_{\bk\uparrow} \Theta(-E_{\bk\uparrow}) 
+ E_{\bk\downarrow} \Theta( -E_{\bk\downarrow})\big)\nonumber\\
&-&\frac{1}{\fermiint}|\Delta_\bq|^2.
\label{Egs}
\end{eqnarray}
In above, we defined the excitation energy $E_{\bk\sigma}$ for spin
state $\sigma$
\bse
\label{eq:Esigma}
\bea
E_{\bk\uparrow} &=& E_k - h - \frac{\bk \cdot \bq}{2m},
\label{ekuparrow}
\\
E_{\bk\downarrow} &=& E_k + h + \frac{\bk \cdot \bq}{2m},
\label{ekdownarrow}
\eea
\ese
with
\bse
\label{eq:energydefs}
\bea
\eps_k &\equiv& \frac{k^2}{2m} - \mu + \frac{q^2}{8m},
\label{eq:tildexi}
\\
E_k &\equiv& (\eps_k^2 +\Delta_\bQ^2)^{1/2},\\
&&\nonumber
\label{eq:excitationenergy}
\eea
\ese
and $\mu =\oh(\mu_\uparrow + \mu_\downarrow)$, $h = \oh(\mu_\uparrow -
\mu_\downarrow)$ the average and difference chemical potentials,
respectively. The introduction of $\Theta$-functions in above
expressions divides the momentum space into three regions, for $h>0$
given by: (1) $\bk\in \bk_1$ such that $E_{\bk \uparrow}>0$ and
$E_{\bk \downarrow}>0$, (2) $\bk\in \bk_2$ such that $E_{\bk
  \uparrow}<0$ and (3) $\bk\in\bk_3$ such that $E_{\bk \downarrow}<0$,
(it is easy to see that the $E_{\bk \sigma}$ cannot both be negative),
and by construction ensures a positive-definite excitation spectrum
$|E_{\bk \sigma}|$ above the FF ground state
\begin{widetext}
\bse
\begin{eqnarray}
|FF_\bQ\rangle
&=&\prod_{\bk \in E_{\bk \sigma}< 0}
\alphah_{\bk\sigma}^\dagger|BCS_\bq\rangle,\\
&=&\prod_{\bk \in \bk_{3}}  
\ch^\dagger_{\bk+\frac{\bQ}{2}\downarrow} 
\prod_{\bk \in \bk_2}  
\ch^\dagger_{-\bk+\frac{\bQ}{2}\uparrow} 
\prod_{\bk \in \bk_1} 
(u_{\bk} + v_{\bk} \ch_{\bk + \frac{\bQ}{2} \downarrow}^\dagger  
\ch_{-\bk+\frac{\bQ}{2}\uparrow}^\dagger)|0\rangle,
\end{eqnarray} 
\label{eq:groundstate}
\ese
\end{widetext}
with $|0\rangle$ the atom vacuum. The first form demonstrates that the
FF ground state can be formally thought of as the Fermi sea of
negative energy Bogoliubov ``quasi-particles'' added to the BCS state
$|BCS_\bq\rangle$ (that is {\em not} the ground state) with the center
of mass momentum $\bq$. The second form, above, shows that
equivalently the FF ground state corresponds to a finite center of
mass momentum ($\bq$) BCS state with unpaired atoms from the two
momenta sets $\bk_2$ and $\bk_3$, defined above.

This analysis thus suggests a possibility of two distinct FF states,
FF$_{fs1}$ and FF$_{fs2}$, that exhibit a single- and two-species
Fermi surfaces, respectively, with their volume difference set by the
imposed species imbalance\cite{SachdevYang}.

\subsubsection{Analysis near $h_{c2}$}
The energetic stability of the state is controlled by
$E_{GS}^{FF}(\Delta_\bq)$, that deep in the BCS limit was first
computed by Fulde and Ferrell\cite{FF}. Its full behavior throughout
the BCS-BEC crossover generally requires a combination of numerical
and analytical analysis. This led to the prediction of a narrow sliver
of stability of the FF state confined to the BCS side of the
crossover, $-1/(k_F\as) \gtrsim 0.46$\cite{SRprl,SRaop}. However, as
discussed in the Introduction, given the expected unfavorable nature
of the FF state (lacking the amplitude nodes necessary to accommodate
the imbalanced atoms) this analysis is unlikely to shed light on the
stability of other (e.g., LO type) states far below the $h_{c2}$
transition to them.  Nevertheless, because the transition at the
upper-critical field $h_{c2}$ (at least in mean-field theory) is
expected to be continuous and therefore $\Delta_\bq$ small, growing
from $0$, $E_{GS}^{FF}(\Delta_\bq)$ can be analyzed analytically by
Taylor-expanding it in $\Delta_\bq$.

Taking advantage of the extensive analysis in\cite{SRprl,SRaop}, near
$h_{c2}$ we obtain
\begin{eqnarray}
E_{GS}^{FF}&\approx&\eps_q|\Delta_\qv|^2 
+ \oh\Vt_{\qv,\qv,\qv,\qv}|\Delta_{\qv}|^4,
\label{Effhc2}
\end{eqnarray}
where
\begin{eqnarray}
\eps_q&\approx& \frac{3n}{4\epsilon_F}\left[-1
+ \frac{1}{2}\ln\frac{v_F^2q^2-4h^2}{\Delta_{BCS}^2}
+ \frac{h}{v_Fq}\ln\frac{v_F q+2h}{v_F q-2h}\right],\nonumber\\
\label{epsSR}
\end{eqnarray}
and
\begin{eqnarray}
\Vt_{\bq,\bq,\bq,\bq}&=&\frac{3n}{4\epsilon_F}\frac{1}{v_F^2q^2-4h^2},
\label{Vt_qqqq}
\end{eqnarray}
that agree with the Larkin-Ovchinnikov result\cite{LO} obtained via a
direct Landau expansion in $\Delta_\bq$. In above we used
\rf{eq:scatt} to eliminate the interaction coupling $\fermiint$ in favor
of the scattering length $\as$ and then reexpressed the latter in
terms of the BCS superfluid gap $\Delta_{BCS}$ according to
$-m/(4\pi\as)=N(\ef)\ln\left(8 e^{-2}/\Delta_{BCS}\right)$, with 3d
density of states
$N(\epsilon)=m^{3/2}\sqrt{\epsilon}/(\sqrt{2}\pi^2)$.  Near its
minimum $\eps_q$ can be approximated quadratically in $q$
\begin{eqnarray}
\eps_q\approx \eps_0 + \oh J(q^2-q^2_0)^2 + \ldots,
\label{epsSRexpand}
\end{eqnarray}
with 
\bse
\begin{eqnarray}
\eps_0&=&\frac{3n}{4\epsilon_F}
\ln\left[2\sqrt{\alpha^2-1}\frac{h}{\Delta_{BCS}}\right]
\equiv\frac{3n}{4\epsilon_F}
\ln\left[\frac{h}{h_{c2}}\right],\ \ \ \ \ \ \ \\
q_0&=&2\alpha\frac{h}{v_F},\\
J&\approx&\frac{1}{\alpha^2-1}
\frac{3n\epsilon_F}{16k_F^2q_0^2}\frac{1}{h^2}.
\end{eqnarray}
\label{eps0Q0J}
\ese
In above, $\alpha$ is a solution of equation arising from the
minimization of the ground-state energy with respect to center of mass
momentum $q$\cite{SRprl,SRaop}
\bse
\begin{eqnarray}
  \alpha &=& \oh\ln\frac{\alpha+1}{\alpha-1},\\
  &\approx&1.200.
\label{alpha}
\end{eqnarray}
\ese
Combining this value with the point at which $\eps_0(h)$ vanishes, we
obtain the N-FFLO upper-critical transition field\cite{FF,LO}
\bse
\begin{eqnarray}
h_{c2}\equiv h_{FFLO} &=& \frac{\Delta_{BCS}}{2\sqrt{\alpha^2-1}},\\
&\approx&0.754\Delta_{BCS}.
\label{h_FFLO}
\end{eqnarray}
\ese
At this transition point we have
\bse
\begin{eqnarray}
q_0&=&\frac{\alpha}{\sqrt{\alpha^2-1}}\frac{\Delta_{BCS}}{v_F},\\
&=&1.81\frac{\Delta_{BCS}}{v_F}=\frac{0.58}{\xi_0},
\ \text{at the $h_{c2}$ transition},\ \ \ \ \ \ \\
J&\approx&\frac{\alpha^2}{\alpha^2-1}
\frac{3n}{16\epsilon_F}\frac{1}{q_0^4},\\
&\approx&0.61\frac{n}{\epsilon_F}\frac{1}{q_0^4},\ \text{at the
    $h_{c2}$ transition},
\end{eqnarray}
\label{eps0Q0J2}
\ese
where the BCS coherence length is given by its standard expression
$\xi_0=v_F/(\pi\Delta_{BCS})$.

Near the $h_{c2}$ transition the FF quartic vertex, \rf{Vt_qqqq} also
simplifies giving 
\bse
\begin{eqnarray}
\hspace{-1cm}
\Vt_{\qv,\qv,\qv,\qv}
&\approx&\frac{3n}{4\epsilon_F}\frac{1}{\alpha^2-1}\frac{1}{4h^2},\\
&\approx&\frac{3n}{4\epsilon_F}\frac{0.57}{h^2},\\
  &\approx&\frac{3n}{4\epsilon_F}\frac{1}{\Delta_{BCS}^2},\ \text{at the
    $h_{c2}$ transition},
\label{Vt_qqqqhc2}
\end{eqnarray}
\ese
which agrees precisely with perturbative LO result\cite{LO}.

As we will see shortly, in single plane-wave FF state the transverse
superfluid stiffness for a supercurrent flowing transversely vanishes
identically, enforced by a Ward identity. As mentioned in the
Introduction this property is a special feature of a single momentum
component FF state and is guaranteed by the underlying rotational
invariance of the spontaneous current to an energy-equivalent ground
state.  Thus, to correctly (even qualitatively) capture a more general
FFLO (e.g., the LO) state requires the inclusion of the multiple
momentum components, as in \rf{eq:Delta_q} and the analysis beyond the
lowest-order LO treatment\cite{LO}. This unfortunately cannot be
calculated analytically through the \BdG\ analysis, although it quite
successfully can
numerically\cite{MachidaNakanishiLO,BurkhardtRainerLO} and through the
Ginzburg-Landau expansion to which we now turn.

\subsection{Ginzburg-Landau expansion near $h_{c2}$}
The analytical treatment of the LO state near $h_{c2}$ relies on the
Ginzburg-Landau expansion in $\Delta_\bq$, that is small near the (in
mean-field) continuous $h_{c2}$ normal-to-FFLO transition
\cite{LO,commentMFTtrans}. This expectation is explicitly supported by
the exact 1d BdG solution\cite{MachidaNakanishiLO} at high fields,
where $\Delta(x)$ is indeed well-approximated by a single harmonic,
with an amplitude $\Delta_q$ that vanishes continuously near $h_{c2}$.

Based on these general arguments, near $h_{c2}$ the Ginzburg-Landau
expansion for the ground-state energy is expected to take a familiar
form
\begin{eqnarray}
\hspace{-0.5cm}\cH&\approx&\sum_\bq\eps_q|\Delta_\qv|^2 
+ \oh\sum_{\bq_1,\bq_2,\bq_3,\bq_4}\Vt_{\qv_1,\qv_2,\qv_3,\qv_4}
\Delta_{\qv_1}^*\Delta_{\qv_2}\Delta_{\qv_3}^*\Delta_{\qv_4},\nonumber\\
\label{HDelta_qhc2}
\end{eqnarray}
where rotational invariance constrains $\eps_q$ to be a function of
the magnitude of $\qv$ only, $\qv_4=\qv_1-\qv_2+\qv_3$, and
$\Delta_\qv$ is a Fourier transform of $\Delta(\rv)$.

It is straightforward to see that translational and rotational
invariances and the quadratic form (in $\Delta_\qv$) of the first term
in $\cH$ guarantee that $\eps_q$ is independent of the type of the
FFLO state, and is therefore identical to that of the FF state,
\rf{epsSR}\cite{SRaop}. This observation guarantees that all harmonics
with magnitude $q_0$ become unstable at the same imbalance field
$h_{c2}$, \rf{h_FFLO}, with the degeneracy only lifted by the quartic
$\Delta_\qv$ term in $\cH$.

Following the standard prescription and guided by the seminal (lowest
order) analysis of Larkin and Ovchinnikov\cite{LO}, we now derive the
quartic vertex function, $\Vt_{\qv_1,\qv_2,\qv_3,\qv_4}$, appearing in
Eq.\rf{HDelta_qhc2}. It, together with $\eps_q$ will then allow us to
derive the key elastic moduli (compression modulus, bending rigidity,
and superfluid stiffnesses) characterizing the Goldstone modes of the
striped FFLO states. As first found in Ref.\onlinecite{RVprl}, one key
observation is that a derivation of a generic Ginzburg-Landau form and
consequently of a generic Goldstone mode theory requires a calculation
that is of higher order than that originally carried out by Larkin and
Ovchinnikov\cite{LO}. In particular, as we will see below, the leading
momentum dependence of the quartic coupling needs to be kept. It
corresponds to an induced short-ranged current-current interaction,
$v_{ij}j_i j_j$, with the supercurrent given by the standard
expression
\begin{eqnarray}
j_i&=&\frac{1}{m}\text{Re}
\left[-\Delta^*(\rv)i\partial_i\Delta(\rv)\right]
\end{eqnarray}

To derive the generic Ginzburg-Landau form, we use the imaginary time
fermionic coherent-state path-integral formulation of the BCS problem,
by computing the partition function,
\bse
\begin{eqnarray}
Z&=&\text{Trace}\left[e^{-\beta H[\psi^*_\sigma,\psi_\sigma]}\right],\\
&=&\int[d\psi^*_\sigma d\psi_\sigma]e^{-S_\tau[\psi^*_\sigma,\psi_\sigma]},
\end{eqnarray}
\ese
where $H$ is the four-Fermi Hamiltonian, \rf{eq:Hfermi},
$\psi_\sigma(\tau,\rv)$ Grassmann (anticommuting) fields, and the
imaginary time action is given by
\begin{eqnarray}
S_\tau[\psi^*_\sigma,\psi_\sigma]&=&\int d\tau d^dr
\left[\psi^*_\sigma\left(\partial_\tau+\hat\xi_\sigma\right)\psi_\sigma
+ \fermiint\psi^*_\uparrow\psi^*_\downarrow\psi_\downarrow
\psi_\uparrow\right],\nonumber\\
\label{Stau}
\end{eqnarray}
where for notational short-hand we defined a space-time coordinate
$\xv\equiv (\tau,\rv)$ and a single particle Hamiltonian operator,
$\hat\xi_\sigma = -\frac{\nabla^2}{2m}-\mu_\sigma$.

We use the Cooper-pair Hubbard-Stratonovich field $\Delta(\xv)$ to
decouple the (quartic) pairing interaction $\fermiint$ and to integrate
out the fermionic atoms:
\bse
\begin{eqnarray}
Z&=&\int[d\psib_\sigma d\psi_\sigma d\Deltab
d\Delta]e^{-S_\tau[\psi^*_\sigma,\psi_\sigma,\Delta^*,\Delta]},\\
&\equiv&\int[d\Delta^* d\Delta]e^{-S_{\text{eff}}[\Delta^*,\Delta]},
\end{eqnarray}
\ese
where $S_\tau=S_0+S_{int}-\frac{1}{\fermiint}\int_\xv|\Delta|^2$,
\bse
\begin{eqnarray}
S_0&=&\int_\xv
\left[\psi^*_\sigma(\partial_\tau+\hat\xi_\sigma)\psi_\sigma
-\frac{1}{\fermiint}|\Delta|^2\right],\\
S_{int}&=&\int_\xv
\left[\Delta\psi^*_\uparrow\psi^*_\downarrow 
+\psi_\downarrow\psi_\uparrow\Delta^*\right],\label{Sint}
\end{eqnarray}
\ese
and we defined the Ginzburg-Landau effective action
\begin{eqnarray}
S_{\text{eff}}[\Delta^*,\Delta]&=&-\ln\left[\int[d\psib_\sigma d\psi_\sigma]
e^{-S_\tau[\psi^*_\sigma,\psi_\sigma,\Delta^*,\Delta]}\right].\hspace{0.7cm}
\end{eqnarray}

Taylor-expanding $S_\tau$ in powers of $S_{int}$ we compute
$S_{\text{eff}}$ in powers of $\Delta$. Relegating all technical
details to Appendix \ref{app:GLexpansion} and focusing on the
time-independent quartic order (in $\Delta(\rv)$) contribution, given
by the connected fourth cumulant $S_4=\int d\tau H_4$, we find
\begin{eqnarray}
H_4&=&\frac{1}{2}\int_{\rv_1\rv_2\rv_3\rv_4}
V(\rv_1,\rv_2,\rv_3,\rv_4)
\Delta^*(\rv_1)\Delta(\rv_2)\Delta^*(\rv_3)\Delta(\rv_4),\nonumber\\
\end{eqnarray}
\begin{widetext}
where
\bse
\begin{eqnarray}
V(\rv_1,\rv_2,\rv_3,\rv_4)&=&
\int\frac{d\omega}{2\pi}
\Gt^0_{\uparrow}(\rv_2-\rv_1,\omega)
\Gt^0_{\downarrow}(\rv_2-\rv_3,-\omega)
\Gt^0_{\uparrow}(\rv_4-\rv_3,\omega)
\Gt^0_{\downarrow}(\rv_4-\rv_1,-\omega),\\
\Vt(\qv_1,\qv_2,\qv_3,\qv_4)&=&
\int\frac{d\omega d^d k}{(2\pi)^{d+1}}
\Gt^0_{\uparrow}(\kv,\omega)
\Gt^0_{\downarrow}(\qv_1-\kv,-\omega)
\Gt^0_{\uparrow}(\kv-\qv_1+\qv_2,\omega)
\Gt^0_{\downarrow}(\qv_4-\kv,-\omega),
\end{eqnarray}
\ese
\end{widetext}
with the noninteracting fermionic Green's function (in Fourier space)
as usual given by
\bse
\begin{eqnarray}
G^0_{\sigma}(\omega_n,q)&=&-\langle\psi_\sigma\psi^*_\sigma\rangle_0,\\
&=&\frac{1}{i\omega_n-\xi_{q\sigma}}.
\end{eqnarray}
\ese

We focus on the FFLO type order parameter
\begin{eqnarray}
\Delta(\rv)&=&\sum_{\Qv_n}\Delta_{\Qv_n}(\rv)e^{i\Qv_n\cdot\rv},
\end{eqnarray}
with Fourier transform given by
\begin{eqnarray}
\Delta(\qv)&=&\sum_{\Qv_n}\Delta_{\Qv_n}(\qv-\Qv_n).
\end{eqnarray}
We note that to go beyond mean-field, in above we included an
additional long-scale positional dependence in the Larkin-Ovchinnikov
order parameters $\Delta_{q_n}(\rv)$ on top of the short-scale
mean-field periodic dependence at the LO wavevectors $q_n$ encoded in
the plane-wave factor.

Substituting this form into $H_4$, gives
\begin{widetext}
\begin{eqnarray}
H_4&=&\frac{1}{2}\sum_{\qv_{n_i}}\int_{\qtv_i}
(2\pi)^d\delta^d(\qtv_1-\qtv_2+\qtv_3-\qtv_4)
\delta_{\qv_{n_1}-\qv_{n_2}+\qv_{n_3}-\qv_{n_4},0}\\
&&\times\Vt(\qv_{n_1}+\qtv_1,\qv_{n_2}+\qtv_2,\qv_{n_3}+\qtv_3,
\qv_{n_4}+\qtv_4)
\Delta^*_{\qv_{n_1}}(\qtv_1)\Delta_{\qv_{n_2}}(\qtv_2)
\Delta^*_{\qv_{n_3}}(\qtv_3)\Delta_{\qv_{n_4}}(\qtv_4),\nonumber
\label{H4general}
\end{eqnarray}
\end{widetext}
with $\qtv_i\equiv\qv_i-\qv_{n_i}\ll \qv_{n_i}$ (due to
$\Delta_{q_{n_i}}(\qtv_i)$ expected to be sharply peaked around
$\qtv_i=0$), allowing us to disentangle the $q$ sums and integrals,
and the corresponding $\delta$-function above, without double-counting
momentum states.

To lowest order in $\qtv_i$, we ignore its dependence in $\Vt$,
equivalent to the LO treatment\cite{LO}, with
$\Vt(\qv_{n_1}+\qtv_1,\qv_{n_2}+\qtv_2,\qv_{n_3}+\qtv_3,
\qv_{n_4}+\qtv_4)\approx
\Vt(\qv_{n_1},\qv_{n_2},\qv_{n_3},\qv_{n_4})$, and thereby obtain the
quartic vertex that determines which set of reciprocal momenta
$\qv_{n_i}$ (satisfying the momentum conservation) minimizes the
interaction energy, thereby defining the structure of the FFLO ground
state. As found by LO\cite{LO}, near $h_{c2}$ it is the striped state
of collinear $\qv_{n_i}$'s that is energetically favored and is well
approximated by the lowest pair of harmonics, $\qv_{n_i}=\pm\qv$.
Focusing on this energetically preferred (near $h_{c2}$) LO state
reduces the general form of $H_4$ in \rfs{H4general} to
\begin{widetext}
\begin{eqnarray}
H_4&=&\frac{1}{2}\int_{\qtv_i}
(2\pi)^d\delta^d(\qtv_1-\qtv_2+\qtv_3-\qtv_4)
\bigg[\sum_{\qv_1=\pm\qv}
\Vt(\qv_1+\qtv_1,\qv_1+\qtv_2,\qv_1+\qtv_3,\qv_1+\qtv_4)
\Delta^*_{\qv_1}(\qtv_1)\Delta_{\qv_1}(\qtv_2)
\Delta^*_{\qv_1}(\qtv_3)\Delta_{\qv_1}(\qtv_4)\nonumber\\
&&+4\Vt(\qv+\qtv_1,\qv+\qtv_2,-\qv+\qtv_3,-\qv+\qtv_4)
\Delta^*_{\qv}(\qtv_1)\Delta_{\qv}(\qtv_2)
\Delta^*_{-\qv}(\qtv_3)\Delta_{-\qv}(\qtv_4)\bigg],
\label{H4VV}
\end{eqnarray}
\end{widetext}
Taylor-expansion of
$\Vt(\qv_{n_1}+\qtv_1,\qv_{n_2}+\qtv_2,\qv_{n_3}+\qtv_3,
\qv_{n_4}+\qtv_4)$ in $\qtv_i$ then gives $H_4=H_4^{(0)}+H_4^{(2)}$,
with
\begin{eqnarray}
H_4^{(0)}&=&\frac{1}{2}\int_{\rv}
\bigg[v_{++}^{(0)}\left(|\Delta_{\Qv}(\rv)|^4+|\Delta_{-\Qv}(\rv)|^4\right)
\nonumber\\
&&+4v_{+-}^{(0)}|\Delta_{\Qv}(\rv)|^2|\Delta_{-\Qv}(\rv)|^2\bigg],
\end{eqnarray}
where
\bse
\begin{eqnarray}
\hspace{-0.3cm}
v_{++}^{(0)}&=&\Vt(\Qv,\Qv,\Qv,\Qv),\\
&=&\int\frac{d\omega d^d k}{(2\pi)^{d+1}}
\Gt^0_{\uparrow}(\kv,\omega)^2
\Gt^0_{\downarrow}(\Qv-\kv,-\omega)^2,\nonumber\\
\hspace{-0.3cm}
v_{+-}^{(0)}&=&\Vt(\Qv,\Qv,-\Qv,-\Qv),\\
&=&\int\frac{d\omega d^d k}{(2\pi)^{d+1}}
\Gt^0_{\uparrow}(\kv,\omega)^2
\Gt^0_{\downarrow}(\Qv-\kv,-\omega)
\Gt^0_{\downarrow}(-\Qv-\kv,-\omega),\nonumber
\end{eqnarray}
\ese
already computed by LO in their mean-field approximation and in
Ref.~\onlinecite{SRaop} from the expansion of the BdG ground state energy
of the FF state discussed in Sec.\ref{sec:BdGff}.

In computing the higher order $H_4^{(2)}$ term, that contains the
essential current-current interaction $v_{ij}j_i j_j$ discussed in the
previous section, we will neglect contributions from the first set of
terms in Eq.\rf{H4VV} ($++$ and $--$ terms), because they lead to
$\jv_\bq^2$ (positive $\bq$ current) and $\jv_{-\bq}^2$ (negative
$\bq$ current) contributions whose coefficients are enforced either by
rotational invariance (which for transverse current each are $1/2$ of
the coefficient of the $\jv_\bq\jv_{-\bq}$ term, required to keep the
smectic phonon field $u$ ``soft''), or for the longitudinal pieces are
small, i.e., higher order, subleading corrections to the nonzero
quadratic (in $\Delta$) terms computed in the previous section.

To proceed we focus on the second term in \rfs{H4VV}, and Taylor
expand it to second order in $\qtv_i$
\begin{eqnarray}
&&\hspace{-1cm}\Vt(\Qv+\qtv_1,\Qv+\qtv_2,-\Qv+\qtv_3,-\Qv+\qtv_4)\approx v^{(0)}_{+-}\\
&&+
\frac{v_{ij}^{(2)}(\qv)}
{4m^2}\left(\qt_{1i}\qt_{4j}+\qt_{1i}\qt_{3j}+\qt_{2i}\qt_{4j}
+\qt_{2i}\qt_{3j}\right),
\nonumber
\end{eqnarray}
taking advantage of momentum conservation $\qv_1-\qv_2+\qv_3-\qv_4=0$
to generate four equivalent terms such that the expression is
explicitly real. In above we defined the current-current coupling
matrix
\begin{widetext}
\bse
\begin{eqnarray}
v_{ij}^{(2)}&=&m^2\int\frac{d\omega d^d k}{(2\pi)^{d+1}}
\Gt^0_{\uparrow}(\kv,\omega)^2
\partial_i\Gt^0_{\downarrow}(\qv-\kv,-\omega)
\partial_j\Gt^0_{\downarrow}(-\qv-\kv,-\omega),\\
&=& g_1\delta_{ij}+g_2\Qh_i\Qh_j,
\end{eqnarray}
\label{vij2}
\ese
\end{widetext}
where
\bse
\begin{eqnarray}
 g_1&=&\frac{N(\epsilon_F)k_F^2}{2v_F^4q_0^4}
  \alpha_1\left(\frac{2h}{v_Fq_0}\right),\\
  &\approx& 2.07\frac{N(\epsilon_F)k_F^2}{2v_F^4q_0^4},
  \ \text{at the $h_{c2}$ transition},\\
  g_2&=&\frac{N(\epsilon_F)k_F^2}{2v_F^4q_0^4}
  \alpha_2\left(\frac{2h}{v_Fq_0}\right),\\
  &\approx& -5.75\frac{N(\epsilon_F)k_F^2}{2v_F^4q_0^4},
  \ \text{at the $h_{c2}$ transition},\ \ \ \
\end{eqnarray}
\label{ab}
\ese
are computed in Appendix \ref{app:GLexpansion}.
Converting this Fourier space expression into real
space gives
\begin{widetext}
\bse
\begin{eqnarray}
H_4^{(2)}&\approx&\frac{1}{2}\int_{\rv}4v_{ij}^{(2)}
\bigg[
\frac{1}{m}\text{Re}\left(-\Delta^*_{\Qv}i\partial_i\Delta_{\Qv}(\rv)\right)
\frac{1}{m}\text{Re}\left(-\Delta^*_{-\Qv}i\partial_j\Delta_{-\Qv}(\rv)\right)\bigg],\\
&\approx&\frac{1}{2}\int_{\rv}4v_{ij}^{(2)}j^i_\qv j^j_{-\qv},\\
&\approx&\frac{1}{2}\int_{\rv}2v_{ij}^{(2)}
(\jv_{\qv}+\jv_{-\qv})_i(\jv_{\qv}+\jv_{-\qv})_j,\\
&\approx&\frac{1}{2}\int_{\rv}8v_{ij}^{(2)} j_i j_j,
\end{eqnarray}
\label{H4jj}
\ese
\end{widetext}
where we reconstructed the $j^i_{\qv}j^j_{\qv}$ and
$j^i_{-\qv}j^j_{-\qv}$ transverse pieces (by rotational invariance)
and defined the total supercurrent $\jv=(\jv_\qv+\jv_{-\qv})/2$.

It is important to note already at this stage that the positivity of
$g_1>0$ in \rfs{ab} generates a positive {\em transverse} superfluid
stiffness, $\rho_s^\perp$ and therefore a well-defined LO
state. Furthermore, we note that (as we will see in Sec.\ref{sec:GM})
the fact that $g_2$ and even $g_1+g_2$ are negative does not cause any
stability difficulties as they lead to
$O[(\Delta_{q_0}/\Delta_{BCS})^4]$ corrections to the longitudinal
superfluid stiffness $\rho_s^\parallel$, a correction that is
subdominant near $h_{c2}$. However, the opposite signs do suggest that
below $h_{c2}$, $\rho_s^\perp$ grows, while $\rho_s^\parallel$
decreases with decreasing chemical potential difference, $h$. It is
thus conceivable that the anisotropy ratio
$\rho_s^\perp/\rho_s^\parallel$ (starting at $0$ just below $h_{c2}$)
may actually grow above $1$, i.e., that the superfluid anisotropy may
reverses, something that would have striking experimental
consequences.

Putting above derived ingredients together we finally obtain the
sought-after Ginzburg-Landau Hamiltonian
\begin{equation}
\cH_{GL} = J\left[|\nabla^2\Delta|^2 - 2q_0^2|\nabla\Delta|^2\right] +
r|\Delta|^2 
+ \frac{1}{2} \lambda_1|\Delta|^4 + \frac{1}{2}\lambda_2 \jv^2 + \ldots,
\label{H_GL} 
\end{equation}
where as just derived, deep in the BCS limit (large positive detuning,
$k_F\as\ll 1$) and near the $h_{c2}$ transition to the polarized
normal state the model parameters are given by

%
\bse
\begin{eqnarray}
J&\approx&\frac{0.61n}{\epsilon_Fq_0^4},\\
q_0&\approx&\frac{1.81\Delta_{BCS}}{\hbar v_F},\\ 
r&\approx&\frac{3n}{4\epsilon_F}\ln\left[\frac{9h}{4h_{c2}}\right],\\
h_{c2}&\approx&\frac{3}{4}\Delta_{BCS},\\
\lambda_1&\approx&\frac{3n}{4\epsilon_F\Delta_{BCS}^2},\\
\lambda_2&\approx&\frac{1.83n m^2}{\epsilon_F\Delta_{BCS}^2q_0^2}.
\end{eqnarray}
\label{Jmoduli}
\ese

Based on the discussion following \rfs{H4jj} we approximated the
anisotropic tensor coupling $v_{ij}^{(2)}$, \rfs{vij2} by an isotropic
one of strength set by the transverse part of $v_{ij}^{(2)}$, namely,
$g_1$, with the difference a subleading correction near $h_{c2}$.  More
generally (away from the weak-coupling BCS limit) these couplings can
be taken as phenomenological parameters to be determined
experimentally.

\section{Theory of Goldstone modes in striped FFLO states}
\label{sec:GM}

\subsection{Landau theory}

The general form of the Ginzburg-Landau model \rf{H_GL} derived in the
previous section has a much broader range of applicability, even if
our derivation and microscopic predictions for the associated coupling
constants in \rfs{Jmoduli} only apply in the BCS regime, near
$h_{c2}$. Independent of the microscopics, the key ingredient of
$\cH_{GL}$ is that it captures the imbalanced atomic Fermi system's
energetic tendency to pair at a finite momentum, and thereby forms a
pair-density wave characterized by a reciprocal lattice vector with
magnitude $q_0$ and spontaneously chosen orientation. More generally,
we expect the quadratic (in $\Delta(\rv)$) part of $\cH_{GL}$
\begin{eqnarray}
\cH_{GL}^0 = \Delta^*\hat\eps\Delta,
\end{eqnarray}
to be characterized by a more generic differential kernel (specialized
to a quadratic form, $\hat{\eps}=J(-\nabla^2-q_0^2)^2 + r-Jq_0^4$ in
\rf{H_GL}), exhibiting a minimum at a finite momentum with a magnitude
$q_0$ and an arbitrary orientation. However, it can be
shown\cite{LRunpublished} that this generalization adds little new
physics to the Lifshitz-like normal-to-FFLO (PDW) transition and to
the emerging Goldstone mode theory, that is our main interest. We will
there work directly with the simpler form in \rf{H_GL}.

With the minimum in the dispersion $\hat{\eps}(q)$ located at a finite
momentum magnitude $q_0$, for $r < r_c$ (that in the mean-field BCS
approximation is given by $r_c= J q_0^4\approx 0.61n/\eps_F$, or
equivalently for at $h < h_{c2}\approx\frac{3}{4}\Delta_{BCS}$) the
gap in $\hat{\eps}(q)$ closes and $\cH_{GL}$ in \rf{H_GL} develops an
instability to a nonzero superconducting (pairing) order parameter
$\Delta(\rv)=\sum_{\qv_n}\Delta_{\qv_n} e^{i\qv_n\cdot\rv}$ at a set
of nonzero wavevectors $\qv_n$ with a magnitude of the fundamental
given by the dispersion minimum $q_0$.

As with other crystallization problems, the nontrivial question of the
choice of the set of momenta $\qv_n$'s is determined by the details of
the interaction (terms higher than quadratic order in $\Delta$) and
will not be re-addressed here. Instead, motivated by the LO findings
and by the more recent analyses
\cite{LO,MachidaNakanishiLO,BurkhardtRainerLO,MatsuoLO,YoshidaYipLO},
we will focus on the {\em unidirectional} pair-density wave
(Cooper-pair stripe) order, characterized by a {\em collinear} set of
$\qv_n$'s.  That is, we will assume that such states are energetically
stable, will develop their Goldstone modes low-energy description and
analyze their stability to fluctuations. 

As we will see below, within this {\em unidirectional} pair-density
wave class of FFLO states, the FF and LO states are representatives of
two qualitatively distinct universality classes and therefore must be
treated separately.  Although (as argued in the Introduction) it is
the latter that is expected to be significantly more stable, for
completeness and potential of other microscopic realizations (where FF
may be stable) we will treat both universality classes.

The low-energy properties of the FF and LO states are described by a
periodically spatially modulated order parameter $\Delta(\rv)$, that
in its simplest form, quantitatively valid near $h_{c2}$ is well
captured with a single $\pm\qv$ pair of ordering momenta
\begin{equation}
  \Delta_{FFLO}(\rv) =\Delta_+(\rv) e^{i\qv\cdot\rv} + 
  \Delta_-(\rv) e^{-i\qv\cdot\rv},
\label{DeltaFFLOcollinear}
\end{equation}
where $\Delta_\pm(\rv)$ are two complex scalar order parameters, the
dominant Fourier coefficients of $\Delta(\rv)$
\begin{eqnarray}
  \Delta_\pm (\rv)= \Delta_\pm^0(\rv) e^{i\phi_\pm(\rv)},
\label{Delta_pm}
\end{eqnarray}
and amplitudes $\Delta_\pm^0$ distinguishing between the FF and LO
states.  We first focus on the amplitudes of these two order
parameters, for now ignoring the corresponding Goldstone modes
$\phi_\pm$. Taking $\Delta_\pm(\rv)$ as spatially independent (to be
justified a posteriori), we use the Ginzburg-Landau theory, \rf{H_GL}
to determine their magnitudes. To this end we find
\begin{widetext}
\begin{equation}
\cH_{Landau} = \tilde{r}\left(|\Delta_+|^2+|\Delta_-|^2\right) 
+\frac{3}{4}\lambda_1\left(|\Delta_+|^2+|\Delta_-|^2\right)^2
+\oh(\lambda_2\frac{q^2}{m^2}-\oh\lambda_1)
\left(|\Delta_+|^2-|\Delta_-|^2\right)^2,
\label{cH_landau} 
\end{equation}
\end{widetext}
where $\tilde{r}=r - J q_0^4 + J(q^2-q_0^2)^2$. The quadratic coupling
$\tilde{r}$ dictates that the most unstable momentum mode is $q=q_0$,
that condenses when $\tilde{r}$ becomes negative, i.e., $r$ falls
below $r_c=J q_0^4$, corresponding to $h=h_{c2}$. Since the first and
second terms are ``rotationally invariant'' in the $\Delta_+ -
\Delta_-$ space, i.e., $O(4)$ invariant, it is the last term that
breaks this symmetry down to the physical $U(1)\otimes U(1)$ and
thereby determines the relative size of these critical $\pm \qv_0$
momenta order parameters, $\Delta_\pm$.

Clearly, for $\lambda_1 > \lambda_2\frac{2q_0^2}{m^2}$, $\cH_{Landau}$
is minimized by the FF state, with only one of the two order
parameters nonzero, with $\Delta^{FF}_-=0$ and 
\begin{eqnarray}
\Delta^{FF}_+&=&\sqrt{\frac{|\tilde{r}|}{\lambda_1+\lambda_2
    q_0^2/m^2}},\ \ \mbox{for $\lambda_1 > \lambda_2\frac{2q_0^2}{m^2}$}.
\end{eqnarray}
In the opposite limit of $\lambda_1 < \lambda_2\frac{q_0^2}{2m^2}$,
the last quartic term in $\cH_{Landau}$ instead selects the LO state,
with the two order parameters equal and nonzero,
\bse
\begin{eqnarray}
\Delta^{LO}_+=\Delta^{LO}_-&=&\sqrt{|\tilde{r}|/(3\lambda_1)},
\ \ \mbox{for $\lambda_1 < \lambda_2\frac{2q_0^2}{m^2}$},\ \ \ \\
&=&\frac{1}{3}\Delta^2_{BCS}\ln(h/h_{c2}),\\
&\equiv&\Delta_{q_0}^{LO}(h).
\end{eqnarray}
\label{DeltaLOmf}
\ese
As first found by LO\cite{LO} and discussed in the previous section,
in the current system the microscopics dictates that it is the latter,
LO state that is the more stable one.

\subsection{Symmetries and order parameters}

The fundamental FFLO order parameter, \rf{DeltaFFLOcollinear}, that
characterizes the FF and LO states clearly distinguishes these two
symmetry-distinct states.

\subsubsection{The Fulde-Ferrell state}

The FF state is characterized by a single (independent) nonzero
complex order parameter, 
\begin{eqnarray}
\Delta_{FF}(\rv)=\Delta_{q_0} e^{i\qv_0\cdot\rv +i\phi},
\end{eqnarray}
that is a plane-wave with the momentum $\qv_0$ and a single Goldstone
mode
\begin{eqnarray}
  \phi=\phi_+,
\end{eqnarray}
corresponding to the local superconducting phase. The state carries a
nonzero, uniform spontaneously-directed supercurrent
\begin{eqnarray}
  \jv_{FF} = \frac{1}{m}|\Delta_{q_0}|^2(\qv_0+\nabla\phi),
\label{jFF}
\end{eqnarray}
and thereby breaks the time-reversal and rotational symmetry, chosen
spontaneously along $\qv_0$, as well as the global gauge symmetry,
corresponding to the total atom conservation.  Although the FF order
parameter itself is not translationally invariant, under translation
by an arbitrary vector ${\bf a}$ it transforms by a multiplication by
a global phase $e^{i\qv_0\cdot{\bf a}}$. It is therefore invariant
under a modified transformation of an arbitrary translation followed
by a gauge transformation. Thus, in the FF state all gauge-invariant
observables and therefore the state are translationally
invariant. Namely, the FF state is a uniform orientationally-ordered
(polar) superfluid.  Under an infinitesimal rotation of the FF current
axis $\qv_0\rightarrow\qv_0 + \delta\qv_0$, its phase transforms as
$\phi\rightarrow\phi + \delta\qv_0\cdot\rv$, costing zero energy. Thus
the underlying rotational symmetry of the FF state requires the
corresponding Goldstone mode Hamiltonian of the superconducting phase
$\phi=\phi_+$ to be invariant under such transformation, that for an
infinitesimal rotation corresponds to a phase shift {\em linear} in
$\rv$ transverse to $\qv_0$. A generic Goldstone-mode Hamiltonian,
that is well-known to satisfy these properties is that of a
smectic\cite{deGennesProst,RVprl}, to harmonic order given by
\begin{eqnarray}
  \cH_0^{FF}
  &=&\oh \tilde K(\nabla_\perp^2\phi)^2 
+ \oh \rho_s^\parallel(\hat\qv_0\cdot\nabla\phi)^2.
\label{Hff0symm}
\end{eqnarray}
Its key qualitative feature is the strict (symmetry-enforced)
vanishing of the $(\nabla_\perp\phi)^2$ stiffness, with $\perp$
designating axes transverse to the spontaneous current ($\hat
r_\parallel\equiv\hat z\equiv\hat\qv_0$) axis. Thus, despite its
uniform density (i.e., it is an orientationally ordered {\em
  superfluid}, rather than a density wave) the FF state is
characterized by a smectic-like Goldstone-mode Hamiltonian, that is
qualitatively distinct from that of a conventional uniform and
isotropic superfluid, described by an xy-model,
\begin{eqnarray}
\cH_{xy}&=&\oh\rho_s(\nabla\phi)^2.
\end{eqnarray}
Namely, as evident from \rfs{Hff0symm} the FF state is infinitely
anisotropic, characterized by an identically vanishing transverse
superfluid stiffness
\begin{eqnarray}
\rho_{s,\perp}^{FF}&=&0,
\end{eqnarray}
a reflection of its underlying rotational symmetry that is {\em
  spontaneously} broken by the ground-state supercurrent
$\jv_{FF}$. Its longitudinal superfluid stiffness $\rho_{s,\parallel}$
is nonzero, measuring the energetic cost of a deviation from the
ground-state current magnitude of $j^0_{FF}
=\frac{1}{m}|\Delta_{q_0}|^2 q_0$.

In the next section (after discussing the LO state), we will support
these symmetry-based arguments by an explicit derivation from the
generic Ginzburg-Landau theory \rf{H_GL} for the FF state.

\subsubsection{The Larkin-Ovchinnikov state}

As illustrated through the Landau analysis, the LO state is described
by a nonzero (standing-wave like) pair-density wave order
parameter. That is, taking the magnitudes of the two order parameters,
$\Delta_+=\Delta_-=\Delta_{q_0}$ in \rf{DeltaFFLOcollinear} to be the same (as
dictated by the last quartic term for $\lambda_1 <
\lambda_2\frac{2q_0^2}{m^2}$) the LO order parameter reduces to a
physically appealing form
\bse
\begin{eqnarray}
\Delta_{LO}(\rv)
&=&2\Delta_{q_0}e^{i\oh(\phi_+ + \phi_-)}
\cos\big[\qv_0\cdot\rv + \oh(\phi_+ - \phi_-)\big],\ \ \ \ \ \ \ \\
&=&2\Delta_{q_0}e^{i\phi}
\cos\big[\qv_0\cdot\rv + \theta\big],
\end{eqnarray}
\label{DeltaLO}
\ese
that is a product of a superfluid and a unidirectional density wave
order parameters, respectively characterized by two Goldstone modes
\bse
\begin{eqnarray}
\phi&=&\oh(\phi_++\phi_-),\\
\theta&=&\oh(\phi_+-\phi_-).
\end{eqnarray}
\label{phitheta}
\ese
%
In a qualitative contrast to the FF state and to a conventional
superfluid or a superconductor, LO state's density and other physical
quantities are periodic along $\hat\qv_0$, exhibiting periodic
uniaxial stripe order. The position of the associated pair-density
wave is characterized by the smectic phonon
\begin{eqnarray}
u = -\theta/q_0,
\end{eqnarray}
giving physical interpretation to the second Goldstone mode $\theta$,
as the phase of the pair-density wave.

We also note that unlike a conventional smectic\cite{deGennesProst}
(e.g., in liquid crystal materials, where one instead is dealing with
a real mass density $\rho(\rv)$ not a pair condensate wavefunction),
here, because $\Delta(\rv)$ is complex, the phases of $\Delta_\pm$ are
independent (though interacting) Goldstone modes\cite{comment2Deltas}.

The mean-field LO order parameter, $\Delta_{LO}$ thus simultaneously
exhibits the ODLRO (superfluid) and the smectic (unidirectional
density wave) orders. It thus spontaneously breaks the rotational,
translational, and global gauge symmetries, and is therefore realizes
a form of a paired supersolid. However, it is distinguished from a
conventional purely bosonic supersolid
\cite{Andreev69,Chester70,Leggett70,KimChan}, where homogeneous
superfluid order and periodic density wave coexist, by the vanishing
of the (``charge''-2 two-atom) zero momentum ($\qv=0$) superfluid
component in the LO condensate\cite{commentNotSS}.

The supercurrent in the LO state is given by
\bse
\begin{eqnarray}
\jv_{LO}&=&\frac{2|\Delta_{q_0}|^2}{m}
\nabla\phi
\left[1+\cos(2\qv\cdot\rv + 2\theta)\right],\\
&\approx&\frac{2|\Delta_{q_0}|^2}{m}\nabla\phi,
\label{jLO}
\end{eqnarray}
\ese
where in the last form we neglected its periodic contribution.  As
expected, in contrast to the FF state, \rf{jFF}, the supercurrent
vanishes in the LO ground state where $\grad\phi=0$.

Similarly to the FF state, the underlying rotational symmetry of the
LO state strongly restricts the form of the Goldstone-mode
Hamiltonian. Under an infinitesimal rotation of $\qv_0$, that defines
the spontaneously-chosen orientation of the pair-density wave, the
phase of the LO state transforms according to
\begin{eqnarray}
\theta\rightarrow\theta + \delta\qv_0\cdot\rv.
\end{eqnarray}
Hence the $\theta=-q_0 u$ sector of the LO Goldstone-mode Hamiltonian
must be invariant under this symmetry and must therefore be described
by a smectic form\cite{deGennesProst,ChaikinLubensky,GP}. On the other
hand because a rotation of the LO state leaves the superconducting
phase, $\phi$ unchanged, the $\phi$ sector of the Hamiltonian
generically does not experience any such restriction. We thus expect
it to be described by a generic anisotropic xy-model form, with the
full harmonic Goldstone-mode Hamiltonian given by
\begin{eqnarray}
  \cH_0^{LO}
  &=&\frac{K}{2}(\nabla_\perp^2u)^2 
+ \frac{B}{2}(\partial_\parallel u)^2\nonumber\\
&& + \oh\rho_s^\perp(\nabla_\perp\phi)^2 
+ \oh\rho_s^\parallel(\partial_\parallel\phi)^2.
\label{Hlo0}
\end{eqnarray}
Thus, unlike the FF state, the LO state is characterized by nonzero,
but unequal superfluid stiffnesses, $\rho_s^\perp\neq
\rho_s^\parallel$.

Another feature of the LO state is that in addition to the primary
order parameter, $\Delta_{LO}$ \rf{DeltaLO}, it is characterized by a
uniform ``charge''-4 superconducting and by a neutral $2\qv_0$-smectic
secondary order parameters,
\bse
\begin{eqnarray}
\Delta_{sc}^{(4)}&=&\Delta_{LO}^2,\label{Deltasc}\nonumber\\
&\approx&2\Delta^2_{q_0}e^{i2\phi},\\
\Delta_{sm}^{(2q)}&=&|\Delta_{LO}|^2,\label{Deltasm}\nonumber\\
&\approx&2|\Delta_{q_0}|^2\cos\big[2\Qv\cdot\rv + 2\theta\big],
\end{eqnarray}
\ese
where in above we neglected the subdominant contributions.  As we will
see in subsequent sections, these order parameters become particularly
important when the primary order parameter $\Delta_{LO}$ vanishes
either due divergent fluctuations (as e.g., for $T>0$ in two and three
dimensions) or via a disordering transition driven by unbinding of
topological defects.

\subsection{Goldstone-mode Hamiltonian}

To support above symmetry-based arguments, we will now use the
Ginzburg-Landau theory \rf{H_GL} to explicitly derive the
Goldstone-mode Hamiltonians for the FF and the LO states.

\subsubsection{The Fulde-Ferrell Hamiltonian}
To this end we use $\Delta_{FF}(\rv)$ inside the $\cH_{GL}$ \rf{H_GL},
but in contrast to the earlier mean-field calculation that determined
the value of $\Delta_{q_0}$, focus on the spatially dependent Goldstone
mode, $\phi(\rv)$. However, we will neglect the spatial
dependence of the amplitude $\Delta_{q_0}$, valid in the ordered phase,
where its deviations from the average condensate value are gapped,
controlled by a finite susceptibility.

Working out the gradients of $\Delta_{FF}(\rv)$ under these conditions
and using the expression for $\jv_{FF}$ inside $\cH_{GL}$, we find
\begin{widetext}
\bse
\begin{eqnarray}
\cH_{FF} &=& J|\Delta_{q_0}|^2
\left[(\nabla^2\phi)^2 + 
\left(2\qv\cdot\nabla\phi + (\nabla\phi)^2\right)^2\right]
+\big(J|\Delta_{q_0}|^2(q^2-q_0^2)+\frac{\lambda_2}{2m^2}|\Delta_{q_0}|^4\big)
\left(2\qv\cdot\nabla\phi + (\nabla\phi)^2\right),
\label{HGMff1}\ \ \ \ \ \ \\
&=& \oh\tilde K (\nabla^2\phi)^2 + 
\oh\rho_s^\parallel\big(\partial_\parallel\phi 
+ \oh q_0^{-1}(\nabla\phi)^2\big)^2.
\label{HGMff2}
\end{eqnarray}
\label{HGMff}
\ese
\end{widetext}
In above, we dropped constant pieces, used the FF amplitude
$\Delta_{q_0}$ (computed in the previous section), defined a
longitudinal derivative
$\partial_\parallel\equiv\hat\qv_0\cdot\nabla$, and in going to the
final form \rf{HGMff2} chose $q$ according to
\bse
\begin{eqnarray}
q^2 &=& q_0^2 - \frac{\lambda_2}{2m^2 J}|\Delta_{q_0}|^2,\\
&\approx& q_0^2, \ \ \ \text{near $h_{c2}$},
\label{qFFchoice}
\end{eqnarray}
\ese
in order to eliminate the terms linear in the fluctuation-current
nonlinear form
\begin{eqnarray}
\delta j_\parallel=\partial_\parallel\phi
+ \oh q_0^{-1}(\nabla\phi)^2,
\label{deltaJparallel}
\end{eqnarray}
as a standard minimization condition for the FF ground state current
$j_{FF}$. This condition is closely analogous to the choice of an
order parameter magnitude to eliminate terms linear in
fluctuations. As expected from the earlier mean-field analysis, near
$h_{c2}$ (where $\Delta_{q_0}$ is small) this corresponds to the
choice of $q\approx q_0$. However, as usual with a magnitude of an
order parameter (here the spontaneous current, $j_{FF}$), the
nonuniversal magnitude of $\qv$ (proportional to $j_{FF}$) will be
modified by fluctuations. Ultimately it is determined by the
requirement that the coefficient of the term linear in $\delta
j_\parallel$ vanishes.

Within this derivation, the Goldstone-mode moduli in $\cH_{FF}$ are
given by
\bse
\begin{eqnarray}
\tilde K&=&2J|\Delta_{q_0}|^2,\\
\rho_s^\parallel&=&8J q_0^2|\Delta_{q_0}|^2,
\label{Krho_par}
\end{eqnarray}
\ese
but more generally are two independent parameters characterizing the
energetics of the single Goldstone mode of the FF state. The
Hamiltonian form, $\cH_{FF}$ (valid beyond its weak-coupling
microscopic derivation) is familiar from studies of conventional
smectic liquid crystals\cite{deGennesProst,ChaikinLubensky,GP}, with
the rotational invariance encoded in two ways. Firstly, to the
quadratic order in $\nabla\phi$ it reduces to the harmonic form
$\cH^0_{FF}$ \rf{Hff0symm}, invariant under an {\em infinitesimal}
rotation of the FF current state. Namely, by the strict vanishing of
the $(\nabla_\perp\phi)^2$ superfluid stiffness, $\rho_s^\perp=0$
(resulting in the ``softer'' transverse Laplacian energetics,
$(\nabla_\perp^2\phi)^2$), it exhibits a vanishing energy cost for
transverse (to $\qv_0$) current fluctuations, with the stiffness for
the change in current {\em magnitude} (along $\qv_0$) controlled by
$\rho_s^\parallel$. Secondly, $\cH_{FF}$ is an expansion in a fully
rotationally-invariant longitudinal current fluctuation, $\delta
j_\parallel$ \rf{deltaJparallel}, whose nonlinearities ensure that it
is fully rotationally invariant even for large reorientations in
$\qv_0$\cite{deGennesProst,ChaikinLubensky,GP} that defines the FF
ground state.

To see the latter we note that under a rotation of $\qv_0=q_0\zh$ by
an angle $\alpha$ in the $\zh-\xh$ plane
\begin{eqnarray}
  q_0\zh\rightarrow\qv = q_0(\zh\cos\alpha + \xh\sin\alpha)
\label{rotate_q0}
\end{eqnarray}
generates a nontrivial, spatially-dependent phase 
\begin{eqnarray}
  \phi^0(\rv)=z(\cos\alpha-1) + x\sin\alpha,
\label{theta0rotate}
\end{eqnarray}
even though the system is clearly in its ground state. Simple algebra
demonstrates that the fully nonlinear form of the longitudinal current
$\delta j_\parallel$ ensures that it and the corresponding energy
$\cH_{FF}$ vanish for $\phi^0(\rv)$, as required by the rotational
invariance.

One might question the necessity of keeping nonlinearities in
$\cH_{FF}$ \rf{HGMff2}. As we will see shortly, because of the
vanishing transverse superfluid stiffness in the FF state, the
fluctuations in the purely harmonic description $\cH_{FF}^0$
\rf{Hff0symm} are infrared-divergent in three and lower
dimensions. Consequently these nonlinearities are in fact absolutely
essential for a well-defined description of such a state.

\subsubsection{The Larkin-Ovchinnikov Hamiltonian}
The derivation of the Larkin-Ovchinnikov Goldstone-mode Hamiltonian
follows a similar route to that for the FF state of the previous
subsection, with many common features, but also some essential
qualitative differences in the results.  To this end, we insert the LO
order parameter $\Delta_{LO}(\rv)$ inside $\cH_{GL}$ \rf{H_GL}, use
the earlier mean-field values of the amplitude $\Delta_{q_0}$,
\rfs{DeltaLOmf} (that vanishes at $h_{c2}$ and grows as
$|h_{c2}-h|^{\beta}$ below $h_{c2}$), ignoring its subdominant spatial
dependence, and track the resulting energetics of two spatially
dependent Goldstone modes, $\phi(\rv)$ and $\theta(\rv)= -q_0
u(\rv)$. We thereby obtain
\begin{widetext}
\bse
\begin{eqnarray}
\cH_{LO} &=&J|\Delta_{q_0}|^2
\left[(\nabla^2\phi_+)^2 + 
\big(2\qv_0\cdot\nabla\phi_+ + (\nabla\phi_+)^2\big)^2 
+(\nabla^2\phi_-)^2 +
\big(2\qv_0\cdot\nabla\phi_- - (\nabla\phi_-)^2\big)^2\right]
+\frac{\lambda_2|\Delta_{q_0}|^4}{2m^2}(\nabla\phi_++\nabla\phi_-)^2
\label{HgmLO1}\nonumber\\
&&\\
&=&\sum_{\alpha=\pm}\left[\frac{1}{4}K(\nabla^2 u_\alpha)^2 + 
\frac{1}{4}B\big(\partial_\parallel u_\alpha 
- \frac{1}{2}(\nabla u_\alpha)^2\big)^2\right] 
+ \frac{1}{8}\rho_s^\perp q_0^2(\nabla u_+ - \nabla u_-)^2,
\label{HgmLO2}\\
&=&\oh K(\nabla^2 u)^2 + 
\oh B\big(\partial_\parallel u - \frac{1}{2}(\nabla u)^2\big)^2
+ \frac{1}{2}\rho_s^\parallel(\partial_\parallel\phi)^2 
+ \frac{1}{2}\rho_s^\perp(\nabla_\perp\phi)^2 + 
\cH_{LO}^{\text{subdom}},
\label{HgmLO3}
\end{eqnarray}
\ese
\end{widetext}
where we dropped the constant and fast oscillating parts that average
away upon spatial integration of the energy density, introduced two
phonon fields
\begin{eqnarray}
u_\pm = \mp\phi_\pm/q_0,
\end{eqnarray}
and chose $q=q_0$ in order to eliminate the term linear in the
nonlinear, rotationally-invariant strain tensor (analog of $\delta
j_\parallel$ in \rf{deltaJparallel})
\begin{equation}
u_{qq}^\pm=\hat{\qv}\cdot\nabla u_\pm -\oh(\nabla u_\pm)^2
\label{u_qq}
\end{equation}
whose nonlinearities in $u_\pm$ ensure that it is fully rotationally
invariant even for large rotations.  We also defined the bend $K$ and
the compressional $B$ smectic elastic moduli
\bse
\begin{eqnarray}
K&=&4J q_0^2|\Delta_{q_0}|^2,\\
&\approx&\frac{0.8 n\Delta_{BCS}^2}{\epsilon_F q_0^2}\ln(h/h_{c2}),\\
B&=&16J q_0^4|\Delta_{q_0}|^2,\\
&\approx&\frac{3.3n\Delta_{BCS}^2}{\epsilon_F}\ln(h/h_{c2}),
\end{eqnarray}
\label{KB} 
\ese
and identified the longitudinal, $\rho_s^\parallel$ and transverse,
$\rho_s^\perp$ superfluid stiffnesses given by
\bse
\begin{eqnarray}
  \rho_s^\parallel&=&B/q_0^2
=\frac{9.8n}{\epsilon_F}\frac{\Delta_{q_0}^2}{q_0^2},\\
&\approx&\frac{3.3n\Delta_{BCS}^2}{\epsilon_F q_0^2}\ln(h/h_{c2}),\\
  \rho_s^\perp &=& \frac{4\lambda_2}{m^2}|\Delta_{q_0}|^4,\\
&\approx&\frac{7.3n}{\epsilon_F}\frac{\Delta_{q_0}^2}{q_0^2}
\left(\frac{\Delta_{q_0}}{\Delta_{BCS}}\right)^2,\\
&\approx&\frac{0.8 n\Delta_{BCS}^2}{\epsilon_F q_0^2}\ln^2(h/h_{c2}).
\end{eqnarray}
\label{rhos_pp}
\ese
A nonzero transverse superfluid coupling, $\rho_s^\perp$ (minimized by
a vanishing supercurrent $\jv\propto\nabla\phi_+ + \nabla\phi_-$)
removes the two independent rotational symmetries, orientationally
locking the two incommensurate ($u_\pm$) smectics. As argued above
based on symmetry, \rf{Hlo0} this leads to the superconducting phase
combination, $\phi=\oh(\phi_+ + \phi_-)$ to be of a conventional xy-
(as opposed to ``soft'' smectic) gradient type, \rf{HgmLO3}.

\begin{figure}[tbp]
\vskip0.25cm 
\epsfig{file=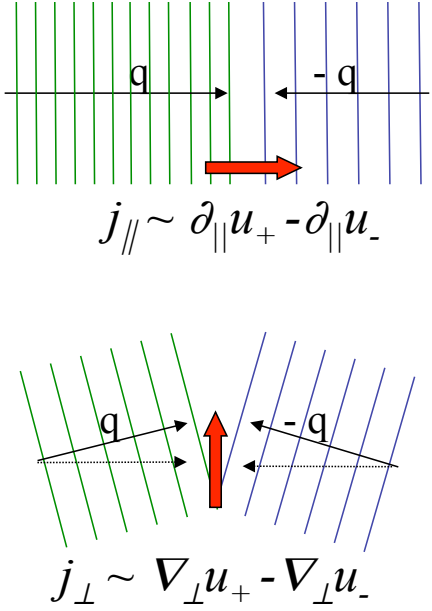,width=6.5cm,angle=0}
\caption{An illustration of longitudinal and transverse (to $\qv$) LO
  nonzero supercurrent configuration controlled by the corresponding
  superfluid stiffnesses, $\rho_s^\parallel$, $\rho_s^\perp$. Because
  there is no change in the effective period associated with the
  transverse current excitation, it is qualitatively energetically
  ``cheaper'', scaling as $\Delta_{q_0}^4$, as compared to the
  longitudinal excitation that scales like $\Delta_{q_0}^2$. As
  indicated in Eq.\rf{derive:ratio}, the ratio therefore vanishes as
  $h$ approaches the transition into a nonsuperfluid phase at $h_{c2}$. }
\label{fig:rhosLO}
\end{figure}

Within the superconductor context, this coupling of $\nabla\phi_\pm$
(xy- rather than smectic-like stiffness of the $\phi$ Goldstone mode)
is straightforward to understand. The LO state effectively corresponds
to two FF states, each carrying a supercurrent along $\qv +
\nabla\phi_+$ and $-\qv + \nabla\phi_-$. A superfluid stiffness
measures the nonzero kinetic energy cost of a nonzero total current
$\jv$. The latter results from either a transverse {\em orientational}
misaligned of $\nabla\phi_\pm$, measured by the $\rho_s^\perp$, or a
mismatch between the $\nabla\phi_+$ and $\nabla\phi_-$ {\em
  magnitudes} (even if directed along $\pm\qv$, respectively),
measured by the $\rho_s^\parallel$. These two distinct LO state
distortions are schematically illustrated in Fig.\ref{fig:rhosLO}.

As advertised in the introduction we thus find from \rf{rhos_pp} that
the LO state is a highly anisotropic superfluid (though less so than
the FF state, where $\rho_s^\perp=0$), with
\begin{equation}
\frac{\rho_s^\perp}{\rho_s^\parallel}=
\frac{3}{4}\left(\frac{\Delta_{q_0}}{\Delta_{BCS}}\right)^2
\approx\frac{1}{4}\ln\left(\frac{h_{c2}}{h}\right)\ll 1,
\label{derive:ratio}
\end{equation}
a ratio that vanishes for $h\rightarrow h_{c2}^-$.  

Because (in contrast to the FF state) the $\rho_s^\perp$ is indeed
nonzero for $h<h_{c2}$, the nonlinearities in $\nabla\phi$ and the
``soft'' term $(\nabla^2\phi)^2$ (contained in the
$\cH_{LO}^{\text{subdom}}$ of \rf{HgmLO3}) are subdominant on long
scales, and can (and will) therefore be neglected\cite{comment_irrel}.

We stress that while the detailed expressions for the moduli above are
specific to the weak-coupling BCS limit near $h_{c2}$ the general form
of $\cH_{LO}$, \rf{HgmLO3}, including the structure of the
symmetry-enforced nonlinearities in the $u$ (smectic) sector is valid
beyond our microscopic derivation, and holds throughout the LO
phase. Specifically, as argued in the previous subsection based on
symmetry, in contrast to the superconducting, $\phi$ sector, the
vanishing of the $(\nabla_\perp u)^2$ coupling is a reflection of the
underlying rotational invariance of the LO striped state. Thus the
``soft'' transverse elastic form of the smectic sector and the form of
the nonlinear strain tensor, $u_{qq}$ (analogous to $\delta
j_\parallel$ \rf{deltaJparallel}) are strictly protected by this
symmetry. As we will see in Sec.\ref{sec:GMfluctuations} (and is
well-known for conventional smectic liquid
crystals\cite{deGennesProst,ChaikinLubensky,GP}), because of this
harmonic elastic ``softness'', in the presence of thermal fluctuations
these smectic nonlinearities are qualitatively important at long
length scales for the description of the FF and LO states.

\subsection{Quantum dynamics}
\label{sec:dynamicsGM}
The bosonic Hamiltonian density $\cH_{GL}$ \rf{H_GL} determines the
classical thermodynamics and equal-time finite temperature correlation
functions of the FFLO system\cite{commentOtherDOF}, that in the
ordered state reduces to $\cH_{FF/LO}$, in \rf{HGMff2} and
\rf{HgmLO3}.

The dynamics is controlled by the Lagrangian density. Focusing on
quantum thermodynamics, the partition function
$Z={\text{Trace}}\left[e^{-\beta
    H[\hat\Delta^\dagger,\hat\Delta]}\right]$ is formulated in a
standard way as a path-integral over coherent states labeled by the
c-fields $\Delta(\tau,\rv)$ and $\Delta^*(\tau,\rv)$
\begin{eqnarray}
Z = \int [d\Delta^* d\Delta] e^{-S[\Delta^*,\Delta]},
\end{eqnarray}
with $\tau = i t$ the imaginary time $0\leq \tau<\beta\equiv1/k_B T$
($\hbar\equiv1$). The coherent state action $S[\Delta^*,\Delta]$ is
given by
\begin{eqnarray}
S[\Delta^*,\Delta]=\int_0^\beta d\tau \int d^dr \cL[\Delta^*,\Delta],
\end{eqnarray}
with the Lagrangian density given by
\begin{eqnarray}
\cL = \Delta^*\partial_\tau\Delta + \cH[\Delta^*,\Delta].
\end{eqnarray}
Using the density-phase representation 
\begin{eqnarray}
\Delta =(n_0+\delta n_+)^{1/2} e^{i\qv\cdot\rv+i\phi_+} 
+ (n_0+\delta n_-)^{1/2}e^{-i\qv\cdot\rv+i\phi_-},\nonumber\\ 
\end{eqnarray}
with $\delta n_\pm$ the fluctuating Cooper-pair density about its
ground state value $n_0=|\Delta_{q_0}|^2$, and at low energies neglecting
the subdominant terms (e.g., fast oscillating pieces or spatial
dependence of gapped amplitude fields) the Lagrangian density reduces
to
\begin{eqnarray}
\cL = i\delta n_+\partial_\tau\phi_+ 
+ i\delta n_-\partial_\tau\phi_- + \cH[\delta n_\pm,\phi_\pm].
\end{eqnarray}
Integrating these Berry's phase
terms\cite{commentBerry,FujimotoFFLOberry} over the ``massive''
(nonzero compressibility, $\chi_0$) $\delta n_\pm$ fluctuations, we
obtain
\begin{eqnarray}
  \cL =\frac{\chi_0}{2}(\partial_\tau\phi_+)^2 
  + \frac{\chi_0}{2}(\partial_\tau\phi_-)^2 + \cH[\phi_+,\phi_-],
\label{Slo}
\end{eqnarray}
with $\cH$ given by the $\cH_{LO}$ in the LO ground state, and FF
state treated similarly using $\cH_{FF}$.  For the LO state, this
analysis then predicts the existence of two anisotropic low-frequency
modes with dispersions
\bse
\begin{eqnarray}
\omega_\phi(\kv) &=&\chi_0^{-1}\sqrt{\rho_s^\perp k_\perp^2 + \rho_s^\parallel
k_z^2},\\
\omega_u(\kv) &=&\chi_0^{-1}\sqrt{K k_\perp^4 + B k_z^2},
\end{eqnarray}
\label{omegasPhiU}
\ese
that can be read off from the analytical continuation of the
Lagrangian into real time, $t = -i\tau$. These modes respectively
correspond to the zeroth sound (the Bogoliubov mode as in a
conventional superfluid) and smectic phonon, unique to the LO
state. In cold atomic gases, these should in principle be measurable
via the Bragg spectroscopy
technique\cite{KetterleBragg,Steinhauer02prl,Papp08prl}

With the Goldstone-mode Lagrangian in hand, we can now calculate the
effects of quantum and thermal fluctuations as well as equilibrium
correlation and response functions\cite{commentBerry}.

\section{Larkin-Ovchinnikov state near $h_{c1}$}
\label{sec:LOhc1}

As emphasized earlier the general form of the action for the
description of the LO state, \rfs{Slo} is expected to hold universally
throughout the ordered state. However, its above derivation and
therefore the expressions for the associated couplings
Eqs.\rf{Jmoduli}, \rf{KB},\rf{rhos_pp} are limited to the weak
coupling BCS regime and near the high chemical potential imbalance
(Zeeman field) normal-to-FFLO transition at $h_{c2}$.

An estimate of these couplings outside 1d throughout the FFLO phases
can only be done numerically. However, in the complementary, low
chemical potential imbalance (Zeeman field) regime, just above the
transition from the fully paired superfluid (BCS-BEC) state to the LO
state at $h_{c1}$, qualitative estimates are possible based on an
analysis of a ``dilute gas'' of fluctuating $\pm \Delta$
domain-walls. We carry out this analysis below by focusing on the LO
state, treating it as a periodic array of fluctuating domain-walls in
$\Delta(\rv)$, akin to the lyotropic phases in soft condensed
matter\cite{deGennesProst,ChaikinLubensky}.

\subsection{Macro- vs micro-phase separation: stability of the LO
  state}

Implicit in our analysis below is the assumption that as the
domain-wall surface energy becomes
negative\cite{MachidaNakanishiLO,BurkhardtRainerLO,MatsuoLO,YoshidaYipLO}
for $h > h_{c1}$, their interaction remains {\em repulsive}, and so
the domain-walls proliferate {\em continuously} as a periodic array
inside the LO state. Under this assumptions (that warrants further
study) the domain-wall density $n_{dw}$ and the associated species
imbalance $P\propto n_{dw}$ ($\approx q_0(h)$) is then set by a
balance between the negative surface energy and the domain-wall
repulsion, growing continuously as a function of $h-h_{c1}$ according
to the Pokrovsky-Talapov's commensurate-incommensurate (CI) transition
phenomenology\cite{PokrovskyTalapov}. This behavior is clearly
exhibited in 1d\cite{MachidaNakanishiLO,YangLL} through an exact
solution and bosonization methods, and has been argued to persist in
higher dimensions
\cite{MachidaNakanishiLO,BurkhardtRainerLO,MatsuoLO,YoshidaYipLO}. The
CI route for a transition to the LO state contrasts sharply with the
Landau theory\cite{LO,SRprl,SRaop} of two independent order parameters
$\Delta_0$, $\Delta_{q}$, that always predicts a first-order BCS-LO
transition. The latter corresponds to the case of an {\em attractive}
domain-wall interaction, that therefore proliferate discontinuously
above $h_{c1}$, leading to the ubiquitous phase separation found in
mean-field theory\cite{SRprl,SRaop}. It is currently unclear what
dimensionless microscopic parameter, analogous to Abrikosov's $\kappa$
(distinguishing between type I and type II
superconductors)\cite{deGennes,Tinkham}, controls these two
alternatives of the macro-phase separation (a first-order transition)
and the micro-phase separated LO state (a continuous transition out of
the gapped SF state)\cite{commentCIcontinuous}.

\subsection{SF-LO transition at $h_{c1}$}

The phenomenology of the domain-wall proliferation above $h_{c1}$ and
the associated (fully gapped, singlet) SF to LO transition can be
captured by a domain-wall energy functional, extended to a general
dimension $d$:
\begin{widetext}
\bse
\begin{eqnarray}
E[n_{dw}]/L^{d-1}&=&\int_z(\eps_{dw}^0-h m_N(h)\xi_{dw})n_{dw} +
\frac{1}{2}\int_{z,z'}
V(z-z')n_{dw}(z)n_{dw}(z'),\\
&=&\int_z m_N(h)\xi_{dw}(h_{c1}- h)n_{dw} + \frac{1}{2}\int_{z,z'}
V(z-z')n_{dw}(z)n_{dw}(z'),
\end{eqnarray}
\ese
\end{widetext}
where $\eps_{dw}^0$ is the domain-wall surface energy at $h=0$, $V(z)$
is the domain-wall interaction energy per unit of area, and
$n_{dw}(z)=\sum_i\tilde{\delta}(z-z_i)$ ($\tilde{\delta}(z)$ and $z_i$
are the profile of width $\xi_{dw}$ and the position (along $z$) of
the $i$th domain-wall.  In above, we used an approximate relation
between the domain-wall 1d density $n_{dw}(h)$ and the magnetization
density $m(h)=n_\uparrow-n_\downarrow$
\begin{eqnarray}
m(h)\approx m_N(h)\xi_{dw} n_{dw}(h),
\label{magnetization_ndw}
\end{eqnarray}
where $m_N(h)$ is the magnetization density (per $d$-dimensional
volume) of the normal state, that is approximately nucleated in the
zeros of the (locally BCS-like) gap function, i.e., on the domain-wall
of width $\xi_{dw}$; $m_N \xi_{dw}$ is the fermion number imbalance
per unit area ($L^{d-1}$) per domain-wall. In the Pauli (weak
imbalance) limit, we expect the former to be approximately given by
$k_F^{d-1}h/h_{c1}$, through a string of relations: $m_N
\xi_{dw}\approx \chi_P h\xi_0\approx (n h_{c1}/\epsilon_F)(\hbar
v_F/\Delta_{BCS})(h/h_{c1})\approx k_F^{d-1}h/h_{c1}\approx k_F^{d-1}$
(using $n\approx k_F^d$ ). This is indeed the case in the
noninteracting Fermi gas limit, where $m_N(h)$ is given as the
solution of
\begin{eqnarray}
  m_N(h) &=& \frac{n}{2}\big[(\muh + \hh)^{\frac{3}{2}}
  \Theta(\muh+\hh) -  (\muh- \hh)^{\frac{3}{2}}\Theta(\muh-\hh)\big],
\nonumber\\
\end{eqnarray}
with the normalized chemical potential $\muh\equiv\mu/\epsilon_F$
determined in terms of the normalized chemical potential difference
(Zeeman energy) $\hh\equiv h/\epsilon_F$ by the number density
equation
\begin{eqnarray}
1 &=& \frac{1}{2}\big[(\muh + \hh)^{\frac{3}{2}}
\Theta(\muh+\hh) +  (\muh- \hh)^{\frac{3}{2}}\Theta(\muh-\hh)\big].
\ \ \ \ \end{eqnarray}
Above set reduces to the Pauli expression $m_N\approx \chi_P h$ for
weak $h\ll\mu$ ($\chi_P=3n/(2\epsilon_F)$) and $m_N\approx n$ for
large $h\gg\mu$. In either case the $h$ dependence of $m_N(h)$ is weak
around $h_{c1}$ and therefore can be neglected, along with other weak
$h$ dependences, such as the soliton width $\xi_{dw}(h)$.

Ignoring fluctuations\cite{QMfluct}, the domain-wall interaction
$V(z)\approx V_0 e^{-|z|/\xi_{dw}}$ is expected to be short-ranged
with the scale set by the soliton width $\xi_{dw}$. Thus, for low
soliton density $n_{dw} \xi_{dw}\ll 1$ its strength is a strong
function of $z-z'$, whose typical value is given by the domain-wall
spacing itself $1/n_{dw}$.  Thus, just above $h_{c1}$ a good
approximation for the above energy density is given by
\begin{eqnarray}
\eps(n_{dw})&\approx& m_N(h)\xi_{dw}(h_{c1}- h)n_{dw} 
+ \frac{\xi_{dw}}{2}V(n_{dw}) n_{dw}^2,\nonumber\\
&\approx& m_N(h)\xi_{dw}(h_{c1}- h)n_{dw} 
+ \frac{\xi_{dw}}{2}V_0e^{-\frac{1}{\xi_{dw} n_{dw}}}n_{dw}^2.\nonumber\\
\label{epsT}
\end{eqnarray}
A minimization then gives the domain-wall density, $n_{dw}(h)$ and
(through Eq.\rf{magnetization_ndw}) the imbalance density $m(h)$ as a
function of chemical potential difference $h$:
\begin{eqnarray}
n_{dw}(h)&\approx&
\xi_{dw}^{-1}\begin{cases} 
\frac{1}{\ln[h_0/(h-h_{c1})]}, & \text{for $n_{dw}\xi_{dw} \ll 1$,}\cr
(h-h_{c1})/h_0, & \text{for $n_{dw}\xi_{dw} \gg 1$,}\cr
\end{cases}\nonumber\\
\label{ndwT0}
\end{eqnarray}
where $\xi_{dw}\approx\xi_0$ and $h_0\approx V_0/(\xi_{dw} m_N)\approx
V_0/(\chi_Ph_{c1}\xi_0)\approx\Delta_{BCS}$ (since $V_0\approx
k_F^{d-1} h_{c1}\approx k_F^{d-1}\Delta_{BCS}$ is the energy lost per
fermion per unit of area [with separation $1/k_F$]; for a 1d exact
solution, one indeed has $h_{c1}=2\Delta_{BCS}/\pi$, consistent with
this estimate).

Note that for large imbalance (high domain-wall density) the
interaction part reduces to a simple quadratic dependence on $n_{dw}$
and thus gives the expected linear growth of the density with
$h-h_{c1}$ (second line above). In contrast, for low density the
interaction is exponentially weak and domain-walls enter the state as
a quickly growing (with a divergent slope) function of $h-h_{c1}$,
though not as discontinuously as in a first-order transition.  These
limits are illustrated in Fig.\ref{fig:CItransition}.

\subsubsection{LO elastic moduli at $T=0$}

For $h>h_{c1}$ a 1d lattice of domain-walls, i.e., the LO state forms.
Because of the underlying rotational invariance its elasticity is
generically given by that of a smectic, Eq.\rf{HgmLO3}. The energy of
the deviation $\delta n_{dw}$ from the minimum domain-wall density
$n_{dw}(h)$, Eq.\rf{ndwT0}, is a quadratic function of $\delta
n_{dw}$, given by 
\bse
\begin{eqnarray}
\eps(\delta n_{dw})&\approx&\eps_0 + \oh \eps''(n_{dw}(h))(\delta 
n_{dw})^2,\\
&\approx&\eps_0 + \oh B(h)(\partial_zu)^2,
\end{eqnarray}
\ese
where the smectic bulk modulus is given by
\bse
\begin{eqnarray}
B(h)&\approx& n_{dw}(h)^2\eps''(n_{dw}(h)),\\
&\approx& (h-h_{c1}) n_{dw}(h) k_F^{d-1}.
\label{Bhc1}
\end{eqnarray}
\ese 
In above, to deduce $B(h)$ we used a relation between the density
$n_{dw}$ and the displacement field $\delta n_{dw} =
-n_{dw}(h)\partial_z u$, a relation $m_N\approx k_F^{d-1}/\xi_{dw}$,
and went to a continuum limit via $\int n_{dw}(h) dz\ldots =
\sum_i\ldots$. As expected, $B(h)$ grows strongly just above $h_{c1}$,
but quickly asymptotes to $B(h\gg h_{c1})\approx \Delta_{BCS}
k_F^{d-1}/\xi_0\approx n \Delta_{BCS}/\epsilon_F$, in scale consistent
with its form found below $h_{c2}$, \rfs{KB}.

A detailed estimate of the bend modulus $K$ is more difficult in this
regime. However, we can take advantage of the relation in Eq.\rf{KB}
to deduce $K$ via $K\approx B/q_0^2$ mean-field relation. Thus we find
\bse
\begin{eqnarray}
  K(h)&\approx& B(h)/n_{dw}(h)^2,\\
  &\approx& \Delta_{BCS} k_F^{d-1}\xi_0\approx\epsilon_F k_F^{d-2},
\ \ \text{for}\ \ h\gg h_{c1}.\ \ \ \ \ \ \ \ \ 
\label{Khc1}
\end{eqnarray}
\ese

From these, we deduce the key smectic penetration length
\bse
\begin{eqnarray}
\lambda(h)&=&\sqrt{K(h)/B(h)},\\
&\approx& 1/n_{dw}(h),\\
&\approx& \xi_0,\ \ \text{for}\ \ h\gg h_{c1}.
\label{lambda_hc1}
\end{eqnarray}
\ese

\subsubsection{LO elastic moduli at $T>0$}

Because the microscopic repulsive domain-wall interaction is
short-ranged (exponentially weak at long scales), sufficiently close
to $h_{c1}$ (where solitons are dilute) thermal fluctuations {\em
  always} qualitatively modify the above $T=0$ predictions.  To
understand this we first derive the Helfrich\cite{Helfrich,PokrovskyTalapov}
interaction between two $(d-1)$-dimensional fluctuating, curvature
dominated (tensionless) domain-walls, separated by an average distance
$z$. The corresponding energy functional of the instantaneous local
domain-wall separation $u(\xv)$ is given by:
\begin{eqnarray}
E_{dw}[u(\xv)]=\oh\kappa\int d^{d-1}x(\nabla_\perp^2 u)^2,
\end{eqnarray}
where the curvature modulus is related to that of the smectic via
$\kappa = K/n_{dw}$. The average transverse domain-wall
spacing is determined by the root-mean squared fluctuations, given by:
\bse
\begin{eqnarray}
\langle u^2\rangle \equiv z_T^2 &=& \frac{T}{\kappa}\int\frac{d^{d-1}
  q_x}{q_x^4},\\
&\sim& \frac{T}{\kappa} x_T^{5-d}.
\end{eqnarray}
\ese 
This gives the thermal collision length $x_T$ as function of
separation
\begin{eqnarray}
x_T\approx \left(\frac{\kappa}{T}\right)^{1/(5-d)} z_T^{2/(5-d)}.
\end{eqnarray}

To deduce the entropic (Helfrich) interaction, we note that
domain-walls separated by distance $z$ reduce each others fluctuation
entropy by an amount $s_0$ (of order $1$) per collision. For a pair of
domain-walls of linear extent $L_x$ there are $(L_x/x_T)^{d-1}$
collisions, and thus the entropic part of the free energy is raised
(relative to the infinite separation) by
\begin{eqnarray}
\delta F_T &\approx& T s_0\left(\frac{L_x}{x_T}\right)^{d-1},
\end{eqnarray}
leading to the Helfrich curvature-controlled interaction 
\bse
\begin{eqnarray}
V_H(z)
&\approx&L_x^{d-1}\frac{T\,^\alpha}{\kappa^\beta}\frac{1}{z^\gamma},\\
&\approx& L_x\frac{T\,^{4/3}}{\kappa^{1/3}}\frac{1}{z^{2/3}},\ \
\text{for 2d LO smectic state},\nonumber\\
\end{eqnarray}
\ese
with
\bse
\begin{eqnarray}
\alpha&=&\frac{4}{5-d},\\
\beta&=&\frac{d-1}{5-d},\\
\gamma&=&\frac{2d-2}{5-d}.
\label{abc}
\end{eqnarray}
\ese

Clearly, asymptotically close to the $h_{c1}$ transition, where the
domain-wall array is sufficiently dilute, the above thermal
fluctuations-induced steric interaction $V_H(z)$ always dominates over
the short-range microscopic interaction. The crossover density
$n_{dw}^T$ is set by a separation $1/n_{dw}^T\equiv z^T_{dw}$ at which
these are comparable, given by
\begin{eqnarray}
n^T_{dw}\approx\frac{1}{\xi_{dw}\ln\left(V_0\kappa^\beta
\xi_{dw}^\gamma/T^\alpha\right)}.
\end{eqnarray}
The corresponding Zeeman field range is $h_{c1}<h<h_T\approx
h_{c1}+T^\alpha/(k_F^{d-1}\xi_{dw}^\gamma\kappa^\beta)$, with the energy
density in this regime given by
\bse
\begin{eqnarray}
\eps_T(n_{dw})&\approx& m_N(h)\xi_{dw}(h_{c1}- h)n_{dw} 
+ V_H(n_{dw}) n_{dw},\nonumber\\
&&\\
&\approx& m_N(h)\xi_{dw}(h_{c1}- h)n_{dw} 
+ \frac{T^\alpha}{\kappa^\beta}n_{dw}^{\gamma+1},\nonumber\\
&&
\label{epsT2}
\end{eqnarray}
\ese
A minimization then gives $n_{dw}(h)$ 
\bse
\begin{eqnarray}
n_{dw}(h,T)&\approx&\frac{(m_N \xi_{dw}\kappa^\beta)^{1/\gamma}}
{T^{\alpha/\gamma}}(h-h_{c1})^{1/\gamma},\nonumber\\
&&\ \ \ \ \ \ \ \text{for}\ \ h_{c1}<h<h_T,\\
&\approx&\frac{k_F^{(5-d)/2}\kappa^{1/2}}
{T^{2/(d-1)}}(h-h_{c1})^{\frac{5-d}{2d-2}},\nonumber\\
&&\ \ \ \ \ \ \ \text{for}\ \ h_{c1}<h<h_T.
\label{ndwT}
\end{eqnarray}
\ese

As expected it shows that a fluctuation-enhanced Helfrich domain-wall
repulsion leads to a significantly slower (than the $T=0$,
Eq.\rf{ndwT0}) power-law, $1/\gamma$ increase in the domain-wall
density and therefore of the species imbalance
$P(h)$. Correspondingly, for this low range of Zeeman field $h < h_T$,
this enhances the smectic bulk modulus, $B(h,T)$ through Eq.\rf{Bhc1}.

For $h>h_T$ the microscopic exponential interaction takes over and the
growth of $n_{dw}(h)$ and $B(h,T)$ crossover to that of the $T=0$
result, Eq.\rf{ndwT0}.

\vspace{0.5cm}
\section{Goldstone modes fluctuations}
\label{sec:GMfluctuations}
\subsection{Gaussian fluctuations}

As an estimate of the role of Goldstone modes fluctuations, we first
study them at the Gaussian level, namely approximate the
Goldstone-mode action $S[u,\phi]$ at the quadratic level,
$S^0[u,\phi]=S^0_{sm}[u] + S^0_{sc}[\phi]$ by dropping the
nonlinearities in $\cH_{LO}$, \rf{HgmLO3}. Combining with \rf{Slo} and
focusing on the LO state (leaving the straightforward extension for
the FF state for later), the LO harmonic action is given by
\begin{widetext}
\begin{eqnarray}
  S^0_{LO} &=&\int_0^\beta d\tau\int dz d^{d-1}r_\perp\left[
    \frac{\kappa}{2}(\partial_\tau u)^2
    + \frac{B}{2}(\partial_z u)^2 + \frac{K}{2}(\nabla_\perp^2 u)^2 
    +\frac{\chi}{2}(\partial_\tau\phi)^2 
    + \frac{1}{2}\rho_s^\parallel(\partial_z\phi)^2 
    + \frac{1}{2}\rho_s^\perp(\nabla_\perp\phi)^2\right],\ \ \ \ 
\label{S0}
\end{eqnarray}
\end{widetext}
where (for later mathematical convenience) we generalized the model to
$d$ dimensions, with a single LO modulation ordering axis
$\zh\equiv\hat r_\parallel$ along $\qv_0$ and $d-1$ space $\rv_\perp$
transverse to $\qv_0$, and introduced $\kappa = 2 q_0^2\chi_0$,
$\chi=2\chi_0$.  The harmonic (imaginary time-ordered) correlations of
the decoupled modes $u(\tau,\rv), \phi(\tau,\rv)$ can be easily
computed exactly, for 2-point correlation functions giving
\bse
\begin{eqnarray}
G_{u}(\tau,\rv)&=&\langle u(\tau,\rv)u(0,0)\rangle_0,\\
&=&\frac{1}{\beta}\sum_{\omega_n}\int^{\Lambda_\perp}\frac{d^dk}{(2\pi)^{d}}
\frac{e^{-i\omega_n\tau+i\kv\cdot\rv}}
{\kappa\omega_n^2 + B k_z^2 + K k_\perp^4},
\label{C0uu}\nonumber\\
G_{\phi}(\tau,\rv)&=&
\langle\phi(\tau,\rv)\phi(0,0)\rangle_0,\\
&=&\frac{1}{\beta}\sum_{\omega_n}\int^{\Lambda}\frac{d^dk}{(2\pi)^{d}}
\frac{e^{-i\omega_n\tau+i\kv\cdot\rv}}
{\kappa\omega_n^2 + \rho_s^\parallel k_z^2 + \rho_s^\perp k_\perp^2}.
\label{C0thetatheta}\nonumber
\end{eqnarray}
\ese
The averaging above was done with the Euclidean probability
distribution, $e^{-S^0_{LO}}/Z_0$, using the harmonic action above,
and $\omega_n = 2\pi n/\beta$ is the standard Matsubara frequency.
Above, $\Lambda$ is the UV cutoff for the Goldstone mode action, set
by the inverse coherence length $1/\xi\approx q_0$.  While the full
expression above can be computed asymptotically in terms of special
functions or numerically, it is more revealing to analyze these in
special limits of interest. The simplest measure of fluctuations is
given by the root-mean-squared (rms) fluctuations of these Goldstone
modes, given by $G_{u}(0,0)=\langle u^2\rangle,
G_{\phi}(0,0)=\langle \phi^2\rangle$.

\subsubsection{$T=0$ quantum fluctuations}

At zero temperature, the Matsubara summations in above expressions
reduce to frequency integrals over $\omega$, in $d$ dimensions giving
\bse
\begin{eqnarray}
\langle u^2\rangle_0^{Q}
&=&\int^{\Lambda_\perp}\frac{d\omega d^dk}{(2\pi)^{d+1}}
\frac{1}
{\kappa\omega^2 + B k_z^2 + K k_\perp^4},\ \ \ \ \\
&\approx&\frac{\Lambda_\perp^{d-1}}{(2\pi)^d\sqrt{\kappa
    B}},\ \ 
\text{for $d>1$},\nonumber\\
\langle\phi^2\rangle_0^{Q}
&=&\int^{\Lambda}\frac{d\omega d^dk}{(2\pi)^{d+1}}
\frac{1}
{\kappa\omega^2 + \rho_s^\parallel k_z^2 + \rho_s^\perp k_\perp^2},\\
&\approx&\frac{\Lambda_\perp^{d-1}}{(2\pi)^d
\sqrt{\kappa\rho_s^\parallel}},\ \ 
\text{for $d>1$},\nonumber
\end{eqnarray}
\ese
Because of the integrand's anisotropies, the details of above
expressions are sensitive to microscopic ultraviolet (UV) and infrared
(IR) cutoffs. However, the key unambiguous finding above is that in
any physical dimension of interest here ($d>1$) quantum Goldstone mode
fluctuations in the LO state are finite, set by the short (UV) length
scale and therefore are qualitatively unimportant within the ordered
LO state. That is, the LO state has a nonzero range of stability to
quantum fluctuations, and therefore at $T=0$ is well-approximated by
its mean-field form\cite{1dLLcomment}. Quantum fluctuations will, of
course, give quantitative corrections to the properties of the LO
state and will be important near quantum transitions to other putative
phases.

Above findings are consistent with known results for mathematically
related systems. The action \rf{S0} for the $d$-dimensional $T=0$ LO
quantum phonon, is isomorphic to a classical Hamiltonian of a
generalized $d+1$-dimensional columnar liquid
crystals\cite{deGennesProst,ChaikinLubensky}, characterized by two
spatial ``stiff'' directions ($z$ and $\tau$) and $d-1$ transverse
``soft'' $\rv_\perp$ axes. The latter is known (by simple
power-counting) to exhibit long-range order down to $d+1=5/2$
dimensions, which is consistent with our finding for $\langle
u^2\rangle$, above.  Similarly, \rf{C0thetatheta} is also consistent
with the well-known property of the xy-model exhibiting long-range
order down to $d+1 = 2$ dimensions.  Thus, at $T=0$ we conclude, as
advertised in the Introduction, that a $d+1$ dimensional superfluid
smectic, i.e. the LO ground state is stable to quantum fluctuations
for $d>1$.

\subsubsection{$T>0$ thermal fluctuations}

At nonzero $T$ the Goldstone modes $u$ and $\phi$ exhibit classical
thermal fluctuations. By the identification of the corresponding
sectors with the well-studied smectic and anisotropic xy models, we
can conveniently take advantage of the large body of literature on
these systems\cite{Caille,deGennesProst,ChaikinLubensky}. However, for
completeness we will work out some of the key findings.

Returning to the full expressions for the harmonic correlation
functions, \rf{C0uu}, \rf{C0thetatheta} at nonzero $T$, we separate
out the dominant classical $\omega_{n=0}=0$ contribution
\bse
\begin{eqnarray}
\langle u^2\rangle_0^{T}
&=&\int^{\Lambda_\perp}_{L_\perp^{-1}}\frac{d^dk}{(2\pi)^{d}}
\frac{T}
{B k_z^2 + K k_\perp^4},\ \ \ \ \\
&\approx&
\left\{\begin{array}{ll}
\frac{T}{2\sqrt{B K}}C_{d-1}L_\perp^{3-d},& d < 3,\\
\frac{T}{4\pi\sqrt{B K}}\ln q_0L_\perp,& d = 3,\\
\end{array}\right.
\label{uuT}\\
\langle\phi^2\rangle_0^{T}
&=&\int^{\Lambda_\perp}_{L_\perp^{-1}}\frac{d^dk}{(2\pi)^{d}}
\frac{T}
{\rho_s^\parallel k_z^2 + \rho_s^\perp k_\perp^2},\ \ \ \ \\
&\approx&
\left\{\begin{array}{ll}
\frac{T}{\sqrt{\rho_s^\parallel\rho_s^\perp}}C_{d}
L_\perp^{2-d},& d < 2,\\
\frac{T}{2\pi\sqrt{\rho_s^\parallel\rho_s^\perp}}\ln q_0L_\perp,& d = 2,\\
\end{array}\right.
\label{ttT}
\end{eqnarray}
\ese
where we neglected the subdominant quantum contribution (worked out
above), defined a constant
$C_d=S_d/(2\pi)^d=2\pi^{d/2}/[(2\pi)^d\Gamma(d/2)]$, with $S_d$ a
surface area of a $d$-dimensional sphere, and introduced an infrared
cutoff by considering a system of finite extent $L_\perp\times L_z$,
with $L_z$ the length of the system along the ordering ($z$) axis and
$L_\perp$ transverse to $z$.  Unless it has a huge aspect ratio, such
that $L_z \sim L_\perp^2/\lambda>> L_\perp$, any large system
($L_\perp,L_z >>\lambda$) will have $\lambda L_z \ll L_\perp^2$.

The key observation here is that the smectic phonons exhibit
fluctuations that diverge, growing logarithmically in 3d and linearly
in 2d with system size $L_\perp$. The superconductor phase
fluctuations $\phi$ exhibit well-known 2d logarithmic
divergences\cite{MerminWagner,Hohenberg,KT}. Because (aside from the
superfluid stiffness anisotropy) these are the same as in an ordinary
superconductor and in the bulk 3d case will be finite, indicating a
long-range off-diagonal order, we will focus on the smectic phonon
fluctuations special to the LO (and FF) states.

The expression for the root-mean squared phonon fluctuations in
\rf{uuT} leads the emergence of important crossover length scales
$\xi_\perp,\xi_z$, related by
\bse
\begin{eqnarray}
  \xi_\perp & = & (\xi_z\sqrt{K/B})^{1/2},\\
  &\equiv&\sqrt{\xi_z\lambda},
\label{xi_pxi_z}
\end{eqnarray}
\ese
that characterize the finite-temperature LO state. These are defined
as scales $L_\perp,L_z$ at which phonon fluctuations are large,
comparable to LO period $a=2\pi/q_0$. Namely, setting
\begin{equation}
\langle u^2\rangle_0^{T}\approx a^2
\end{equation}
in \rfs{uuT} we find
\bse
\begin{eqnarray}
\xi_\perp&\approx&
\left\{\begin{array}{ll}
\frac{a^2\sqrt{B K}}{T}\sim\frac{K}{T q_0},& d = 2,\\
a e^{4\pi a^2\sqrt{B K}/T}\sim a e^{\frac{c K}{T q_0}},& d = 3,\\
\end{array}\right.
\end{eqnarray}
\label{xiperp}
\ese
where in the second form of the above expressions we took the simplest
approximation for the smectic anisotropy length $\lambda=\sqrt{K/B}$
to be $\lambda = a \sim 1/q_0$, and introduced an order $1$ Lindemann
constant $c$\cite{Lindemann}, that depends on the somewhat arbitrary
definition of ``large'' phonon rms fluctuations.

The thermal connected correlation function of LO phonons
\begin{equation}
C_{u}(\rv_\perp,z)
=\langle\left[u({\bf r_\perp},z)-u({\bf 0},0)\right]^2\rangle_0\;.
\label{C_T}
\end{equation}
is also straightforwardly worked out, in 3d giving the logarithmic
Caill\'e form\cite{Caille}
\bse
\begin{eqnarray}
C^{3d}_{u}(\rv_\perp,z)&=&2T\int{d^2{q_\perp}d
q_z\over(2\pi)^3}{1-e^{i{\bf q}\cdot{\bf r}}\over K  q_\perp^4 + B
q_z^2}\;,\nonumber\\
&=&{T\over2\pi\sqrt{K  B}}\;g^{3d}_T\left({z\lambda\over r_\perp^2},
{r_\perp\over a}\right)\;,\nonumber\\
&=&{T\over2\pi\sqrt{K  B}}
\left[\ln\left({r_\perp\over a}\right)-
\frac{1}{2}{\text{Ei}}\left({-r_\perp^2\over 4\lambda|z|}\right)\right],
\ \ \ \ \ \ \ \ \ \\
&\approx&{T\over2\pi\sqrt{K  B}}\left\{\begin{array}{lr}
\ln\left({r_\perp\over a}\right),&r_\perp\gg\sqrt{\lambda|z|}\;,\\
\ln\left({4\lambda z\over a^2}\right),&r_\perp\ll\sqrt{\lambda|z|}\;,\\
\end{array}\right.
\label{Cuu3dT0}
\end{eqnarray}
\ese
where $\text{Ei}(x)$ is the exponential-integral function.  As
indicated in the last form, in the asymptotic limits of
$r_\perp\gg\sqrt{\lambda z}$ and $r_\perp\ll\sqrt{\lambda z}$ above 3d
correlation function reduces to logarithmic growth with $r_\perp$ and
$z$, respectively.

In 2d we instead have\cite{TonerNelsonSm}
\bse
\begin{eqnarray}
C^{2d}_{u}(x,z)&=&2T\int{d{q_x}d
q_z\over(2\pi)^2}{1-e^{i{\bf q}\cdot{\bf r}}\over K  q_x^4 + B
q_z^2}\;,\nonumber\\
&=&{T\over2\pi\sqrt{K  B}}\;g^{2d}_T\left({z\lambda\over x^2},
{x\over a}\right)\;,\nonumber\\
&=&{2T\over B}
\bigg[\left(\frac{|z|}{4\pi\lambda}\right)^{1/2}
  e^{-x^2/(4\lambda|z|)}\nonumber\\
&&+ \frac{|x|}{4\lambda}\mbox{erf}
\big(\frac{|x|}{\sqrt{4\lambda|z|}}\big)\bigg]\ \ \ \ \ \ \ \\
&\approx&{2T\over B}\left\{\begin{array}{lr}
\left(\frac{|z|}{4\pi\lambda}\right)^{1/2},
&x\ll\sqrt{\lambda|z|}\;,\ \ \ \ \\
\frac{|x|}{4\lambda}, &x\gg\sqrt{\lambda|z|}\;,\ \ \ \ \\
\end{array}\right.
\label{Cuu2dT0}
\end{eqnarray}
\ese
where $\text{erf}(x)$ is the Error function. 

Above finding of the divergence of smectic phonon fluctuations at
nonzero temperature have immediate drastic implications for the
properties of the LO (and FF) states. As emphasized in the
Introduction, the most important of these is that the thermal average
of the Landau's LO order parameters \rf{DeltaLO} vanishes in
thermodynamic limit
%
\begin{eqnarray}
\langle\Delta_{LO}(\rv)\rangle_0
&=&2\Delta_{q_0}\langle e^{i\phi}
\cos\big[\qv_0\cdot\rv + \theta\big]\rangle_0,\nonumber\\
&=&2\Delta_{q_0}e^{-\oh\langle\phi^2\rangle_0-\oh q_0^2\langle u^2\rangle_0}
\cos\big(\qv_0\cdot\rv),\nonumber\\
&=&2\tilde\Delta_{q_0}(L_\perp)\cos\big(\qv_0\cdot\rv),\nonumber\\
\label{DeltaLOave2b}
\end{eqnarray}
%
with the thermally suppressed order parameter amplitude given by
\bse
\begin{eqnarray}
\hspace{-1cm}
\tilde\Delta_{q_0}(L_\perp)&=&
\Delta_{q_0}e^{-\oh\phi^2_{rms}}
\left\{\begin{array}{ll}
e^{-L_\perp/\xi_\perp},& d = 2,\\
\left(\frac{a}{L_\perp}\right)^{\eta/2},& d = 3,\\
\end{array}\right.\\
&\rightarrow &0,\ \ \mbox{for $L_\perp\rightarrow\infty$},
\label{DeltaR}
\end{eqnarray}
\ese 
where $\phi_{rms}^2\equiv\langle\phi^2\rangle_0$. In above we used
results for the phonon and phase fluctuations, \rf{uuT}, \rf{ttT} and
defined the Caill\'e exponent
\begin{eqnarray}
  \eta&=&\frac{q_0^2 T}{8\pi\sqrt{B K}}.
\end{eqnarray}
We also neglected the subdominant quantum phonon fluctuations and
included all finite (quantum and thermal) superconducting phase
fluctuations inside the nonzero Debye-Waller factor,
$e^{-\oh\phi^2_{rms}}$. Thus, in qualitative contrast to its
mean-field description, at long scales (longer than $\xi_{\perp,z}$) a
LO state is characterized by a {\em uniform} superconducting order
parameter and density.

To further characterize a finite-temperature LO state, we compute
the Cooper-pair momentum distribution function, 
\begin{widetext}
\bse
\begin{eqnarray}
  n^{LO}_\kv &=&\int_{\rv,\rv'}
\langle\Delta^*_{LO}(\rv)\Delta_{LO}(\rv')\rangle
  e^{i\kv\cdot\rv - i\kv\cdot\rv'},\\
  &\approx& \sum_{q_n = n q_0}|\Delta_{q_n}|^2\int_{\rv,\rv'}
  \langle e^{i(\phi_\rv - \phi_\rv')}e^{i q_n (u_{\rv'} - u_{\rv})}\rangle
  e^{i\kv\cdot(\rv - \rv')+i q_n (z-z')},\\
  &\approx& \sum_{q_n = n q_0}V|\Delta_0|^2\int_{\rv}
   e^{-\oh C_\phi(\rv)}  e^{-\oh q_n^2 C_{u}(\rv)}e^{i(\kv-q_n\zh)\cdot\rv},
\end{eqnarray}
\label{nk}
\ese
\end{widetext}
where $V$ is the system's volume and we generalized the LO order
parameter to include high harmonics $q_n$.  To lowest order we
neglected the subdominant coupling between the superfluid phase and
the LO phonons, ignored the nonlinear effects, approximating $u$ and
$\phi$ as Gaussian fields. The asymptotic form of $n_\kv^{LO}$ can
then be readily obtained analytically. As we saw above, in 3d the
superfluid phase, $\phi$ exhibits only finite fluctuations about an
ordered state, with
\begin{eqnarray}
C_{\phi}^{3d}(r\gg\xi)&\approx&\frac{T}{2\xi\sqrt{\rho_s^\parallel\rho_s^\perp}},
\end{eqnarray}
where we took the coherence length $\xi$ as the short-scale cutoff. 
These phase fluctuations then simply give a finite Debye-Waller factor
suppression of the momentum distribution function amplitude,
corresponding to the usual fluctuation-driven condensate depletion,
with a fraction of atoms pushed out of the $\qv_0$ condensate.

In contrast, the 3d LO phonon fluctuations diverge logarithmically,
\rf{Cuu3dT0}, strongly modifying $n_\kv$ from its mean-field
$\delta$-function form. We thus find $n_\kv$ to exhibit a power-law
peak around the ordering wavevector $\qv_0$ (and its harmonics,
$\qv_n$), reminiscent of (1+1)d Luttinger liquids and two-dimensional
crystals\cite{Landau1dsolid,Peierls1dsolid,MerminWagner,KT}
\begin{eqnarray}
  n^{LO}_\kv &\approx& \sum_{q_n\neq0}\frac{n_{q_n}}{|k_z - n
    q_0|^{2-n^2\eta}},\ \ \mbox{for $d=3$},
\label{nkresult}
\end{eqnarray}
where for simplicity we specialized to $\kv = k_z\zh$. The form-factor
amplitude is approximately given by $n_{q_n}\approx V|\Delta_{q_n}|^2
e^{-\frac{T}{4\xi\sqrt{\rho_s^\parallel\rho_s^\perp}}}$.

\begin{figure}[tbp]
\vskip0.25cm 
\epsfig{file=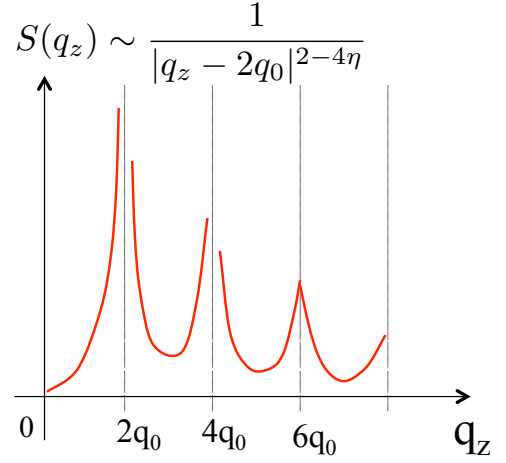,width=6.5cm,angle=-90}
\caption{The structure function, $S(\qv)$ for the 3d LO state,
  displaying power-law (as opposed to $\delta$-function) Bragg peaks,
  characteristic of the LO superfluid's spatial quasi-long-range
  order.}
\label{fig:Sq}
\end{figure}

An additional characterization of the LO state is through a
structure function, $S(\qv)$, a Fourier transform of the density
correlation function, that in 3d is given by
\bse
\begin{eqnarray}
S^{LO}(\qv)&=&\int d^3r\langle\delta\rho(\rv)\delta\rho(0)\rangle e^{-i\qv\cdot\rv},\\
&\approx&\int d^3r\langle|\Delta_{LO}(\rv)|^2|\Delta_{LO}(0)|^2\rangle
e^{-i\qv\cdot\rv},\\
&\approx& \sum_{q_n}|\Delta_{q_n}|^4\int_{\rv}
\langle e^{i 2 q_n (u_0 - u_{\rv})}\rangle_0
  e^{-i(\qv - 2q_n\zh)\cdot \rv},\ \ \ \ \ \ \ \ \ \\
&\approx&\sum_{n}\frac{|\Delta_{q_n}|^4}{|q_z - 2n q_0|^{2-4n^2\eta}}, 
\ \ \mbox{for $d=3$},
\end{eqnarray}
\label{Sq}
\ese
where we approximated phase and phonon fluctuations by Gaussian
statistics (in 3d valid up to weak logarithmic corrections\cite{GP})
and replaced atomic density fluctuations by (twice) the LO condensate
density. The latter neglects a contribution from the imbalanced atoms,
without qualitatively modifying the result (since atomic density is
locked to the condensate density) and is furthermore quantitatively
subdominant for $h\ll h_{c2}$, where the imbalance is low. As for
$n_{\kv}$, we find that the logarithmically divergent 3d phonon
fluctuations lead to a structure function, with highly anisotropic
($q_z\sim q_\perp^2/\lambda$) quasi-Bragg peaks (see Fig.\ref{fig:Sq})
replacing the true Bragg peaks characteristic of the mean-field
long-range periodic order. These predictions are a reflection of the
well-known\cite{deGennesProst,ChaikinLubensky} and experimentally
tested\cite{SqSmExp} behavior of conventional smectic liquid crystals.

In two dimensions, the LO order is even more strongly suppressed
thermal fluctuations. The linear growth of the 2d phonon fluctuations
leads to exponentially short-ranged correlations in the LO order
parameter and in the density. Because it is the ``soft'' smectic
Goldstone mode that is responsible for these interesting properties
they are necessarily also shared by the FF state\cite{Shimahara98}.

As discussed in the Introduction, another fascinating consequence of
the thermal vanishing of $\langle\Delta_{LO}\rangle$, \rf{DeltaR} is
that the leading nonzero Landau order parameter characterizing the LO
state is the translationally-invariant ``charge''-4 (4-atom pairing)
superconducting order parameter, $\Delta_{sc}$, introduced in
\rf{Deltasc}. Thus in the presence of thermal fluctuations the LO
phase corresponds to an exotic state in which the off-diagonal order
is exhibited by pairs of Cooper pairs, i.e., a bound quartet of atoms,
rather than by the conventional 2-atom Cooper
pairs\cite{Berg09nature}. In 2d and 3d this higher order pairing is
driven by arbitrarily low-$T$ fluctuation, rather than by a fine-tuned
attractive interaction between Cooper pairs, and therefore has no
mean-field description. We will discuss these and other
fluctuation-induced phases as well as transitions between them in
Sec.\ref{sec:transitions}.

\subsection{Nonlinear elasticity: beyond Gaussian fluctuations}
\label{elastRG}
\subsubsection{Perturbation theory}
As is clear from the derivation of the previous subsection, the
restoration of the translational symmetry (a uniform LO state with a
vanishing $\langle\Delta_{LO}(\rv)\rangle$, etc.) by thermal
fluctuations is a robust prediction of the quadratic theory, that
cannot be overturned by the left-out nonlinearities. However, the
asymptotic form of the correlation functions computed within the
harmonic approximation only extends out to the nonlinear length scales
$\xi_{\perp,z}$, beyond which the divergently large LO phonon
fluctuations invalidate the neglect of the nonlinear phonon operators
\begin{eqnarray}
  \cH_{\text{nonlinear}}&=&-\oh B (\partial_z u)(\nabla u)^2 
  +\frac{1}{8} B (\nabla u)^4.\ \ \ \
\label{Hnonlin}
\end{eqnarray}
These will necessarily qualitatively modify predictions
\rf{Cuu3dT0},\rf{Cuu2dT0}, \rf{nkresult}, and \rf{Sq} on scales longer
than the crossover scales $\xi^{NL}_{\perp,z}$, that we compute next.
\begin{figure}[htbp]
\vskip0.25cm 
\epsfig{file=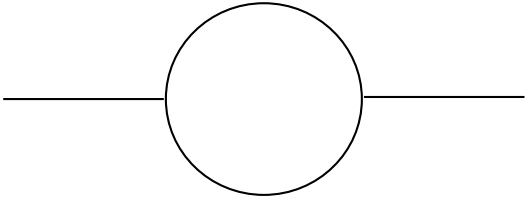,width=7.cm,angle=0}
\caption{Feynman graph that renormalizes the elastic moduli $K$, $B$
  of the LO superfluid.}
\label{fig:deltaBloop}
\end{figure}

To see this, we use a perturbative expansion in the nonlinear
operators \rf{Hnonlin} to assess the size of their contribution to
e.g., the free energy. Following a standard field-theoretic analysis
these can be accounted for as corrections to the compressional $B$ and
bend $K$ elastic moduli, with the leading contribution to $\delta B$,
summarized graphically in Fig.\ref{fig:deltaBloop}, and given by
\bse
\begin{eqnarray}
\delta B&=&-\oh T B^2\int_{\bf q} q_\perp^4
G_{u}({\bf q})^2\;,\label{deltaBa}\\
&\approx&-\oh T B^2\int_{-\infty}^{\infty}{d q_z\over2\pi}
\int_{L^{-1}_\perp}{d^{d-1}q_\perp\over(2\pi)^{d-1}}
{q_\perp^4\over(K q_\perp^4+B q_z^2)^2}\;,\label{deltaBb}\nonumber\\
&\approx&-{1\over8} {C_{d-1}T\over 3-d}\;
\left({B\over K^3}\right)^{1/2} L_\perp^{3-d}B\;.\label{deltaBc}
\end{eqnarray}
\ese
In above, we neglected the subdominant contribution from quantum
fluctuations, i.e., used thermal equal-time correlator, $G_{u}(\qv)$,
focused on $d\leq3$ (which allowed us to drop the uv-cutoff
($\Lambda$) dependent part that vanishes for
$\Lambda\rightarrow\infty$), and cutoff the divergent contribution of
the long wavelength modes via the infra-red cutoff $q_\perp>1/L_\perp$
by considering a system of a finite extent $L_\perp$.  Clearly the
anharmonicity become important when the fluctuation corrections to the
elastic constants (e.g., $\delta B$ above) become comparable to the
bare microscopic values. The divergence of this correction as
$L_\perp\rightarrow\infty$ signals the breakdown of the conventional
harmonic elastic theory on length scales longer than a crossover scale
$\xi^{NL}_\perp$
\begin{eqnarray}
\xi^{NL}_\perp&\approx&
\left\{\begin{array}{ll}
\frac{1}{T}\left(\frac{K^3}{B}\right)^{1/2},& d = 2,\\
a e^{\frac{c}{T}\left(\frac{K^3}{B}\right)^{1/2}},& d = 3,
\end{array}\right.
\label{xiNL}
\end{eqnarray}
which we define here as the value of $L_\perp$ at which $|\delta
B(\xi^{NL}_\perp)|=B$. Within the approximation of the smectic
screening length $\lambda = a$, these nonlinear crossover lengths
reduce to the phonon disordering lengths \rf{xiperp},\rf{xi_pxi_z},
defined by a Lindemann-like criterion.  Clearly, on scales longer than
$\xi^{NL}_{\perp,z}$ the perturbative contributions of nonlinearities
diverge and therefore cannot be neglected. Their contribution are thus
expected to qualitatively modify the harmonic predictions of the
previous subsection.

\subsubsection{Renormalization group analysis in $d=3-\epsilon$ dimensions} 
\label{sec:RG}

To describe the physics beyond the crossover scales,
$\xi^{NL}_{\perp,z}$ -- i.e., to make sense of the infra-red divergent
perturbation theory found in Eq.\ref{deltaBc}\ -- requires a
renormalization group analysis. This was first performed in the
context of conventional liquid crystals and Lifshitz points in a
seminal work by Grinstein and Pelcovits (GP)\cite{GP}. For
completeness, we complement GP's treatment with Wilson's
momentum-shell renormalization group (RG) analysis, extending it to an
arbitrary dimension $d$, so as to connect to the behavior in 2d, that
has an exact solution\cite{GW}.

To this end we integrate (perturbatively in $\cH_{\text{nonlinear}}$)
short-scale Goldstone modes in an infinitesimal cylindrical shell of
wavevectors, $\Lambda e^{-\delta\ell}<q_\perp<\Lambda$ and
$-\infty<q_z<\infty$ ($\delta\ell\ll 1$ is infinitesimal).  The
leading perturbative momentum-shell coarse-graining contributions come
from terms found in direct perturbation theory above, but with the
system size divergences controlled by the infinitesimal momentum
shell. The thermodynamic averages can then be equivalently carried out
with an effective coarse-grained Hamiltonian of the same form
\rf{HgmLO3}, but with all the couplings infinitesimally corrected by
the momentum shell. For smectic moduli $B$ and $K$ this gives
\bse
\begin{eqnarray}
\delta B&\approx&-\frac{1}{8} g B\delta\ell,\\
\delta K&\approx&\frac{1}{16} g K\delta\ell,
\end{eqnarray}
\label{deltaBdeltaK}
\ese
where dimensionless coupling is given by
\bse
\begin{eqnarray}
g&=&C_{d-1}\Lambda_\perp^{3-d}T\left({B\over K^3}\right)^{1/2}\;,\\
&\approx&\frac{T}{2\pi}\left({B\over K^3}\right)^{1/2}\;,
\label{g}
\end{eqnarray}
\ese
and in the second form we approximated $g$ by its value in 3d.
Because it is only the smectic nonlinearities that are qualitatively
important, all other stiffnesses do not experience corrections that
accumulate at long scales. Eqs.\rf{deltaBdeltaK} show that $B$ is
softened and $K$ is stiffened by the nonlinearities in the presence of
thermal fluctuations, making the system effectively more isotropic.

\begin{figure}[bth]
\vspace{0.5cm}
\centering
\setlength{\unitlength}{1mm}
\begin{picture}(30,30)(0,0)
\put(-30,50){\begin{picture}(0,0)(0,0)
\includegraphics{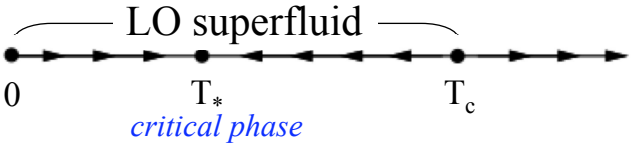}
\end{picture}}
\end{picture}
\vspace{-.5cm}
\caption{Renormalization group flow for a LO state in
  $d<3$-dimensions, illustrating that at low T it is a ``critical
  phase'' displaying universal power-law phenomenology, controlled by
  a nontrivial infrared stable fixed point.}
\label{fig:RGflow}
\end{figure}

For convenience we then rescale the lengths and the remaining long
wavelength part of the fields $u^<({\bf r})$ according to
$r_\perp=r_\perp'e^{\delta\ell}$, $z=z'e^{\omega\delta\ell}$ and
$u^<({\bf r})= e^{\phi\delta\ell}u'({\bf r'})$, so as to restore the
ultraviolet cutoff $\Lambda_\perp e^{-\delta\ell}$ back up to
$\Lambda_\perp$. The underlying rotational invariance insures that the
graphical corrections preserve the rotationally invariant strain
operator $\big(\partial_z u - \oh({\nabla}_\perp u)^2\big)$,
renormalizing it as a whole. It is therefore convenient (but not
necessary) to choose the dimensional rescaling that also preserves
this form. It is easy to see that this choice leads to
\begin{equation}
\phi=2-\omega\;.\label{chi_choice}
\end{equation}
The leading (one-loop) changes to the effective coarse-grained and
rescaled action can then be summarized by differential RG flows 
\bse
\begin{eqnarray}
\frac{d B(\ell)}{d\ell}&=&(d+3-3\omega-{1\over8}g(\ell))B(\ell)
\;,\label{Bflow}\\
\frac{d K(\ell)}{d\ell}&=&(d-1-\omega+{1\over16}g(\ell))K(\ell)
\;.\label{Kflow}
\end{eqnarray}
\ese 
From these we readily obtain the flow of the dimensionless coupling
$g(\ell)$
\begin{eqnarray}
\frac{d g(\ell)}{d\ell}&=&(3-d)g-\frac{5}{32}g^2\;,\label{g_flow}
\end{eqnarray}
whose flow for $d<3$ away from the $g=0$ Gaussian fixed point encodes
the long-scale divergences found in the direct perturbation theory
above. As summarized in Fig.\ref{fig:RGflow} for $d<3$ the flow
terminates at a nonzero fixed-point coupling
$g_*=\frac{32}{5}\epsilon$ (with $\epsilon\equiv 3-d$), that
determines the nontrivial long-scale behavior of the system (see
below). As with treatments of critical points\cite{WilsonKogut}, but
here extending over the whole LO phase, the RG procedure is
quantitatively justified by the proximity to $d=3$, i.e., smallness of
$\epsilon$.

We can now use a standard matching calculation to determine the
long-scale asymptotic form of the correlation functions on scales
beyond $\xi^{NL}_{\perp,z}$. Namely, applying above coarse-graining RG
analysis to a computation of correlation functions allows us to relate
a correlation function at long length scales of interest to us (that,
because of infrared divergences is impossible to compute via a direct
perturbation theory) to that at short scales, evaluated with
coarse-grained couplings, $B(\ell)$, $K(\ell)$,\ldots. In contrast to
the former, the latter is readily computed via a perturbation theory,
that, because of shortness of the length scale is convergent. The
result of this matching calculation to lowest order gives correlation
functions from an effective Gaussian theory
\begin{eqnarray}
G_{u}(\tau=0,\kv)
&\approx&\frac{T}{B(\kv) k_z^2 + K(\kv) k_\perp^4},
\label{Guu}
\end{eqnarray}
with moduli $B(\kv)$ and $K(\kv)$ that are singularly
wavevector-dependent, latter determined by the solutions $B(\ell)$ and
$K(\ell)$ of the RG flow equations \rf{Bflow} and \rf{Kflow} with
initial conditions the microscopic values $B$ and $K$, e.g., as given
by the BCS predictions, \rfs{KB}.

{\em 2d analysis:} In $d=2$, at long scales $g(\ell)$ flows to a
nontrivial infrared stable fixed point $g_*=32/5$, and the matching
analysis predicts correlation functions characterized by anisotropic
wavevector-dependent moduli
\bse
\begin{eqnarray}
K({\bf k})&=&K\left(k_\perp\xi^{NL}_\perp\right)^{-\eta_K}
f_K(k_z\xi^{NL}_{z}/(k_\perp\xi^{NL}_\perp)^\zeta)\;,\label{Kg}
\ \ \ \ \ \ \ \ \ \\
&\sim& k_\perp^{-\eta_K},\nonumber\\
B({\bf k})&=&B\left(k_\perp\xi^{NL}_\perp\right)^{\eta_B}
f_B(k_z\xi^{NL}_{z}/(k_\perp\xi^{NL}_\perp)^\zeta)\;,\label{Bg}\\
&\sim& k_\perp^{\eta_B}.\nonumber
\end{eqnarray}
\label{KgBg}
\ese
Thus, on scales longer than $\xi^{NL}_{\perp,z}$ these qualitatively
modify the real-space correlation function asymptotics of the harmonic
analysis in the previous subsection.  In Eqs.\rf{KgBg} the universal
anomalous exponents are given by
\bse
\begin{eqnarray}
\eta_B&=&{1\over8}g_*={4\over 5}\;\epsilon\;,\label{etaB2}\nonumber\\
&\approx&{4\over5}\;,\;\;\;\mbox{for}\; d=2\;,\label{etaB3}\\
\eta_K&=&{1\over16}g_*={2\over 5}\;\epsilon\;,\label{etaK2}\nonumber\\
&\approx&{2\over5}\;,\;\;\;\mbox{for}\; d=2\;,\label{etaK3}
\end{eqnarray}
\ese 
determining the $z-\rv_\perp$ anisotropy exponent via \rf{Guu} to be
\bse
\begin{eqnarray}
\zeta&\equiv& 2-(\eta_B+\eta_K)/2\;,\\
&=&\frac{7}{5},
\end{eqnarray}
\label{zeta}
\ese
as expected reduced by thermal fluctuations down from its harmonic
value of $2$. The $\kv_\perp-k_z$ dependence of $B(\kv),K(\kv)$ is
determined by universal scaling functions, $f_B(x),f_K(x)$ that we
will not compute here. The underlying rotational invariance (special
to a LO state realized in an isotropic trap) gives an {\em exact}
relation between the two anomalous $\eta_{B,K}$
exponents\cite{commentWI}
\bse
\begin{eqnarray}
3-d &=& {\eta_B\over 2} + {3\over  2}\eta_K\;,
\label{WI}\\
1  &=& {\eta_B\over 2} + {3\over 2}\eta_K\;,\ \ \mbox{for
  $d=2$},
\label{WI2d}
\end{eqnarray}
\ese
which is obviously satisfied by the anomalous exponents,
Eqs.\rf{etaK2},\rf{etaB2}, computed here to first order in
$\epsilon=3-d$\cite{commentWI}.

Thus, as advertised in the Introduction, we find that a finite
temperature 2d LO state is highly nontrivial and qualitatively
distinct from its mean-field perfectly periodic form. In addition to a
vanishing LO order parameter and associated fluctuation-restored
translational symmetry, it is characterized by a universal nonlocal
length-scale dependent moduli, \rfs{KgBg}. Consequently its Goldstone
mode theory and the associated correlations are not describable by a
local field theory, that is an analytic expansion in local field
operators.  Instead, in 2d, on length scales beyond
$\xi^{NL}_{\perp,z}$ thermal fluctuations and correlations of the LO
state are controlled by a nontrivial fixed point, characterized by
universal anomalous exponents $\eta_{K,B}$ and scaling functions
$f_{B,K}(x)$ defined above.

Above we obtained this nontrivial structure from an RG analysis and
estimated these exponents within a controlled but approximate
$\epsilon$-expansion. Remarkably, in 2d an exact solution of this
problem was discovered by Golubovic and Wang\cite{GW}. It predicts an
anomalous phenomenology in a qualitatively agreement with the RG
predictions above, and gives exact exponents
\bse
\begin{eqnarray}
\eta_B^{2d}&=&1/2,\\
\eta_K^{2d}&=&1/2,\\
\zeta^{2d}&=&3/2.
\end{eqnarray}
\label{etaGW2d}
\ese 

{\em 3d analysis:} In $d=3$, the nonlinear coupling $g(\ell)$ is
marginally irrelevant, flowing to $0$ at long scales. Despite this,
the marginal flow to the Gaussian fixed point is sufficiently slow
(logarithmic in lengths) that (as usual at a marginal
dimension\cite{WilsonKogut}) its power-law in $\ell$ dependence leads
to a universal, asymptotically {\em exact} logarithmic wavevector
dependence\cite{GP}
\bse
\begin{eqnarray}
  K({\kv_\perp,k_z=0})&\sim&K|1+
\frac{5g}{64\pi}\ln(1/k_\perp a)|^{2/5}\;,\label{K3d}\\
  B({\kv_\perp=0,k_z})&\sim&B|1+\frac{5g}{128\pi}
\ln(\lambda/k_za^2)|^{-4/5}.\ \ \ \ \ \ \ \ \ \ \label{B3d}
\end{eqnarray}
\label{KB3d}
\ese
This translates into an equal-time LO order parameter correlations
given by 
\bse
\begin{eqnarray}
n(z,\rv_\perp=0)&=& \langle\Delta^*_{LO}(\rv)\Delta_{LO}(0)\rangle,\\
&\sim&e^{-c_1(\ln z)^{6/5}}\cos(q_0 z),
\end{eqnarray}
\label{SrGP}
\ese
($c_1$ a nonuniversal constant) as discovered in the context of
conventional smectics by Grinstein and Pelcovits\cite{GP}.  Although
these 3d anomalous effects are less dramatic and likely to be
difficult to observe in practice, theoretically they are quite
significant as they represent a qualitative breakdown of the
mean-field and harmonic descriptions, that respectively ignore
interactions and thermal fluctuations.


We conclude this section by noting that all of the above analysis is
predicated the validity of the purely elastic model, \rfs{HgmLO3}, that
neglects topological defects, such as vortices and dislocations. If
these unbind (as they undoubtedly do in 2d at any nonzero
temperature\cite{TonerNelsonSm}), then our above prediction only hold
on scales shorter than the separation $\xi_v$, $\xi_d$ between these
defects.

\section{Topological defects in a Larkin-Ovchinnikov state}
\label{sec:defects}

We now turn to the discussion of the topological defects, followed in
the subsequent section by an analysis of phases and transitions
accessible by their unbinding. As discussed in Sec.\ref{sec:GM} the LO
superfluid is distinguished by two independent order parameter
components $\Delta_{\pm\qv_0}$, corresponding to $\pm\qv_0$ finite
center of mass momentum pairing\cite{comment2Deltas}. These complex
order parameters, $\Delta_{\pm\qv_0}=|\Delta_{q_0}|e^{i\phi_\pm}$ in
turn lead to two independent phase Goldstone modes $\phi_\pm$ (or
equivalently $\phi,\theta=-q_0 u$, \rf{phitheta}), controlled by a
long-scale Hamiltonian, that at harmonic level is given by
\begin{widetext}
\bse
\begin{eqnarray}
H_{LO}^0&=&\int dz d^{d-1}r_\perp\left[\sum_{\sigma=\pm}
\left(\frac{\rho_s^\parallel}{4}(\partial_z\phi_\sigma)^2 + \frac{K}{4q_0^2}(\nabla_\perp^2\phi_\sigma)^2\right)
 + \frac{\rho_s^\perp}{8}(\nabla_\perp\phi_++\nabla_\perp\phi_-)^2
\right],\ \ \ \ \\
  &=&\int dz d^{d-1}r_\perp\left[
\frac{B}{2}(\partial_z u)^2 + \frac{K}{2}(\nabla_\perp^2 u)^2 
    + \frac{1}{2}\rho_s^\parallel(\partial_z\phi)^2 
    + \frac{1}{2}\rho_s^\perp(\nabla_\perp\phi)^2\right],\ \ \ \ 
\label{H0loSum}
\end{eqnarray}
\label{H0lo}
\ese
\end{widetext}
with the couplings given in Eqs.\rf{KB},\rf{rhos_pp} and in obtaining
\rfs{H0lo} we dropped the subdominant higher gradient ($K$) term in
$\phi$. The analysis of this Hamiltonian in the absence of topological
defects was discussed in the previous section. We now use it to
understand the energetics of defects beyond that ``spin-wave''
approximation.

As in an ordinary superfluid, because $\phi_\pm$ are {\em compact}
phase fields ($\phi_\pm$ and $\phi_\pm + 2\pi$ are physically
identified), in addition to their smooth configurations, there are
vortex topological excitations, corresponding to nonsingle-valued
configurations of $\phi_\pm(\rv)$. These are defined by two
corresponding integer-valued closed line integral enclosing a vortex
line
\begin{eqnarray}
\oint d\vec{\ell}\cdot\vec{\nabla}\phi_\pm = 2\pi n_\pm,
\end{eqnarray}
that we collectively designate by a two component integer-valued
vector $\Nv_v = (n_+,n_-)$, with $n_\pm\in \mathbb{Z}$.  These integer
vector defects, $\Nv_v$ are associated with the fundamental group
$\Pi_1$ of the torus $U(1)\otimes U(1)$,\cite{commentZ2} that
characterizes the low-energy manifold of Goldstone modes of the LO
state. In this respect the LO superfluid has similarities to other
$U(1)\otimes U(1)$ systems, such as easy-plane spinor-1
condensates\cite{PodolskyS1prb} and two-gap superconductors, e.g.,
MgB$_2$\cite{Babaev02prl}.

In a differential form, the line defects are equivalently encoded as
\begin{equation}
\nabla\times \nabla\phi_\pm={\bf m}_\pm\;,\label{vortexpm}
\end{equation}
with vortex line topological ``charge'' density given by
\begin{equation}
{\bf m}_\pm(\rv)=2\pi\sum_i\int n^i_\pm \hat{\bf t}_i(s_i)
\delta^3({\bf r}-{\bf r}_i(s_i)) d s_i\;,
\label{m_pm}
\end{equation}
where $s_i$ parameterizes the $i$'th vortex line (or loop), ${\bf
  r}_i(s_i)$ gives its positional conformation, $\hat{\bf t}_i(s_i)$
is the local unit tangent, and vortex ``charges'' $n^i_\pm$ are
independent of $s_i$, since the charge of a given line is constant
along the defect. Furthermore,
\begin{equation}
\nabla\cdot{\bf m}({\bf r})=0
\label{continuity}
\end{equation}
enforces the condition that vortex lines cannot end in the bulk of the
sample; they must either form closed loops or extend entirely through
the system.

\subsection{Vortices and dislocations}

\begin{figure}[bth]
\centering
\setlength{\unitlength}{1mm}
\begin{picture}(40,80)(0,0)
\put(75,-3){\begin{picture}(30,80)(0,0)
\includegraphics{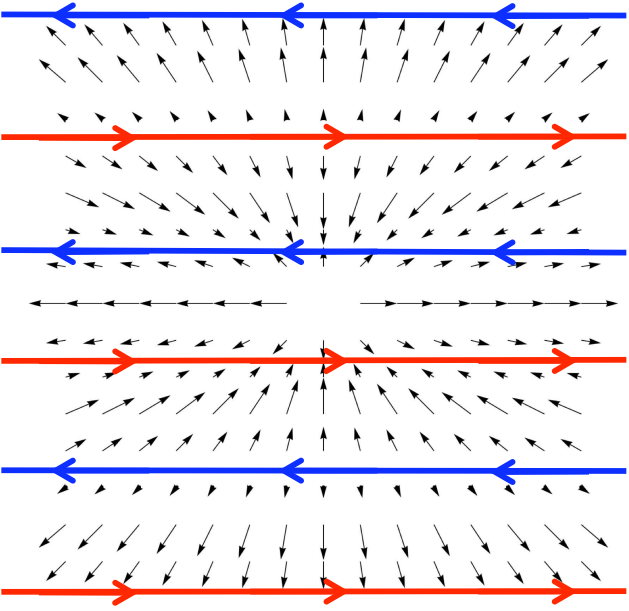}
\end{picture}}
\end{picture}
\caption{A $2\pi$-vortex defect in a LO state. The 2d arrows indicate
  the corresponding complex condensate LO wavefunction,
  $\Delta_{LO}^v(\rv)$, with orientations characterizing the phase
  $\phi(\rv)$.}
\label{fig:vortex}
\end{figure}

As with the Goldstone modes in \rf{DeltaLO}, where it was more
convenient to work with the more physical sum and difference modes,
$\phi(\rv)=\oh(\phi_++\phi_-)$, $\theta(\rv)=\oh(\phi_+-\phi_-)$,
Eqs.\rf{phitheta},\rf{H0loSum} we consider topological defects
associated with singularities in $\phi(\rv)$ and
$\theta(\rv)$,\cite{commentZ2} defined by
\bse
\begin{eqnarray}
\oint d\vec{\ell}\cdot\vec{\nabla}\phi &=& 2\pi n_v,\\
\oint d\vec{\ell}\cdot\vec{\nabla}\theta &=& 2\pi n_d.
\end{eqnarray}
\ese
Given the definitions of $\phi,\theta$, the corresponding vortex
(v) and dislocation (d) defects ``charges'', $n_{v,d}$ are related to
the integer-valued $n_\pm$ according to
\bse
\begin{eqnarray}
n_v&=&\oh(n_++n_-),\\
n_d&=&\oh(n_+-n_-),
\end{eqnarray}
\ese
and therefore admit half-integer and integer topological ``charges'',
collectively designated by a vortex flavor vector $\cNv =
(n_v,n_d)$. The minimal values of $n_+=\pm 1$, $n_-=\pm 1$, lead to
four fundamental defects ($8$ counting the overall $\pm$ signs of each)
\bse
\begin{eqnarray}
\cNv_v=(\pm 1,0)&\leftrightarrow& \Nv_{v,v}=(\pm 1,\pm 1),\ \ \ \ \ 
\label{Nv1}\\
\cNv_d=(0,\pm 1) &\leftrightarrow& \Nv_{v,-v}=(\pm 1,\mp 1),\ \ \ \ \ 
\label{Nv2}\\
\cNv_{v-d}=(\pm 1/2,\pm 1/2) &\leftrightarrow&
\Nv_{v,0}=(\pm 1,0),\ \ \ \ \ \label{Nv3}\\
\cNv_{v-d}^*=(\pm 1/2,\mp 1/2)&\leftrightarrow&
\Nv_{0,v}=(0,\pm 1),\ \ \ \ \ \label{Nv4} 
\end{eqnarray}
\label{Nv}
\ese
where the first two defects, $\cNv_v,\cNv_d$ are ordinary
$2\pi$-vortices in $\phi_+$ {\em and} in $\phi_-$, that respectively
correspond to a $2\pi$-vortex in $\phi$ superimposed on defect-free LO
layers (Fig.\ref{fig:vortex}) and an integer $a$-dislocation in the
periodic LO smectic state.

\begin{figure}[tbp]
\vskip0.25cm 
\epsfig{file=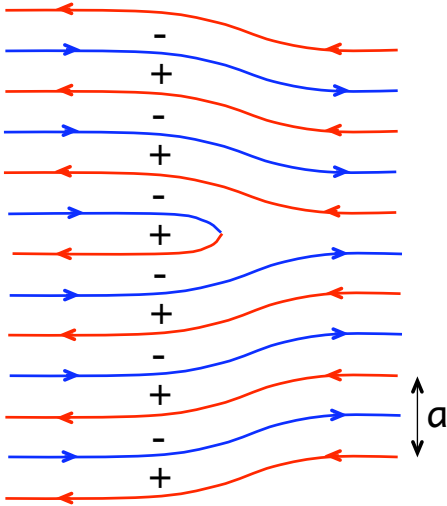,width=6.5cm,angle=0}
\caption{An integer $a$-dislocation defect in the LO layered
  structure, with no defects in its superfluid phase.}
\label{fig:dislocation}
\end{figure}

As illustrated in Fig.~\ref{fig:dislocation} for a 2d state, the
latter of these two corresponds to a LO state with an edge dislocation
of a smectic layer, or equivalently, two missing adjacent $\pm$
domain-walls, with no additional phase winding in $\phi$. In 2d the LO
order parameter in the presence of these point defects is given by
\bse
\begin{eqnarray}
  \Delta_{LO}^v &=& 2e^{i\varphi(y,z)}\cos(q_0 z),
\ \ \mbox{for $\cNv_v=(1,0)$},\ \ \ \\
  \Delta_{LO}^d &=& 2\cos\big[q_0 z + \theta_{d}(y,z)\big],
\ \ \mbox{for $\cNv_d=(0,1)$},\ \ \ \ \ \ \ 
\end{eqnarray}
\ese
where $\varphi(y,z) = \tan^{-1}(z/y)$ is the azimuthal coordinate
angle creating the singular vortex, and
$\theta_d(y,z)=\varphi(y,x)+\theta_0(y,x)$ the corresponding
dislocation angle with a nonsingular part $\theta_0(y,x)$ accounting
for smectic anisotropic elasticity. As can be seen from their form in
terms of the $\Nv_v = (n_+,n_-)$ description, these two seemingly
simpler integer defects in Eqs.\rf{Nv1},\rf{Nv2} are actually
composites of the fundamental $\cNv_{v,0} = (1,0)$ and $\cNv_{0,v} =
(0,1)$ defects.
%

\begin{figure}[tbp]
\vskip0.25cm 
\epsfig{file=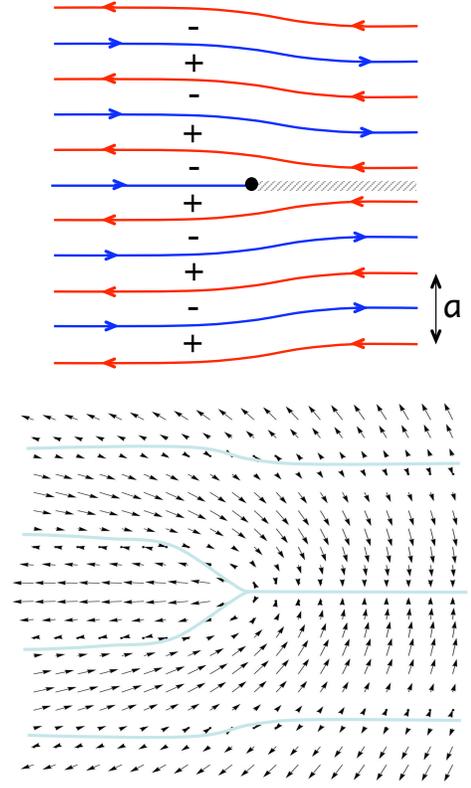,width=8.5cm,angle=0}
\caption{Two complementary forms illustrating a half-integer
  vortex-dislocation $(\pi,a/2)$ defect in the LO state.  The 2d
  arrows indicate the corresponding complex condensate LO
  wavefunction, $\Delta_{LO}^{v-d}(\rv)$, with orientations
  characterizing the phase $\phi(\rv)$.}
\label{fig:vortexdislocation}
\end{figure}

The novel $\cNv_{v-d}=(\pm 1/2,\pm 1/2)$, $\cNv_{v-d}=(\pm 1/2,\mp
1/2)$ defects are half-integer vortex-dislocation composites, where a
$\pm\pi$-vortex is bound to a half of single domain-wall, an
$a/2$-dislocation, illustrated in Fig.\ref{fig:vortexdislocation}.  In
terms of the two coupled phase fields, $\phi_+,\phi_-$ these
half-integer defects correspond to an {\em integer} vortex in one, but
not both $\phi_\pm$ phases.  In 2d the LO order parameter in the
presence of such a half-integer defect is given by
\begin{eqnarray}
  \Delta_{LO}^{v-d} &=& 2e^{i\varphi(y,z)/2}\cos\big[q_0 z+\varphi(y,z)/2\big],\\
&&\ \ \ \ \ \ \mbox{for $\cNv_v=(1/2,1/2)$}.\nonumber
\end{eqnarray}


In contrast to a conventional uniform superfluid or a standard smectic
density wave, here the product form of the pair-density wave LO order
parameter \rf{DeltaLO} allows this composite half-integer
defect. Namely, although the superfluid and the smectic density wave
order-parameter components each change by a minus sign (wind by a
phase of $\pi$), the product LO order-parameter $\Delta_{LO}$,
\rfs{DeltaLO} remains single-valued in the presence of such $(\pm
1/2,\pm 1/2)$, $(\pm 1/2,\mp 1/2)$ fractional composite defects.

\subsection{Energetics}
\label{sec:defectsE}

The energy cost of above described vortices and dislocations can be
computed with straightforward extensions of standard
analyses\cite{ChaikinLubensky,TonerNelsonSm}.

\subsubsection{$2\pi$-vortex: $\cNv_v=(1,0)$}
\begin{figure}[bth]
\centering
\setlength{\unitlength}{1mm}
\begin{picture}(40,80)(0,0)
\put(-21,-2){\begin{picture}(30,80)(0,0)
\includegraphics{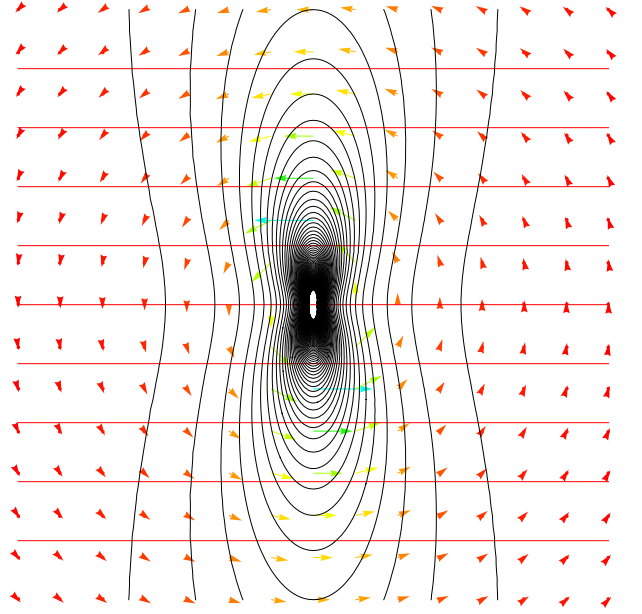}
\end{picture}}
\end{picture}
\caption{Anisotropic superfluid flow around a $2\pi$-vortex
  near $h_{c2}$, predicted to be characterized by anisotropic
  superfluid ratio $\rho_s^\parallel/\rho_s^\perp$ that diverges at
  $h_{c2}$.}
\label{fig:vortexflow}
\end{figure}

The $2\pi$-vortex in the LO superfluid is obtained via a solution of
the Euler-Lagrange equation for $\phi(\rv)$
\begin{equation}
\rho_s^\parallel\partial_z^2\phi^v 
+\rho_s^\perp\nabla_\perp^2\phi^v=0\;,
\label{ELphi}
\end{equation}
under the condition of a $2\pi$-vortex line singularity 
\begin{eqnarray}
  \nabla\times\nabla\phi^v&=&2\pi\int\hat{\bf t}(s)
  \delta^3({\bf r}-{\bf r}(s)) d s\;,\label{vortexEq}
\end{eqnarray}
located at $\rv(s)$ with a unit tangent $\hat{\bf t}(s)$. The smectic
layers remained undistorted at long scales, $u=0$ (equivalently
$\phi_+=\phi_-$).  The solution displays a qualitatively standard
form, but with distortions associated with a large anisotropy ratio
$\rho_\parallel/\rho_\perp$, that, as we demonstrated dramatically
diverges at the upper-critical field $h_{c2}$. This is reflected in
the anisotropy of the mass current flow around a vortex line running
within the LO (smectic, xy-) layers, (a mass flow with velocity
component along the smectic layer normal, $\hat{\qv}_0=\zh$),
illustrated in Fig.\ref{fig:vortexflow}.  For concreteness we look at
two straight vortex line configurations.

{\em $2\pi$-vortex line $|| \qv_0$:} 

For a straight vortex line running along $\qv_0$ (z-axis) above
equations reduce to
\bse
\begin{eqnarray}
\nabla_\perp^2\phi^v&=&0\;,\\
\nabla\times\nabla\phi^v&=&2\pi\zh
\delta^2({\bf r}_\perp),
\end{eqnarray}
\ese
which gives $\phi^v(\rv)=\varphi(x,y)$, where $\varphi(x,y) =
\tan^{-1}(y/x)$ is the azimuthal coordinate angle within a smectic
layer (xy-plane).  The corresponding superfluid ``velocity''
$\vv_v=\nabla\phi^v$ for such a $2\pi$-vortex line directed along
$\hat{\qv}_0=\zh$ (flowing in the xy $\perp$-plane) is isotropic,
given by a standard $\hat\varphi/r_\perp$ form
\begin{equation}
\vv_v^{\parallel}=\frac{1}{r^2_\perp}
(-y, x, 0).
\end{equation}
Integrating the kinetic energy density, \rfs{H0loSum} in a system of
dimensions $L_\perp\times L_\perp\times L_z$ we readily find the
energy to be given by the familiar 3d (linear-logarithmic) form,
\begin{equation}
E_v^{\parallel}=\pi\rho_s^\perp
L_z\ln\left(L_\perp/a\right),
\end{equation}
diverging linearly with vortex line's length, $L_z$ and
logarithmically with systems in-plane extent $L_\perp$.

{\em $2\pi$-vortex line $\perp\qv_0$:} 

In contrast, a vortex line directed along the smectic layers (taken
here to be along the $x$-axis, with flow confined to the anisotropic
$y-z$ plane) characterized by
\begin{eqnarray}
\nabla\times\nabla\phi^v&=&2\pi\xh \delta(y)\delta(z)
\end{eqnarray}
is described by
$\phi(\rv)=\varphi(\sqrt{\rho_s^\parallel}y,\sqrt{\rho_s^\perp}z)=
\tan^{-1}\left(\frac{z}{y}\sqrt{\frac{\rho_s^\perp}{\rho_s^\parallel}}
\right)$ and a velocity field
\begin{equation}
\vv_v^{\perp}=\frac{\sqrt{\rho_s^\parallel\rho_s^\perp}}
{\rho_s^\parallel y^2 +\rho_s^\perp z^2 }
(0, -z, y),
\label{anisotropicCore}
\end{equation}
illustrated in Fig.\ref{fig:vortexflow}.

As expected from the superfluid-stiffness anisotropy near $h_{c2}$,
where $\rho_s^\perp/\rho_s^\parallel \ll 1$, the superflow around a
vortex falls off much more slowly across the LO layers (along $z$)
than within them (along $\rv_\perp$). This leads to (near $h_{c2}$ a
highly) anisotropic elliptic-like superflow with the major axis
oriented along the LO layers normal (along $\qv_0$) and the velocity
flow that is {\em not} everywhere tangent to the equipotential flows,
as illustrated in Fig.\ref{fig:vortexflow}.

Substituting this flow inside the Hamiltonian \rf{H0loSum}, the
corresponding vortex energy is then readily shown to be given by
\begin{equation}
  E_v^{\perp}=\pi\sqrt{\rho_s^\parallel\rho_s^\perp}
  L_\perp\ln\left[(\rho_s^\parallel L_\perp^2+
\rho_s^\perp L_z^2)^{1/2}/a\right],
\label{E2pi}
\end{equation}
In 2d it reduces to a familiar logarithmic form by replacing the
$L_\perp$ factor by the sample thickness.

\subsubsection{$a$-dislocation: $\cNv_d=(0,1)$}

An integer dislocation defect in the LO superfluid is similarly
obtained via a solution of the Euler-Lagrange equation for
$u=-\theta/q_0$, obtained by minimizing Eq.\ref{H0loSum}, with a
condition of $\phi=0$ (equivalently $\phi_+=-\phi_-$)
\begin{equation}
B\partial_z^2u_d - K\nabla_\perp^4u_d=0\;,
\label{ELu}
\end{equation}
and the singularity and continuity conditions
\bse
\begin{eqnarray}
\nabla\times{\bf v}_d&=&{\bf m}_d\;,\label{dislocationEqi}\\
\nabla\cdot{\bf m}_d&=&0\;,
\end{eqnarray}
\ese
with the singular strain
\begin{equation}
{\bf v}_d=\nabla u_d\;,\label{dislocationEqii}
\end{equation}
and the dislocation ``charge'' density
\begin{equation}
{\bf m}_d({\bf r})=a\sum_i\int n^i_d \hat{\bf t}_i(s_i)
\delta^3({\bf r}-{\bf r}_i(s_i)) d s_i\;.
\label{m_d}
\end{equation}
As for the vortex, $s_i$ parameterizes the $i$'th dislocation loop,
${\bf r}_i(s_i)$ is the conformation of that loop, $\hat{\bf
  t}_i(s_i)$ is its local unit tangent, and $n_d^i$ is the number of
excess layers associated with the dislocation.  Note that $n_d^i$ is
independent of $s_i$, since the charge of a given dislocation line is
constant along the line defect.

The dislocation solution is most straightforwardly obtained in 
Fourier space, where equations reduce to:
\bse
\begin{eqnarray}
q_z v^z_{d}+ \lambda^2 q_\perp^2{\bf q}_\perp\cdot {\bf v}_{d}
&=&0,
\label{ELfourier}\\
i{\bf q}\times{\bf v}_d&=&{\bf m}_d\;,
\label{dislocationF}
\end{eqnarray}
\ese
and the general solution is given by
\begin{equation}
{\bf v}_d={i{\bf q}\times{\bf m}_d\over q^2}+i{\bf q}\chi(\qv)\;,
\label{chi}
\end{equation}
with $\chi(\qv)$ a smooth elastic distortion around the dislocation
line obtained via the Euler-Lagrange equation, \rf{ELu}, and is given by
\begin{equation}
\chi=-{q_z(1-\lambda^2 q_\perp^2)\over \Gamma^{sm}_qq^2}\epsilon_{z i j} 
q_i m_j^d.
\label{chii}
\end{equation}
The superfluid phase remains uniform at long scales, $\phi=0$
(equivalently $\phi_+=-\phi_-$).  In above we defined the inverse of
the LO phonon propagator
\begin{equation}
\Gamma^{sm}_\qv\equiv q_z^2 + \lambda^2 q_\perp^4\;,
\label{Gamma_sm}
\end{equation}
with $\lambda = \sqrt{K/B}$, a length characterizing the LO
domain-wall deformations.  Inserting above expression for ${\bf v}_d$
into the original elastic Hamiltonian Eq.\ref{H0loSum} gives the
Coulomb gas-like Hamiltonian that determines the energy of an
arbitrary dislocation configuration
\begin{equation}
H_d=\oh \int_{\bf q}\left[{K q_\perp^2\over\Gamma^{sm}_\qv} P_{i j}^\perp(\qv)
 +{\cal E}_c\delta_{ij}\right]m_i^d({\bf q})m_j^d(-{\bf q}),
\label{Hd}
\end{equation}
where $P_{i j}^\perp({\bf q})=\delta_{i j}^\perp - q_i^\perp
q_j^\perp/q_\perp^2$ is the in-(xy) plane transverse projection
operator, with $\perp$ indicating projection onto the smectic layers.
In above we also added dislocation core energy determined by its
short-scale ($a=2\pi/q_0$) configuration, at which above continuum
analysis no longer applies. A simplest estimate (up to possibly
nontrivial dimensionless factors set by $\lambda/a$, that can only be
determined through a microscopic calculation) of this core energy
density ${\cal E}_c$ is given by $K$.

In 3d, two limiting dislocation line configurations are the screw and
edge dislocations,
\bse
\begin{eqnarray}
{\bf m}_d^{screw}(\rv)&=&a\zh \delta^2(\rv_\perp),\label{mScrew}\\
{\bf m}_d^{edge}(\rv)&=&a\xh \delta(y)\delta(z),\label{mEdge}
\end{eqnarray}
\ese
running perpendicular to and along smectic layers,
respectively. Within above harmonic continuum approximation, the
long-scale elastic contribution (first term in \rf{Hd}) to the energy
of a screw dislocation vanishes identically (due to $P_{i
  j}^\perp({\bf q})$), physically because of smectic's soft in-plane
curvature elasticity and due to the absence of long-scale
compressional distortion in a screw dislocation, as can be readily
verified. Thus its energy is determined by short-scale core energetics
inaccessibility to harmonic elastic analysis.

For a single edge dislocation along $\xh$ \rf{mEdge} $\vec{m}^d(\qv)
= a\xh 2\pi\delta(q_x)$, giving for the elastic part of the energy
\bse
\begin{eqnarray}
  E_{d,el}^{3d-edge}&=& L_\perp a^2\int_{\bf q}{K q_\perp^2\over\Gamma^{sm}_\qv}
    (1-q_x^2/q_\perp^2)2\pi\delta(q_x),\ \ \ \ \ \ \\
    &=& L_\perp a^2\int \frac{dq_ydq_z}{(2\pi)^2}
    {K q_y^2\over q_z^2 +\lambda^2 q_y^4},\\
    &=& \frac{K a}{2\lambda} L_\perp,
\end{eqnarray}
\ese
where the factor of $L_\perp$ came from the regularized identity
$2\pi\delta(q_x)\delta(q_x) = L_\perp\delta(q_x)$.  Although the first
``elastic'' contribution in \rfs{Hd} no longer vanishes, the above
energy is dominated by the short cutoff scale, $a$ and therefore in
principle indistinguishable from the core energy. Thus, based on this
analysis we can only provide an estimate of the 3d dislocation
energies,
\bse
\begin{eqnarray}
E_d^{3d-screw}&\approx& K L_z,\\
E_d^{3d-edge}&\approx& \frac{K a}{\lambda} L_\perp\approx (B K)^{1/2} a
L_\perp\approx K L_\perp,\ \ \ \ \ 
\end{eqnarray}
\label{Ea}
\ese
diverging linearly with its length. 

Correspondingly, as first demonstrated by Toner and
Nelson\cite{TonerNelsonSm}, in two dimensions the energy of a smectic
edge dislocation is constant, i.e., system length independent, 
given by
\begin{eqnarray}
E_d^{2d}&\approx& a\sqrt{B K}\approx K.
\end{eqnarray}
As we will see in the next section, this observation has crucial
implications for the instability of a finite temperature 2d LO
state\cite{TonerNelsonSm}.

\subsubsection{half-integer vortex-dislocations defect:
  $\cNv_{v-d}=(1/2,1/2)$}

As discussed above\cite{RVprl}, on general grounds a LO (but not a FF)
superconductor allows novel half-integer defects that are a composite
of a $\pi$-vortex and $a/2$-dislocation, illustrated in
Fig.\ref{fig:vortexdislocation}. The form and the energy associated
with this defect requires a solution of Euler-Lagrange equations for
$\phi_\pm$
\begin{equation}
-\rho_s^\parallel\partial_z^2\phi_\pm 
+\frac{K}{q_0^2}\nabla_\perp^4\phi_\pm 
-\oh\rho_s^\perp\nabla_\perp^2\phi_\pm-\oh\rho_s^\perp\nabla_\perp^2\phi_\mp 
= 0,
\label{ELu_pm}
\end{equation}
with singularity conditions
\bse
\begin{eqnarray}
\nabla\times\nabla\phi_+&=&2\pi\xh\delta(y)\delta(z),\\
\nabla\times\nabla\phi_-&=&0,
\end{eqnarray}
\ese
for concreteness taken to be a straight vortex line in $\phi_+$
running parallel to LO layers, along the $\xh$ axis, and $\phi_-=0$.

Solving above equations gives the defect's shape from which the energy
is readily computed
\begin{equation}
E_{\pi-a/2}\approx\frac{1}{4}L_\perp
\left[\pi\sqrt{\rho_s^\parallel\rho_s^\perp}
\ln\big[(\rho_s^\parallel L_\perp^2+\rho_s^\perp L_z^2)^{1/2}/a\big] 
+ K\right].
\label{Epi}
\end{equation}
Because the energy scales as a square of the defect's topological charge
$n_{v,d}$, as expected, the above energy of the half
vortex-dislocation defect is one-quarter of the sum of the energies of
the unit vortex and dislocation.

\subsection{Dual description: coupled xy-smectic models}

In characterizing phases in terms of topological defects, it is often
useful to have a continuum field-theoretic description of defects and
of the corresponding transitions associated with their unbinding. Such
a description, that complements the LO Goldstone model, \rf{H0lo} is
obtained through a duality transformations that we develop here.  In
3d the analysis is somewhat complicated and results in two coupled
$U(1)$ gauge theories, whose derivation and analysis we leave to a
future publication\cite{LRunpublished}.  Here, for simplicity we will
exclusively focus on two dimensions.

To this end, starting with $H_{LO}$, \rf{H0loSum}, allowing for singular
configurations of $\phi$ (vortices) and $u$ (dislocations) discussed
in the previous subsection, and integrating out the smooth parts of
Goldstone modes, we obtain a Coulomb gas description of a finite
density of topological defects
\begin{eqnarray}
H^{xy-sm}_{CG}&=&
\oh\int_{\bf  q}\left[\frac{\sqrt{\rho^\perp_s\rho^\parallel_s}}
{\Gamma_\qv^{xy}}|m_{\qv,v}|^2
+\frac{K q_\perp^2}{\Gamma_\qv^{sm}}|m_{\qv,d}|^2\right]\nonumber\\
&&+\sum_{\rv_i}\left(E^v_c n_{\rv_i,v}^2 + E^d_c n_{\rv_i,d}^2\right),
\end{eqnarray}
where 
\bse
\begin{eqnarray}
\Gamma_\qv^{xy}&=& q_\perp^2\sqrt{\rho_s^\perp/\rho_s^\parallel}
+  q_z^2\sqrt{\rho_s^\parallel/\rho_s^\perp},\\
\Gamma_\qv^{sm}&=& q_z^2 + \lambda^2 q_\perp^4.
\end{eqnarray}
\ese
In above, $m_{\qv,v}, m_{\qv,d}$ are the Fourier transforms of the
vortex and dislocation densities $m_{v}(\rv)=\sum_i2\pi
n_{\rv_i}^{v}\delta^2(\rv - \rv_i^v)$, $m_{d}(\rv)=\sum_i a
n_{\rv_i}^{d}\delta^2(\rv - \rv_i^d)$, with
$n^{v,d}_{\rv_i}=\oh(n^+_{\rv_i} \pm n^-_{\rv_i})$, and
$n^\pm_{\rv_i}\in \Z$ the independent elementary integer defects,
defined in the previous subsection. In above we also introduced the
vortex and dislocation core energy $E_c^{v,d}$ to account for short
scale core physics not accounted for by the above continuum
description. Although by time-reversal symmetry the core energies of
the elementary defects $n^\pm$ are identical, those of the vortex and
dislocation composites will generically be distinct with $E_c^{v} \neq
E_c^{d}$.

To obtain a continuum description of these integer-valued fields, it
is convenient to decouple interaction of these defects using a
Hubbard-Stratonovich transformation by introducing a pair of dual real
fields (corresponding potentials) $\tphi,\ttheta$ with the Hamiltonian
\begin{eqnarray}
\tilde H[\tphi,\ttheta,n_{v,d}]&=& \tilde H_0[\tphi,\ttheta]
+i\int_{\rv}\left(\tphi(\rv) m_v(\rv) +q_0\ttheta(\rv) m_d(\rv)\right)
\nonumber\\
&&+\sum_{\rv_i}\left(E^v_c n_{\rv_i,v}^2 + E^d_c n_{\rv_i,d}^2\right),
\label{Hmmphitheta}
\end{eqnarray}
where
\begin{eqnarray}
\tilde H_0[\tphi,\ttheta]=\oh\int_{\bf q}\left[
\frac{\Gamma^{xy}_\qv}{\sqrt{\rho^\perp_s\rho^\parallel_s}}|\tphi_\qv|^2
+\frac{\Gamma^{sm}_\qv q_0^2}{K q_\perp^2}|\ttheta_\qv|^2\right].
\end{eqnarray}

The thermodynamics is characterized by a partition function 
\begin{widetext}
\bse
\begin{eqnarray}
Z &=& \int[d\tphi d\ttheta]\prod_{\rv_{i}^{v,d}}\sum_{n_{\rv_{i}}^{v,d}}
e^{-\tilde H[\tphi,\ttheta,n_{\rv_{i}}^{v,d}]}\\
&=&\int[d\tphi d\ttheta]e^{-\tilde H_0[\tphi,\ttheta]}
\prod_{\rv_{i}^{v,d}}\sum_{n_{\rv_{i}}^{v,d}}
e^{i2\pi n_{\rv_{i,v}}\tphi(\rv_{i,v})
+ i2\pi n_{\rv_{i,d}}\ttheta(\rv_{i,v})
- E^v_c n_{\rv_i,v}^2 - E^d_c n_{\rv_i,d}^2}
\end{eqnarray}
\ese
\end{widetext}
where we chose to measure all energies in units of $k_B T$.  The
summation over vortex and dislocation charges $n_{\rv_i}^{v,d}$ can be
readily done in the dilute defect limit, corresponding to large core
energies $E_c$. In this limit the summations can be limited to nine
lowest order terms, corresponding to no vortices present, $\Nv_{0-0}$,
one vortex of either species and sign at a site, $\Nv_{\pm 1,0},
\Nv_{0,\pm 1}$, or two vortices of distinct species sitting at a
single site, $\Nv_{\pm 1,\pm 1}$. Equivalently, in terms of $\cNv$
vortex-dislocation nomenclature, these respectively correspond to
\begin{eqnarray}
&&\cNv:\ (0,0),\ \pm(\pi,a/2),\ \pm(\pi,-a/2),\ \pm(2\pi,0),\ \pm(0,a).
\nonumber\\
&&
\end{eqnarray}

This gives
\begin{widetext}
\begin{eqnarray}
Z &\approx&\int[d\tphi d\ttheta]e^{-\tilde H_0[\tphi,\ttheta]}
\left[1+2\sum_{\rv_1}
\left(e^{-\frac{1}{4}E_c^v-\frac{1}{4}E_c^d}\cos\big[\pi(\tphi+\ttheta)\big]
+e^{-\frac{1}{4}E_c^v-\frac{1}{4}E_c^d}\cos\big[\pi(\tphi-\ttheta)\big]
\right.\right.\nonumber\\
&&\left.\left.+
e^{-E_c^v}\cos(2\pi\tphi)+e^{-E_c^d}\cos(2\pi\ttheta)
\right)\right].
\end{eqnarray}
\end{widetext}
All other vortex configurations (for example two defects of either
species sitting on {\em distinct} sites) are not included as they
correspond to higher order contributions, suppressed at low fugacity.
After re-exponentiating, above leads to a generalized sine-Gordon
model for the coupled $\tphi,\ttheta$ fields, with the dual
Hamiltonian
\begin{widetext}
\begin{eqnarray}
\tilde H&=&
\oh\int_{\bf q}\left[
\frac{\Gamma^{xy}_\qv}{\sqrt{\rho^\perp_s\rho^\parallel_s}}|\tphi_\qv|^2
+\frac{\Gamma^{sm}_\qv q_0^2}{K q_\perp^2}|\ttheta_\qv|^2\right]
-\int_\rv\left[g_{\pi,a/2}\cos(\pi\tphi)\cos(\pi\ttheta)
+ g_{2\pi,0}\cos(2\pi\tphi) + g_{0,a}\cos(2\pi\ttheta)\right]
\label{HdualSG}
\end{eqnarray}
\end{widetext}
where the couplings are
$g_{\pi,a/2}=\frac{4}{a^2}e^{-\frac{1}{4}E_c^v-\frac{1}{4}E_c^d}$,
$g_{2\pi,0}=\frac{2}{a^2}e^{-E_c^v}$ and
$g_{0,a}=\frac{2}{a^2}e^{-E_c^d}$, and we converted the sum over
$\rv_1$ into an integral over $\rv$ via a lattice constant $a$.

In the opposite limit of a small core energy, $E_c^{v,d}$ the defect
density is high and the summation over the integer charges
$n^\pm_{\rv}$ can be carried out utilizing the Poisson summation
formula, that leads to a replacement of the e.g., $\cos(2\pi\tphi)$
potentials above by the Villain potential, defined by: 
\bse
\begin{eqnarray}
e^{-\beta V_V[\tphi]}&=&\sum_n e^{i2\pi n\tphi - E_c n^2},\\
&=&\sum_p e^{-\frac{1}{4 E_c}(2\pi\tphi - 2\pi p)^2}.
\end{eqnarray}
\ese 
The latter can be approximated by a single harmonic with an
effective coupling $g^{v,d}_{\text{eff}}=1/(2E_c^{v,d} a^2)$.

The resulting generalized sine-Gordon model is convenient for
analyzing the effects of defects on the LO state, particularly for the
computation of their screening on long scales, unbinding, and for the
analysis of the resulting disordered state. From the form \rf{HdualSG}
it is clear that (aside from an inconsequential anisotropy) the dual
vortex sector described by $\tphi$ has a standard sine-Gordon form.
In contrast, the dual dislocation sector, described by $\ttheta$ is
qualitatively modified by the highly nonlocal and qualitatively
anisotropic smectic kernel, $\Gamma^{sm}_\qv$\cite{BergDual}.

We next turn to a detailed discussion of implication of these defects
for possible phases and transitions associated with the LO state.

\section{Phases and Phase transitions: Larkin-Ovchinnikov liquid
  crystals}
\label{sec:transitions}

We now turn to a discussion and characterization of a class of phases
emerging from the Larkin-Ovchinnikov state by a partial disordering of
it. Because these phases partially break spatial symmetries, we will
refer to them collectively as LO liquid crystals.  We first discuss
general symmetry based possibilities and then explore their concrete
realization in terms of fluctuation-driven proliferation of
topological defects.

\subsection{Landau's order parameters and spontaneously broken
  symmetries}

A conventional characterization of phases is through the Landau's
classification, where phases are distinguished by non-vanishing order
parameters and corresponding symmetries that they break. From this
prospective a zero-temperature LO (smectic pair-density wave)
superfluid ($SF_{Sm}$) breaks three symmetries: the translations
$T_{\qv_0}$ along $\qv_0$ ($\zh$), the rotations $R_{\qv_0}$ about the
axis transverse to $\qv_0$ (of the full 3d Euclidean group $E(3)$),
and the $U(1)$ symmetry associated with atom conservation.  The
corresponding non-vanishing order parameter, that transforms
non-trivially under these group of symmetries is the LO pair-density
wave \rf{DeltaLO}.

By considering all possible basic combinations of spontaneously broken
subset of these symmetries, we uncover five additional atomic liquid
crystal phases, that are descendants of the smectic LO
state\cite{comment_morephases}. To enumerate these systematically we
begin with the zero-temperature LO state, where all three above
symmetries are broken and partially restore them by progressively
disordering the state. Restoring the translational symmetry
$T_{\qv_0}$, while keeping $U(1)$ and $R_{\qv_0}$ leads to a state
with orientational and off-diagonal orders, that is a nematic
superfluid, $SF_N$. By analogy with the conventional (non-superfluid)
nematics the resulting state can be characterized by a complex
traceless symmetric second-rank tensor $Q_{ij}$.

Subsequently disordering the orientational order and thereby also
restoring the rotational symmetry $R_{\qv_0}$ leads to an isotropic
superfluid, $SF_I$, that exhibits a finite species imbalance.
Symmetry-wise the resulting state is isomorphic to the polarized
superfluid, $SF_M$ predicted\cite{SRprl,SRaop} and observed to appear
at a nonzero imbalance on the BEC side of the BCS-BEC crossover. In
contrast, (as a descendant of the LO state expected to be stabilized
by Fermi surfaces imbalance) here the $SF_I$ state is realized in an
imbalanced superfluid on the BCS side, something that has been
searched for dating back to Sarma\cite{Sarma}, but has not been
possible within mean-field treatments, that instead predict an
instability to phase separation\cite{Bedaque03,SRprl,SRaop}.

As summarized by Tables \ref{table:phasesRelate},\ref{table:phases}
the additional three phases, smectic, nematic and isotropic Fermi
liquids, $FL_{Sm}, FL_N, FL_I$ are the non-superfluid counterparts of
the three discussed above, obtained by first restoring the $U(1)$
symmetry by disordering the off-diagonal long-range order.  The fully
disordered $FL_I$ state is simply the normal state of the polarized
Fermi gas, albeit strongly interacting. Together these intermediate
fluctuation-induced phases (along with a number of other possible ones
that we discuss below) naturally interpolate between the fully gapped
singlet (homogeneously and isotropic) BCS superconductor at zero
imbalance and low temperature, and the normal polarized Fermi liquid
at large imbalance and/or high temperature.

\begin{table}[t]
\begin{tabular}{ c  c  c  c  c  } 
$FL_{Sm}$&$\rightarrow$ &$FL_N$ &$\rightarrow$ & $FL_I$\\[0.2cm]
$\ \ \ \ \big\uparrow$\small{U(1)}&    
&$\ \ \ \ \ \big\uparrow$\small{U(1)}& 
&$\ \ \ \ \ \big\uparrow$\small{U(1)} \\[0.2cm]
$SF_{Sm}$&$\rightarrow$ &$SF_N$ &$\rightarrow$ & $SF_I$ \\[0.5cm]
\end{tabular}
\caption{Five phases that naturally emerge as disordered descendants
  of the LO (superfluid smectic, $SF_{Sm}$) state.}
\label{table:phasesRelate}
\end{table}

\begin{table}[t]
\begin{tabular}{| c | c | c | c | } 
\hline
phases & $U(1)$ & $T_{\qv_0}$ & $R_{\qv_0}$ \\
\hline
$FL_I$ &   $\surd$       &  $\surd$        &   $\surd$      \\
\hline
$FL_N$ &  $\surd$        &  $\surd$        & X \\
\hline
$FL_{Sm}$ & $\surd$      & X & X \\
\hline
$SF_I$ & X & $\surd$       &  $\surd$      \\
\hline
$SF_N$ & X &  $\surd$       & X \\
\hline
$SF_{Sm}$ & X & X & X \\
\hline
\end{tabular}
\caption{A summary of LO liquid crystal Fermi-liquid (FL) and
  superfluid (SF) phases, and corresponding order parameters and
  broken symmetries, indicated by X's. Unbroken
  symmetries are marked by check marks. The subscripts $I, N, Sm$
  respectively indicate the Isotropic, Nematic and Smectic orders.}
\label{table:phases}
\end{table}

\subsection{Larkin-Ovchinnikov liquid crystals via topological defects
  unbinding}

A complementary way to characterize phases and phase transitions
between them is in terms of topological defects
proliferation\cite{KT}. For conventional phases and transitions (e.g.,
3d xy or Ising models) this is simply a complementary description that
is sometimes convenient. However, for topological phases, where Landau
order parameter is unavailable or just insufficient to distinguish two
phases, this defect-proliferation approach is indispensable and
therefore superior to the order-parameter Landau's ``soft-spin''
description.  One prominent and familiar example is the description of
the low-temperature quasi-long-range ordered phase of the 2d xy model
and its disordering. In this case, both the low- (``ordered'') and
high-temperature (disordered) phases exhibit a vanishing xy order
parameter, and are respectively only topologically (not symmetry)
distinguished by bound and unbound vortex defects and by the
associated behavior of the correlation functions (power-law and
exponential, respectively)\cite{KT}.

With this in mind, we characterized the LO and its descendant states
in terms of a proliferation of the four types of topological defects
$\cNv=(0,a), (2\pi,0), (\pi,\pm a/2)$, introduced and analyzed in
Sec.\ref{sec:defects}.

As discussed above, at $T=0$ the LO phase is characterized by
long-range off-diagonal and translational order and is thereby
distinguished by a nonzero LO order parameter $\Delta_{LO}$. In this
ground state all of the above topological defects are absent, confined
into topologically neutral pairs.

\subsubsection{3d phases and transitions}
Although, (as we explicitly demonstrated in
Sec.\ref{sec:GMfluctuations}) at nonzero temperature the LO order
parameter vanishes in 3d (and in 2d), the 3d LO phase is distinguished
from its more disordered descendants by the absence of unbound
topological defects, in direct analogy with the quasi-long-range
ordered state of the 2d xy model. Thus, in this same sense at nonzero
$T$ the 3d LO smectic is a topologically ordered (but phonon,
elastically disordered) phase.

Upon increasing temperature, or decreasing the stiffnesses (e.g., by
tuning the strength of the Feshbach-resonant interactions, or by
adjusting the fermionic species imbalance), one or more of the four
topological defects will proliferate, thereby leading to a transition
of the LO smectic $SF_{Sm}$ (PDW) into one of its descendants. The
actual sequence of defects unbinding is determined by the relative
energetics, given by Eqs.\rf{E2pi},\rf{Ea},\rf{Epi}. Lacking a
reliable calculation (throughout the $T-P-1/k_Fa$ phase diagram) of
dependences of the stiffnesses on the experimentally tunable
parameters ($T$, imbalance $P$, interactions $1/k_Fa$, and the number
of atoms), the phase diagram can only be qualitatively mapped out in
terms of the three effective stiffnesses
$K,\rho_s^{\parallel,\perp}$. In the thermodynamic limit
($L_{\perp,z}\rightarrow\infty$, not necessarily relevant to atomic
traps; see below), the relative defects energetics is quite
unambiguous:
\begin{widetext}
\begin{eqnarray}
  &&E^d_{(0,a)}\sim K L \ll 
E^{v-d}_{(\pi,a/2)}\sim\frac{\rho_s}{4}L\ln L + \frac{K}{4}L
\ll E^{v}_{(2\pi,0)}\sim \rho_s L\ln L,\ \ \ \mbox{for
    $L_{\perp,z}\sim L\rightarrow\infty$},
\label{eq:defectsE}
\end{eqnarray}
\end{widetext}
where for simplicity we ignored anisotropies in $\rho_s^i$.  Based on
this energetics one may be tempted to conclude that in this limit
(unless preempted by a first-order transition) it is the integer
dislocation loop defects that proliferate first and the LO smectic
preferentially disorders into a nematic superfluid, $SF_{N}$. However,
in contrast to the 2d KT mechanism\cite{KT}, the 3d disordering
transitions take place when the relevant stiffness, renormalized by
quantum and thermal fluctuations is continuously driven to zero at the
transition, or takes place at a finite (rather than a vanishing)
fugacity. For a thermal transition this roughly corresponds to a
transition temperature set by the corresponding stiffnesses,
$\overline{\rho}_s=\sqrt{\rho_s^\parallel\rho_s^\perp}$ and $K, B$.

Lacking a detailed quantitative theory of such 3d transitions we can
only construct a qualitative phase diagram, that we display in
Fig.\ref{fig:phasediagramLO3d}.  It summarizes all basic phases that
naturally appear upon disordering of the LO smectic superfluid by
unbinding four fundamental type of topological defects discussed
above. One important physical input is the observation that increasing
the Zeeman energy, $h$ (or equivalently, the species imbalance,
$m=n_\uparrow-n_\downarrow$) toward the normal Fermi liquid state,
$FL_I$ at $h_{c2}$, predominantly leads to a suppression of the
superfluid stiffness and therefore to a destruction of the SF
order. Conversely, a reduction of the Zeeman field (and species
imbalance) toward a conventional isotropic and homogeneous superfluid
$SF_I$ at $h_{c1}$, primarily leads to a reduction of the elastic
moduli of the smectic pair-density wave by increasing its period
$1/q_0$ and thereby weakening the interaction between the LO
domain-walls (see \rfs{Bhc1}). Thus lowering of $h$ is expected to
predominantly suppress, i.e., melt the positional smectic order.

Thus, starting with the LO $SF_{Sm}$ state and {\em decreasing} $h$
leads to the unbinding of the integer dislocations $(0,a)$, and a
transition to an orientationally-ordered, i.e., a nematic ``charge''-4
superfluid, $SF^{4}_N$. The later ``charge''-4 feature of $SF^{4}_N$
naturally appears as the remaining secondary order parameter
$\Delta_{sc}^{(4)}=\Delta_{LO}^2$ once the LO positional order
$\Delta_{LO}$ is destroyed by unbinding of integer dislocation loops.
General arguments predict a subleading singularity $\sim (h -
h_{N-Sm})^{1-\alpha}$ of $\Delta_{sc}^{(4)}(h)$ at the continuous
transition $h_{N-Sm}$ between $SF^{4}_N$ and $SF_{Sm}$. Since in
contrast, the ``charge''-2 SF nematic order vanishes in the LO state,
a direct transition to it from the LO state can generically only
proceed through a first-order transition.

Conversely, we expect that {\em increasing} $h$ starting with the LO
$SF_{Sm}$ will lead to a suppression of $\rho_s$, a proliferation
$(2\pi, 0)$, and a transition to a $2q$-smectic Fermi liquid,
$FL^{2q}_{Sm}$, a non-superfluid periodic state with a wavevector that
is twice the LO state. Alternatively, as suggested by the energetics
in \rfs{eq:defectsE}, it maybe that the lower-energy half-vortex
dislocation defects $(\pi, a/2)$ (or the $(\pi, -a/2)$, but not both)
unbind first, in which case a transition to a nematic Fermi liquid,
$FL^{**}_N$ (with the restored translational and U(1) charge
symmetries) takes place. The resulting state is qualitatively distinct
from the more conventional nematic (orientationally ordered) $FL_N$
phase in which {\em both} $(\pi, a/2)$ and $(\pi, -a/2)$ are
proliferated. Both are also distinct from the nematically ordered
$FL^{*}_N$ state, in which only integer dislocations, $(0,a)$ and
integer vortices, $(2\pi, 0)$ are unbound.  One can envision a number
of other states and phase transitions at low $h$ by further
considering the disordering of the nematic superfluid, $SF_N^{4}$ by
unbinding various patterns of disclinations and $\pi$-vortices. We
leave a more detailed analysis of these to a future study.

\subsubsection{2d phases and transitions}

The nature of the phase diagram changes qualitatively in two
dimensions at nonzero temperature. Based on the work of
Ref.\onlinecite{TonerNelsonSm} (in the context of conventional smectic
liquid crystals) and from the analysis in Sec.\ref{sec:defectsE}, that
shows a finite energy cost of a 2d smectic dislocation, we conclude
that at any nonzero $T$ in 2d, the dislocations proliferate, thereby
destroying the LO ($SF_{Sm}$) phase. It is replaced by a homogeneous,
but quasi-orientationally-ordered, ``charge''-4 superfluid nematic,
$SF_N^{4}$. Upon rotation this superfluid will display $1/4$
``charge'' vortices ($\oint{\nabla\phi}\cdot d{\bf l}=\pi\hbar/2m$),
that, because of its nematic order we expect to form a uniaxially
distorted hexagonal lattice. Upon changing $T$, $h$, $k_F a$, the
nematic superfluid can then undergo further disordering transitions
toward a polarized Fermi liquid and an isotropic (e.g., BCS)
superfluid. In particular, we expect two Kosterlitz-Thouless
transitions associated with the loss of 2d superfluid and
orientational (nematic) quasi-long-range orders. Because $\Delta_q$
and therefore $\sqrt{\rho_\perp^s\rho_\parallel^s}\sim \Delta_q^3$
vanishes strongly near $h_{c2}$, we expect the superfluid KT
transition to precede the nematic-isotropic one.

\begin{widetext}
\begin{figure}[tbp]
\epsfig{file=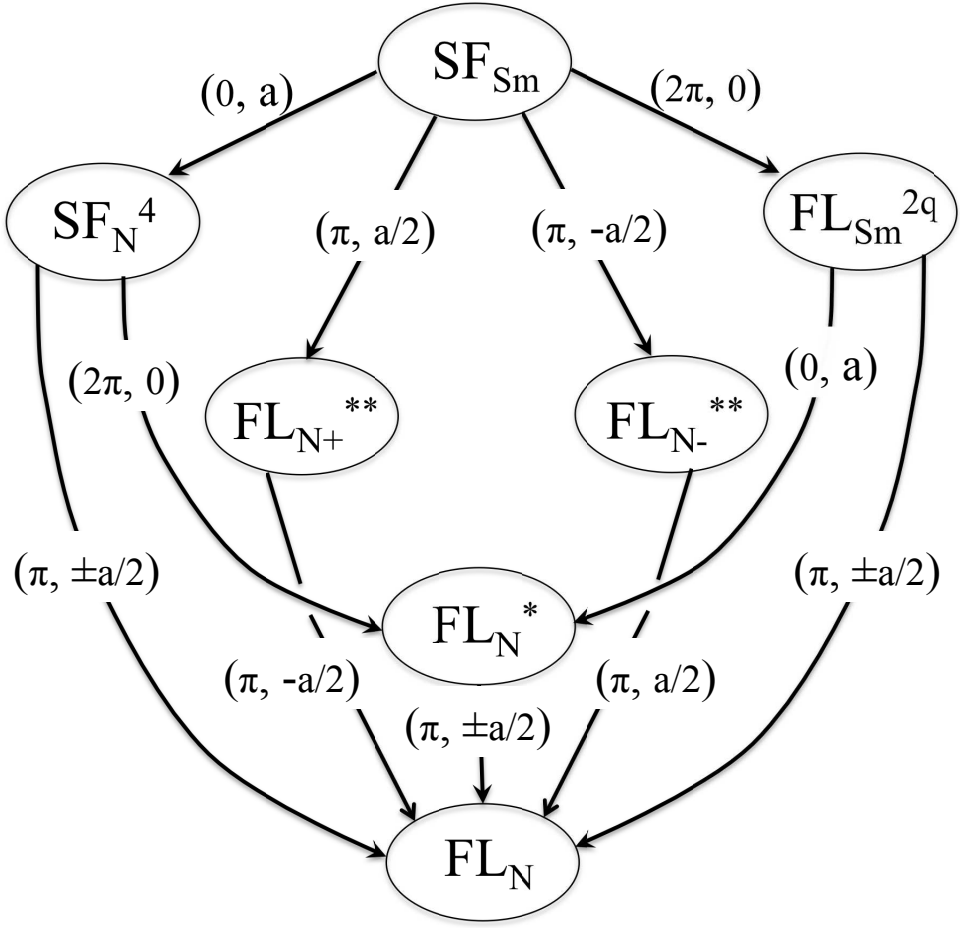,width=13cm,angle=-90}
\caption{A flowchart of superfluid ($SF$) and nonsuperfluid ($FL$)
  phases, exhibiting smectic ($Sm$) and nematic ($N$) conventional
  orders as well as topological orders (indicated by $*$ and $**$),
  induced by a proliferation of various combination of topological
  defects, $(0,a)$, $(2\pi,0)$, and $(\pi,\pm a/2)$.}
\label{fig:flowchart}
\end{figure}

\end{widetext}



\subsection{Fractionalized phases and topological order}

In above discussion we argued for the existence of at least three
topologically distinct Fermi liquid phases that naturally emerge from
disordering of the LO ($SF_{Sm}$) phase by unbinding different
combinations of allowed defects. Because, as demonstrated above the
conventional vortex $(2\pi, 0)$ and a conventional dislocation $(0,
a)$ are composites of the fundamental defects $(\pi, \pm a/2)$ the
nonsuperfluid states $FL^*_N$, $FL^{**}_N$ and their isotropic cousins
$FL_I^*$, $FL_I^{**}$ (in which disclinations are also unbound) are
expected to be ``fractionalized''\cite{SenthilFisher}, topologically
distinct from their conventional Fermi liquid analogs, where $(\pi,
\pm a/2)$ are also unbound.

These novel phases are analogous to the putative phase-disordered
fractionalized states obtained by unbinding double ($hc/e$) vortices,
studied extensively by Sachdev, and by Balents, Senthil, Fisher, and
collaborators\cite{SachdevDoubleVortex,Balents,SenthilFisher} in the
context of high temperature superconductors.
The resulting nonsuperfluid phase is distinguished from a conventional
Fermi liquid by a gapped ``vison'', a $\Z_2$ defect that is a remnant
of the fundamental $hc/2e$ vortex after the composite $hc/e$ (double)
vortices proliferate.  These states are also characterized by
``fractionalized'' charge-neutral spin-1/2 bosonic and charge-e
spinless fermionic excitations.

The states $FL^*_N$, $FL^{**}_N$ also bare a close relation to the
collective mode fractionalization discussed by Sachdev, et
al. \cite{SachdevBook,SachdevMI,NussinovZaanen} in the context
of quantum paramagnetic phases, emerging from disordering a collinear
spin-density wave.  As with the $\left(U(1)\otimes U(1)\right)/\Z_2$ LO
state, where the order parameter is a product of the superfluid and
smectic order parameters, Eq.\rf{DeltaLO}, there too the order
parameter is of $\left(S_2\otimes U(1)\right)/\Z_2$ product form,
encoding spatial modulation of the spin density, and therefore admits
half-integer (vison-like) defects.

As we have seen in the previous subsection, a characterization of both
the conventional and the fractionalized (topologically ordered but
otherwise disordered) phases can be faithfully formulated directly in
terms of distinct patterns of proliferation of the four types of
vortex and dislocation defects, listed in \rfs{Nv}. It can also
equivalently be done in terms of a dual sine-Gordon (and in 3d $U(1)$
gauge-theory) model $\tilde H$, \rfs{HdualSG}. We now present a
complementary effective Ising gauge theory description, that can
sometimes be convenient. In the absence of an underlying rotational
invariance (that otherwise has been our focus throughout) and for
simplicity ignoring anisotropy of the striped LO state, all the phases
and transitions can be captured by a Euclidean action (in space-time
$\vx = (\tau,\rv)$)
\begin{widetext}
\begin{eqnarray}
S = - t_\phi\sum_{\langle\vx,\vx'\rangle}\sigma_{\vx,\vx'}
\cos(\phi_\vx - \phi_{\vx'})
 - t_\theta\sum_{\langle\vx,\vx'\rangle}\sigma_{\vx,\vx'}
\cos(\theta_\vx - \theta_{\vx'}) - K_\sigma\sum_\square\prod\sigma_{\vx,\vx'}
\label{IsingGaugeS}
\end{eqnarray}
\end{widetext}
where the Ising gauge field, $\sigma_{\vx,\vx'}$ couples to the
$U(1)\otimes U(1)$ bosonic (rotor) matter fields $e^{i\phi_\vx},
e^{i\theta_\vx}$, and its nonzero $\Z_2$ flux through a plaquette
encodes the presence of a half-integer $(\pi,a/2)$ defect. The gauge
field $\sigma_{\vx,\vx'}$ encodes the local Ising redundancy of
splitting the LO order parameter \rf{DeltaLO} into a ``charge''-2
boson, $b_\rv^\dagger=e^{-i\phi_\rv}$, that creates a zero-momentum
Cooper-pair (diatomic molecule) and a neutral boson,
$\rho_{q,\rv}^\dagger=e^{-i\theta_\rv}$, that creates a density wave
at a LO wavevector $q$. Formally, the gauge field $\sigma$ can be
introduced as a Hubbard-Stratonovic field that decouples the LO order
parameter $\Delta_q = b\rho_q$ into its charge and density wave parts.
The variety of phases and transitions between them are summarized in
the phase diagram, Fig.\ref{fig:phasediagramLO3d} and a flow-chart,
Fig.\ref{fig:flowchart}.

In above formulation, the LO superfluid ($SF_{Sm}$) is a state at
large $t_\phi, t_\theta$ and arbitrary $K_\sigma$, in which both $b$
and $\rho_q$ are Bose-condensed, and $\sigma$ is gapped through a
Higg's mechanism. In this state, gapped $2\pi$ vortices in $\phi$ and
$\theta$ respectively correspond to the $(2\pi,0)$ superconducting
vortex and the $(0,a)$ integer dislocation, that we discussed in the
previous subsection.

For small $t_\phi$ and large $t_\theta$ $2\pi$ vortices in $\phi$
proliferate, driving $b$ normal and restoring the $U_\phi(1)$ (atom
conservation) symmetry, while keeping $\rho_q$ condensed. A large
$K_\sigma$ forces a vanishing Ising flux with $\sigma=1$, that
corresponds to a gapped vison. The resulting nonsuperfluid state is
deconfined in a sense that that it exhibits a gapped bosonic $b$
excitation carrying an Ising charge and thereby acquiring a phase $\pi$
upon encircling a vison. In our earlier notation this is the
nonsuperfluid periodic state we dubbed $FL_{Sm}^{2q}$, in which
$(2\pi,0)$ vortices have proliferated, but dislocations remain
bound. Lowering $K_\sigma$ drives visons gapless, corresponding to a
proliferation of the $(\pi,a/2)$ fractional defects that induces a
transition to the homogeneous but orientationally ordered (nematic)
nonsuperfluid state, $FL_N$. We note that condensation of $(\pi,a/2)$
in the presence of unbound $(2\pi,0)$ defects automatically also leads
to a proliferation of $(\pi,-a/2)$ defects and therefore (aside from
the nematic conventional order) the resulting $FL_N$ is fully
disordered.

In the opposite regime of large $t_\phi$ and small $t_\theta$ $2\pi$
vortices in $\theta$ proliferate, driving $\rho_q$ normal and
restoring the $U_\theta(1)$ translational symmetry, while keeping $b$
condensed. For large $K_\sigma$ vison remains gapped and the resulting
superfluid homogeneous state exhibits gapped density excitations
$\rho_q$, carrying an Ising charge and nontrivial statistics with the
vison. The resulting state is the charge-4 nematically ordered
superfluid, $SF^4_N$.  Upon reducing $K_\sigma$ visons proliferate,
driving a transition to $FL_N$ through this alternate route.

In contrast to lowering $K_\sigma$, the transitions out of the
$FL^{2q}_{Sm}$ and $SF^4_N$ states at large $K_\sigma$ can be driven
by respectively lowering $t_\phi$ and $t_\theta$ and thereby unbinding
the second set of integer defects, $(0,a)$ and $(2\pi,0)$,
respectively. Since visons remain gapped, the resulting nonsuperfluid
nematic state is the topologically ordered $FL_N^*$, qualitatively
distinct from $FL_N$. The deconfinement transition $FL_N^*$-$FL_N$
is then driven by lowering $K_\sigma$ through a condensation of visons
and is in the inverted Ising universality class.

A naive attempt at a generalization of the above Ising gauge theory
action, \rf{IsingGaugeS} to the rotationally invariant smectic form
suggests a replacement of the $t_\theta$ operator by a gauge-invariant
lattice Laplacian. However, on general grounds, without fine-tuning
such Ising lattice form appears to preclude a fully
rotationally-invariant formulation necessary for a fully-rotationally
invariant LO (superfluid smectic) state.

Finally, we note that above orientationally ordered state can further
disorder into isotropic states by proliferation of disclinations. We
leave a more detailed study of these phases and the corresponding
transitions to future research.

\section{Fermions}
\label{sec:fermions}

So far all of our discussion following the defining microscopic model
in Sec.\ref{sec:microscopics} has been confined to the bosonic sector
of the Larkin-Ovchinnikov state. The resulting low-energy quantum
thermodynamics is encoded in the Hamiltonian, $H_{LO}$, Eq.\rf{HgmLO3}
(and the Lagrangian $\cL_{LO}$, Eq.\rf{Slo}) for the superfluid phase
$\phi$ and smectic phonon $u$ Goldstone modes.  In contrast to the
fully gapped superconductors where this is sufficient at low energies,
the gapless nature of the LO state also requires the inclusion of
gapless fermionic excitations for a complete description of the state.
While a detailed analysis of these is beyond the scope of the present
manuscript, below we comment on a few a key features of the fermionic
sector of the LO state.

\subsection{Gapless fermionic excitations in the Larkin-Ovchinnikov
  state near $h_{c2}$}
As we have seen in our discussion of the Goldstone modes, there are
two complementary descriptions of the FFLO states, respectively valid
near $h_{c2}$ and $h_{c1}$.  Just below $h_{c2}$ at the continuous
FL-FFLO phase the pairing order parameter, $\Delta(\rv)$ is small with
a weak sinusoidal modulation, and the momentum-space description is
most appropriate. The corresponding mean-field many-body wavefunction
for e.g., the FF state is given by Eq.\rf{eq:groundstate}. It is of
the BCS form with a range of $\kv$ over which the energy of the
corresponding Bogoliubov quasi-particles is driven negative by the
Zeeman field. These therefore form a Fermi sea with all the
phenomenology associates with the gapless fermionic excitations at the
Fermi surface.

For a more general FFLO state defined by a set of reciprocal lattice
vectors $\qv_n$, no exact analytical solution of the BdG Hamiltonian
is available. The periodic $\Delta(\rv)$ couples all particle (atoms)
and hole states connected by the reciprocal lattice $\qv_n$, leading
to an anomalous $\Delta(\rv)$-dependent band structure of Bogoliubov
quasi-particles.

However, some progress can be made for a purely sinusoidal LO state
with only $\pm \qv$ reciprocal lattice vectors. As discussed earlier a
BCS-like mean-field Hamiltonian for such a state is given by 
\begin{widetext}
\bea
&&H^{LO} =  \sum_{\bk,\sigma} (\epsilon_k -\mu_\sigma) 
\ch_{\bk\sigma}^{\dagger} \ch_{\bk\sigma}^{\phdag}
+\sum_{\bk}\Big(
\Delta^*_\bQ 
\ch_{-\bk\downarrow}^{\phdag}
\ch_{\bk+\bQ\uparrow}^{\phdag}
+ \Delta^*_{-\bQ} 
\ch_{-\bk\downarrow}^{\phdag}
\ch_{\bk-\bQ\uparrow}^{\phdag} + h.c.\Big).
\label{eq:loBCS}
\eea
\end{widetext}
The difference from the FF BCS Hamiltonian, analyzed exactly in
Sec.\ref{sec:microscopics} is in the appearance of both $\Qv$ and
$-\Qv$ Fourier components of $\Delta(\rv)$, taken to be equal for the
simplest cosine LO form. These couple each creation operator
$\ch_{-\bk\downarrow}^{\dagger}$ to two fermionic atom annihilation
operators $\ch_{\bk+\Qv\uparrow}^{\phdag}$ and
$\ch_{\bk-\Qv\uparrow}^{\phdag}$, with each in turn coupling to
another set of two operators at different $\kv$'s, thereby generating
an infinite-dimensional space that needs to be diagonalized.  Focusing
on the three-dimensional subspace for each $\kv$, the Hamiltonian can
be rewritten in the following \BdG\ form:
\bea
H^{LO} &=& \sum_{\bk} 
\Psih^\dagger_\bk
\begin{pmatrix}
\xi_{\bk+\bQ\uparrow} & \Delta^\phstar_\bQ & 0\cr 
\Delta^*_\bQ  &-\xi_{-\bk\downarrow} & \Delta^*_{-\bQ}\cr
0 & \Delta^\phstar_{-\bQ} & \xi_{\bk-\bQ\uparrow}\end{pmatrix}
\Psih_\bk + \sum_\bk \xi_{-\bk\downarrow},\nonumber\\
&=& \sum_{\bk} 
\Psih^\dagger_\bk\hat{H}_{\text{BdG}}\Psih_\bk 
+ \sum_\bk \xi_{-\bk\downarrow},
\label{eq:BdGlo}
\eea
with the three component generalization of the Nambu spinor given by
\be
\label{eq:nambu}
\Psih_\bk\equiv\begin{pmatrix}
\ch_{\bk+\bQ\uparrow}^{\phdag} \cr 
\ch_{-\bk\downarrow}^{\dagger} \cr
\ch_{\bk-\bQ\uparrow}^{\phdag} 
\end{pmatrix},
\ee
and $\hat{H}_{\text{BdG}}$ the corresponding \BdG\ Hamiltonian matrix.
The nontrivial eigenstates encode that the excitation in the LO state
is a linear combination of a hole $(-\kv,\downarrow)$, an atom
$(\kv+\Qv,\uparrow)$, and an atom $(\kv-\Qv,\uparrow$).

Although it appears that the problem reduces to a diagonalization in
this three-dimensional subspace, in fact (in contrast to the single
Fourier component FF state, where $\hat{H}_{\text{BdG}}$ is
block-diagonal, here) there is a coupling between a bottom component
of a spinor $\Psih_\bk$ and the top component of the spinor
$\Psih_{\bk'=\bk-2\bQ}$,
\be
\label{eq:nambuCouple}
\Psih_{\bk-2\bQ}=\begin{pmatrix}
\ch_{\bk-\bQ\uparrow}^{\phdag} \cr 
\ch_{-\bk+2\bQ\downarrow}^{\dagger} \cr
\ch_{\bk-3\bQ\uparrow}^{\phdag}\cr
\end{pmatrix}.
\ee
This leads to the aforementioned infinite-dimensional space to
diagonalize, corresponding to the band structure of the Bogoliubov
quasi-particles, analogous to a system with a diagonal periodic
potential.

Despite these complications, for a large enough $q$ (such that $q^2/2m
\gg \Delta_q$, well satisfied near $h_{c2}$ where $q\approx 1/\xi$) an
approximate treatment is possible because the coupling is dominated by
the degenerate particle-hole states near the Fermi surface, which
reduces the problem to only a {\em pair} of states for every value of
$\kv$. For positive (negative) $\kv$ the pair is the top (bottom) two
components of the spinor $\Psih_\bk$, leading to a Cooper-pair with
$\Qv$ ($-\Qv$) center of mass momentum.

Diagonalizing the BdG Hamiltonian then leads to two sets of the
Bogoliubov quasi-particle operators $\alphah_{\bk\sigma\Qv}$,
$\alphah_{\bk\sigma -\Qv}$ 
\bse\label{eq:alphasLO}
\bea
\alphah_{\bk\uparrow\Qv} 
&\approx& u_{\kv,\bQ} \ch_{\bk+\frac{\bQ}{2}\uparrow}^{\phdag} + 
v_{\kv,\bQ}  \ch_{-\bk+\frac{\bQ}{2}\downarrow}^{\dagger},
\label{alpha_up1}\\
\alphah_{-\bk\downarrow\Qv}^\dagger 
&\approx& -v_{\kv,\bQ}^*  \ch_{\bk+\frac{\bQ}{2}\uparrow}^{\phdag} +
u_{\kv,\bQ}^*  \ch_{-\bk+\frac{\bQ}{2}\downarrow}^{\dagger},
\label{alpha_down1}\\
\alphah_{\bk\uparrow -\Qv} 
&\approx& u_{\kv,-\bQ}  \ch_{\bk-\frac{\bQ}{2}\uparrow}^{\phdag} + 
v_{\kv,-\bQ}  \ch_{-\bk-\frac{\bQ}{2}\downarrow}^{\dagger},
\label{alpha_up2}\\
\alphah_{-\bk\downarrow -\Qv}^\dagger 
&\approx& -v_{\kv,-\bQ}^*  \ch_{\bk-\frac{\bQ}{2}\uparrow}^{\phdag} +
u_{\kv,-\bQ}^*\ch_{-\bk-\frac{\bQ}{2}\downarrow}^{\dagger},
\label{alpha_down2}
\eea
\ese
with the corresponding four branches of excitation spectrum,
$E_{\bk\sigma\Qv_i}$
\bse
\label{eq:EsigmaQpm}
\bea
E_{\bk\uparrow\Qv} &\approx& (\varepsilon_k^2 +\Delta_\bQ^2)^{1/2} 
- h + \frac{\bk \cdot \bQ}{2m},
\label{ekuparrowLO}\\
E_{\bk\downarrow\Qv} &\approx& (\varepsilon_k^2 +\Delta_\bQ^2)^{1/2}
+ h - \frac{\bk \cdot \bQ}{2m},
\label{ekdownarrowLO}\\
E_{\bk\uparrow-\Qv} &\approx& (\varepsilon_k^2 +\Delta_\bQ^2)^{1/2}
- h - \frac{\bk \cdot \bQ}{2m},
\label{ekuparrowQmLO}\\
E_{\bk\downarrow-\Qv} &\approx& (\varepsilon_k^2 +\Delta_\bQ^2)^{1/2} 
+ h + \frac{\bk \cdot \bQ}{2m},
\label{ekdownarrowQmLO}
\eea
\label{EkLO}
\ese
$\varepsilon_k = \frac{k^2}{2m} - \mu + \frac{q^2}{8m}$, and the
coherence factors $u_k$, $v_k$, approximately given by the FF
expressions in \rfs{uvFF}. In the first [second] pair of equations,
Eqs.\rf{alpha_up1},\rf{alpha_down1}
[Eqs.\rf{alpha_up2},\rf{alpha_down2}], the particle-hole hybridization
is via $\Delta_q e^{i\qv\cdot\rv}$ [$\Delta_q e^{-i\qv\cdot\rv}$], as
for the FF state.

\begin{figure}[bth]
\centering
\setlength{\unitlength}{1mm}
\begin{picture}(40,62)(0,0)
\put(-15,-5){\begin{picture}(30,30)(0,0)
\includegraphics{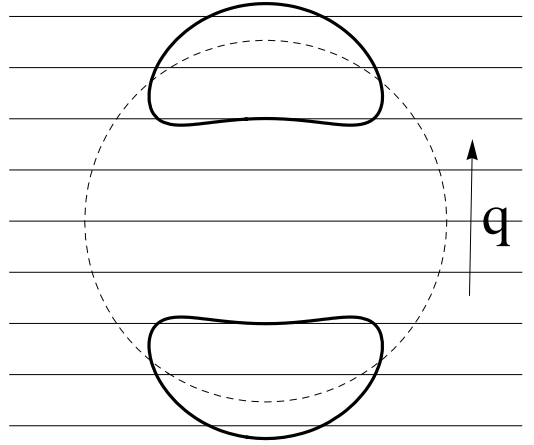}
\end{picture}}
\end{picture}
\caption{An illustration of Fermi pockets (full curve) of the gapless
  Bogoliubov quasi-particles characteristic of the Larkin-Ovchinnikov
  ground state. The periodic array of domain-walls in
  $\Delta_{LO}(z)$, the associated wavevector $\qv$, and the
  Fermi-surface of the underlying normal state (dashed circle) are
  also indicated.}
\label{fig:fermiPockets}
\end{figure}

As demonstrated for the FF state in Sec.\rf{sec:microscopics}, it is
clear from the spectra $E_{\kv\sigma\pm\Qv}$, \rfs{EkLO}, that the LO
state exhibits $\kv$ regions of both gapped and gapless fermionic
excitations, with closing of the gap driven by a combination of the
Zeeman (imbalance) energy $h$ and the Doppler shift $\kv\cdot\Qv$. For
pairing driven by $\Delta_q e^{i\qv\cdot\rv}$ ($\Delta_q
e^{-i\qv\cdot\rv}$) the gapless states appear at the minimum of the
$E_{\bk\uparrow\Qv}$ ($E_{\bk\uparrow-\Qv}$) located in a wedge
$-\theta_{m}<\theta<\theta_m$ around $\kv\parallel-\Qv$
($\kv\parallel\Qv$), where for positive $h$, $E_{\bk\uparrow\Qv}$
($E_{\bk\uparrow-\Qv}$) is driven negative. The LO ground state thus
takes the form given in Eq.\rf{eq:groundstate}, exhibiting Fermi
pockets of Bogoliubov quasi-particles, with a Fermi surface of gapless
excitations defined by $E_{\tilde{\bk}_F\sigma,\pm\Qv}=0$, as
illustrated in Fig.\ref{fig:fermiPockets}.  

In the complementary wedge of $\kv$ around $\kv\cdot\qv=0$, i.e.,
running along the LO stripes given by the nodes of the LO order
parameter, the superconducting gap remains finite, growing to its
maximum value $\Delta_\Qv-h > 0$ (suppressed by $h$) as
\begin{eqnarray}
E_{gap}&\approx& E^{max}_{gap}-\frac{q k_z}{2m},\nonumber\\
&\approx& E^{max}_{gap}-\frac{q k_F}{2m}\sqrt{1-k_\perp^2/k_F^2}.
\end{eqnarray}
The quadratic $k_\perp$ dispersion along the stripes is characterized
by an enhanced effective mass $m_{eff}\approx
m\epsilon_F/\Delta_{BCS}\gg m$.  

While above analysis gives some understanding of the LO states, it is
approximate and does not address the detailed nature of a more generic
LO-like states. These may be understood by self-consistently
diagonalizing the BdG equations in real space. An estimate of the
mean-field BCS-LO and LO-FL phase boundaries requires a computation of
the total energy as was done in Ref.\onlinecite{SRaop} for the FF
state. A complete solution for the LO state requires a numerical
analysis\cite{MachidaNakanishiLO,BurkhardtRainerLO,MatsuoLO}.

\subsection{Gapless fermionic excitations in the Larkin-Ovchinnikov
  state near $h_{c1}$}

The above calculation of the LO quasi-particle spectrum near $h_{c2}$
can be complemented by a treatment valid near $h_{c1}$.  At low $h$,
the LO order parameter $\Delta_{LO}(z)$ is given by a periodic (period
$2\pi/q$) array of $\pm\Delta_0\approx\pm\Delta_{BCS}$ domain-walls of
width
$\xi$\cite{MachidaNakanishiLO,BurkhardtRainerLO,MatsuoLO,YoshidaYipLO},
rather than a single harmonic.

The form of the quasi-particle excitations is determined by a
two-component Nambu wavefunction
$\vec\psi=(\psi_\uparrow(\xv,z),\psi_\downarrow(\xv,z))$ satisfying
the BdG equation
\bea
\begin{pmatrix}
\frac{\hat{p}^2}{2m}-\epsilon_F & \Delta(z)\cr 
\Delta(z) &-\frac{\hat{p}^2}{2m}+\epsilon_F\cr\end{pmatrix}
\begin{pmatrix}
\psi_\uparrow\cr
\psi_\downarrow\end{pmatrix}_\alpha
=E_\alpha
\begin{pmatrix}
\psi_\uparrow\cr
\psi_\downarrow\end{pmatrix}_\alpha,
\label{eq:BdGeqn}
\eea
where $\hat{p}\equiv\nabla$ and we took the order parameter to be real
(characteristic of the mean-field LO state) stripe domain-walls lying
in the $\xv$ plane, with the normal along $\zh$, and approximated the
chemical potential by a Fermi energy $\epsilon_F$, valid deep in the
BCS regime. In a matrix form, the BdG equation is given by
\begin{eqnarray}
\left[\left(\frac{\hat{p}^2}{2m}-\epsilon_F\right)\sigma_z +\Delta(z)\sigma_x\right]
\vec{\psi}(\xv,z) = E\vec{\psi}(\xv,z).
\end{eqnarray}
For a striped LO state we can utilize translational invariance along
the stripes, $\xv$, taking
$\vec{\psi}(\xv,z)=\vec{\psi}_{\kv_\perp}(z)e^{i\kv_\perp\cdot\xv}$, with
$\vec{\psi}_{\kv_\perp}(z)$ satisfying
\begin{eqnarray}
\left[\left(\frac{\hat{p}_z^2}{2m}-\tilde\epsilon_F(k_\perp)\right)\sigma_z 
+\Delta(z)\sigma_x\right]
\vec{\psi}_{\kv_\perp}(z) = E_{k_\perp}\vec{\psi}_{\kv_\perp}(z),\nonumber\\
\end{eqnarray}
with $\tilde\epsilon_F(k_\perp)=\epsilon_F-k_\perp^2/2m$ the effective
1d Fermi energy.

The low-energy spectrum is determined by the excitations near the
Fermi energy
\begin{eqnarray}
  \vec{\psi}_{\kv_\perp}(z)\approx \vec\phi^+_{\kv_\perp}(z) e^{i\tilde{k}_F z} +
   \vec\phi^-_{\kv_\perp}(z) e^{-i\tilde{k}_F z}
\end{eqnarray}
with $\tilde{k}_F^2/2m \equiv \tilde\epsilon_F(k_\perp)$ and the
envelope wavefunctions $\vec\phi^\pm_{\kv_\perp}(z)$ satisfying
\begin{eqnarray}
&&\hspace{-.5cm}\left[\left(\pm\tilde{v}_F\hat{p}_z
      +\frac{\hat{p}_z^2}{2m}
      -\tilde\epsilon_F\right)\sigma_z 
    +\Delta(z)\sigma_x\right]
  \vec{\phi}^\pm_{\kv_\perp}(z)\nonumber\\
&&\hspace{6cm}= E_{k_\perp}\vec{\phi}^\pm_{\kv_\perp}(z),\nonumber\\
\label{phiBdG}
\end{eqnarray}
For $\tilde{\epsilon}_F(k_\perp)$ sufficiently large, so that we can
linearize around the Fermi points $\pm\tilde{k}_F$ by neglecting the
quadratic correction to the dispersion,
$H_{p^2}=\frac{\hat{p}_z^2}{2m}\sigma_z$, we obtain
\begin{eqnarray}
\left[\left(\mp i\tilde{v}_F\partial_z
-\tilde\epsilon_F\right)\sigma_z 
    +\Delta(z)\sigma_x\right]
  \vec{\phi}^\pm_{\kv_\perp}(z) =
E_{k_\perp}\vec{\phi}^\pm_{\kv_\perp}(z).\nonumber\\
\end{eqnarray}
For a single ``- to +'' domain-wall the solution can be readily
found\cite{ZagoskinBook,AshvinThesis} and exhibits two normalizable
Andreev zero-energy bound states,
\bse
\begin{eqnarray}
\vec{\phi}^\pm_{\kv_\perp}(z) &=& A
\begin{pmatrix}
1\cr
\mp i
\end{pmatrix}
e^{-\frac{1}{\tilde v_F}\int_0^z\Delta(z')dz'},\\
&\approx&\frac{1}{\sqrt{2\tilde\xi}}
\begin{pmatrix}
1\cr
\mp i
\end{pmatrix}
e^{-|z|/\tilde\xi},
\end{eqnarray}
\label{phipm}
\ese
where $\tilde{\xi}\equiv \tilde v_F/\Delta_0$ is the effective
coherence length (setting the width of the bound state) and in the
second line we approximated the ``- to +'' domain-wall by a step
function $\Delta(z)=\Delta_0\text{sgn}(z)$ valid for
$z\gg\tilde{\xi}$. For the opposite sign ``- to +'' domain-wall,
$\Delta(z)=-\Delta_0\text{sgn}(z)$ the normalizable Andreev zero modes
are given by identical expressions, but with the reversed sign in
front of the $i$ in the second component of the Nambu spinor.

The minimum excitation gap is determined by the nonlinear correction
to the free spectrum, the perturbation
$H_{p^2}=\frac{\hat{p}_z^2}{2m}\sigma_z$, that we incorporate through
a degenerate perturbation theory in the two-component degenerate
subspace of Andreev zero modes,
$\vec{\phi}^\pm_{\kv_\perp}(z)$. A simple computation gives the
off-diagonal matrix elements $H_\pm = \langle
\vec{\phi}^+|H_{p^2}|\vec{\phi}^-\rangle\approx\frac{1}{2m\tilde\xi^2}$,
splitting the Andreev zero modes to $E^\pm_{k_\perp}=\pm E_{k_\perp}$, with
\bse
\begin{eqnarray}
  E_{k_\perp}&\approx&\frac{1}{2m\tilde\xi^2}
  =\frac{\Delta_0^2}{4\tilde\epsilon_F(k_\perp)},\ \ \mbox{for
    $k_\perp^2/2m\ll\epsilon_F$},\\
  &=&\frac{\Delta_0^2}{4(\epsilon_F-k_\perp^2/2m)}
  \approx\frac{\Delta_0^2}{4\epsilon_F}+
  \left(\frac{\Delta_0}{2\epsilon_F}\right)^2\frac{k_\perp^2}{2m}. 
\hspace{1cm}
\label{Ekper2}
\end{eqnarray}
\label{Ekper}
\ese
The maximum of the gap can be estimated through a variational solution
of the Eq.\rf{phiBdG} for the nondegenerate case of $\tilde
v_F=\tilde\epsilon_F=0$, giving
\bse
\bea
\xi_{min} &\approx&\xi_0(\Delta_0/\epsilon_F)^{1/3},\\
E_{gap}^{max} &\approx&\Delta_0(\Delta_0/\epsilon_F)^{1/3}
\eea
\ese
for the width of the state localized on the domain-wall and the
maximum of the gap.  

For a periodic array of domain-walls with period $a=2\pi/q_0$ (the LO
state), the coupling between the Andreev states localized on each
domain wall splits their energy into bands separated by gaps and
dispersing with $-\pi/a < k_z\le \pi/a$. The eigenstate for the bottom
of the band, $k_z=\tilde{k}_F$ can be obtained by generalizing the
single domain-wall zero-energy state, Eq.\rf{phipm} to a periodic
array $\Delta_{LO}(z)$. The solution for the bottom of the Andreev
band is simply a sum of two periodic arrays of states, localized on
``- to +'' (at $z=0$) and on ``+ to -'' (at $z=a/2$) domain-wall
arrays, respectively
\begin{widetext}
\begin{eqnarray}
\vec{\phi}^\pm_{\kv_\perp}(z) &=& \frac{1}{\sqrt{4N\tilde\xi}}\left[
\begin{pmatrix}
1\cr
\mp i
\end{pmatrix}
e^{-\frac{1}{\tilde v_F}\int_0^z\Delta_{LO}(z')dz'}
+\begin{pmatrix}
1\cr
\pm i
\end{pmatrix}
e^{\frac{1}{\tilde v_F}\int_{a/2}^z\Delta_{LO}(z')dz'}\right].
\end{eqnarray}
\end{widetext}

For a nonzero chemical potential difference $h$ (Zeeman energy) to
impose the spin imbalance, the spin up excitation spectrum \rf{Ekper2}
simply uniformly shifts down by $h$.  $h$ exceeding the minimum gap
$E_{gap}^{min}=\Delta_0^2/4\epsilon_F$, Eq.\rf{Ekper}, generically
induces Fermi pockets (Fig.\ref{fig:fermiPockets}) of the Bogoliubov
quasi-particles centered around $\pm\qv$ out to $k_\perp^\pm
=\pm\frac{2\epsilon_F}{\Delta_0}
\sqrt{2m(h-\Delta_0^2/4\epsilon_F)}$\cite{comment1dPockets}. As
expected (from the fact that these are different regimes of the same
LO phase) the above outlined spectrum is in a qualitative agreement
with that near $h_{c2}$, found in the previous subsection.

\subsection{Fermion-Goldstone modes coupling in the Larkin-Ovchinnikov
  state}

\begin{figure}[bth]
\centering
\setlength{\unitlength}{1mm}
\begin{picture}(100,25)(0,0)
\put(15,-20){\begin{picture}(110,25)(0,0)
\includegraphics{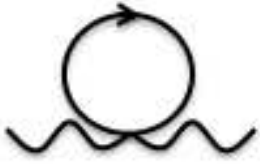}
\end{picture}}
\end{picture}
\caption{Lowest order Feynman diagram of a fermionic contribution, that
  corrects the superfluid stiffness.}
\label{fig:correct_rhos}
\end{figure} 

\begin{figure}[bth]
\centering
\setlength{\unitlength}{1mm}
\begin{picture}(100,25)(0,0)
\put(17,-20){\begin{picture}(110,25)(0,0)
\includegraphics{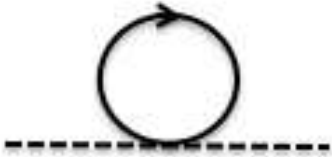}
\end{picture}}
\end{picture}
\caption{Lowest order Feynman diagram of a fermionic contribution,
  that corrects the smectic compressional modulus.}
\label{fig:correctB}
\end{figure} 

\begin{figure}[bth]
\centering
\setlength{\unitlength}{1mm}
\begin{picture}(100,25)(0,0)
\put(15,-27){\begin{picture}(110,25)(0,0)
\includegraphics{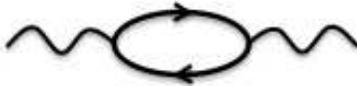}
\end{picture}}
\end{picture}
\caption{Lowest order subdominant Landau-damping contribution to the
  superfluid stiffness coming from finite fermion density.}
\label{fig:LandauDamping}
\end{figure} 

As discussed above, the LO state is characterized by a simultaneous
presence of gapless Goldstone modes (particularly ``soft'' for the
fully rotationally invariant geometry of an isotropic trap) and
gapless fermions. Consequently, for a complete description a coupling
between these must be included. These can in principle be derived from
the microscopic BCS Hamiltonian evaluated inside the LO state beyond a
conventional mean-field treatment. Alternatively, the
fermion-Goldstone mode couplings can be simply deduced based on
symmetry considerations. The leading ones include, the coupling of the
supercurrent and of the LO phonon to the quasi-particle current and
number density, and are given by
\bse
\begin{eqnarray}
H_{j_s,j}&\sim&\nabla\phi\cdot \psi^\dagger i\nabla\psi + h.c.\\
H_{j_s,n}&\sim&(\nabla\phi)^2\psi^\dagger\psi\\
H_{a-p}&\sim&\Big(\partial_z u + \frac{1}{2}(\nabla u)^2\Big)
\psi^\dagger\psi + (\nabla u\cdot\psi^\dagger i\nabla\psi)^2 + h.c..\nonumber\\
\label{couplingsFB}
\end{eqnarray}
\ese 
Because atom number conservation is effectively ``broken'' in the
superfluid LO state, above couplings must be supplemented by the
anomalous (number violating) operators, such as e.g.,
$(\nabla\phi)^2\psi\psi+h.c.$.  Effects of these interactions on the
fermionic and collective bosonic ($\phi,u$) spectral functions
requires a detailed analysis, that parallels studies of gauge
fields\cite{Pincus,Reizer,HLR,Polchinski,AIM,SSLee}, Goldstone modes
\cite{OKF01prb} and critical modes\cite{EAKim,MetlitskiSachdev}
coupling to electrons in gapless superconductors and metals.  A
preliminary analysis suggests that in the LO state these derivative
couplings (enforced by the underlying gauge and spatial symmetries)
only lead to a finite renormalization of model's parameters as well as
Landau-like damping of the Goldstone modes. In addition to the above
symmetry-dictated couplings, the presence of gapless fermions can
generate Berry's phase terms\cite{commentBerry}, that can
qualitatively modify the conventional LO phonon dynamics derived in
Sec.\ref{sec:dynamicsGM}. We leave a detailed study of these
interesting questions to future research.

\section{Larkin Ovchinnikov states in a trap}
\label{sec:LDA}

The primary experimental application of our results is to polarized
paired superfluidity in trapped degenerate atomic gases.  It is thus
crucial to extend our bulk analysis to take into account the effect of
the trapping potential $V_t(\br)$, that in a typical experiment is
well-approximated by a harmonic-oscillator potential.  While a full
analysis of the effect of the trap is beyond the scope of this
manuscript, in the present section we study this problem within the
well-known local density approximation (LDA).  We note that several
recent studies (e.g., Refs.~\onlinecite{Mizushima,
  Pieri,SRprl,SRaop,Torma,Yi,Chevy,DeSilva,HaqueStoof}) have also
addressed polarized superfluidity in a trap.

\subsection{The local density approximation}

Much like the WKB approximation, LDA corresponds to using expressions
for the bulk system, but with an effective local chemical potential
$\mu(r)=\mu-V_t(r)$ in place of $\mu$. The validity of the LDA
approximation relies on the smoothness of the trap potential, with the
criterion that $V_t(r)$ varies slowly on the scale of the {\it
  longest\/} physical length $\lambda$ (the Fermi wavelength,
scattering length, effective range, etc.) in the problem, i.e.,
$(\lambda/V_t(r)) dV_t(r)/dr\ll 1$. Its accuracy can be equivalently
controlled by a small parameter that is the ratio of the single
particle trap level spacing $\delta E$ to the smallest characteristic
energy $E_c$ of the studied phenomenon (e.g, the chemical potential,
condensation energy, etc.), by requiring $\delta E/E_c \ll 1$.  Within
the LO state the longest length is clearly the LO period,
$a=2\pi/q_0$, that near $h_{c2}$ is bounded by the coherence length
(that near unitarity can be as short as inter-atomic spacing $\sim
R/N^{1/3}$, where $R$ is the trapped condensate radius and $N$ is the
total number of atoms), and thus $\ll R$. Thus, in this regime, away
from $h_{c1}$ (where for a continuous BCS-LO transition the period $a$
is expected to diverge) the effects of the trap can be safely treated
within the LDA.

The generalization of the model of an imbalanced resonant Fermi gas to
a trap is straightforward:
\begin{eqnarray}
\hspace{-0.7cm}H &=& \int
d^3r\Big(\hat{\psi}^\dagger_\sigma\frac{-\nabla^2}{2m}\hat\psi_\sigma 
+ (V_t(\br) - \mu_\sigma(\rv))\hat{\psi}^\dagger_\sigma\hat\psi_\sigma 
\nonumber\\
&+&\fermiint\hat\psi^\dagger_\uparrow\hat\psi^\dagger_\downarrow\hat\psi_\downarrow
\hat\psi_\uparrow\Big),
\label{eq:singlechannellda}
\end{eqnarray}
where $\hat\psi_\sigma(\br)$ is a fermionic field operator with a
Fourier transform $\ch_{\bk\sigma}$.  Henceforth, to be concrete, we
shall focus on an isotropic harmonic trap (although this
simplification can easily be relaxed) with
\bse
\begin{eqnarray}
V_t(\br) &=& \oh m \omega_t^2 r^2,\\
&\equiv&\mu\frac{r^2}{R^2},
\end{eqnarray}
\ese
latter expression defining the cloud size $R$.  Within LDA (valid for
a sufficiently smooth trap potential $V_t(r)$, see above), locally the
system is taken to be well-approximated as {\it uniform\/}, but with a
local chemical potential given by
\bse
\begin{eqnarray}
\label{eq:mulda}
\mu(r) &\equiv& \mu - \frac{1}{2} m \omega_t^2 r^2,\\
&=& \mu\left(1-\frac{r^2}{R^2}\right),
\end{eqnarray}
\ese
where the constant $\mu$ is the true chemical potential (a Lagrange
multiplier) enforcing the total atom number $N$.  The
spatially-varying spin-up and spin-down local chemical potentials are
then:
\bse
\begin{eqnarray}
\mu_\uparrow(r) &=& \mu(r) +h ,
\\
\mu_\downarrow(r) &=& \mu(r) - h,
\end{eqnarray}
\ese
with the chemical potential difference $h$ {\it uniform\/}.

Consequently, within LDA the system's energy density is approximated
by that of a uniform system, \rf{Egs}, with the spatial dependence
(via the trap) entering only through $\mu(r)$.  The ground state
energy is then simply a volume integral of this energy density.  Thus,
the phase behavior of a uniform system as a function of chemical
potential, $\mu$, translates into a spatial cloud profile through
$\mu(r)$, with the critical phase boundaries $\mu_c$ corresponding to
critical radii defined by $\mu_c = \mu(r_c,h)$.~\cite{SRprl,SRaop} As
first predicted\cite{SRprl}, this leads to a shell-like cloud
structure that has subsequently been observed
experimentally~\cite{Zwierlein06Science,Partridge06Science,Shin2006prl,
  Navon2009prl}.

Within the LDA, we can furthermore deduce the effects of the trap on
the structure of the LO state. We turn to this analysis next.
\begin{figure}[bth]
\centering
\setlength{\unitlength}{1mm}
\begin{picture}(100,35)(0,0)
\put(-5,-45){\begin{picture}(110,35)(0,0)
\includegraphics{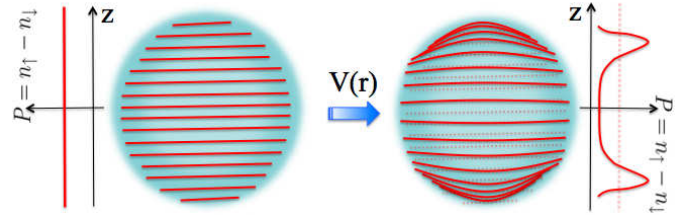}
\end{picture}}
\end{picture}
\caption{The collinear Larkin-Ovchinnikov state, compressed by an
  isotropic trap, $V_t(r)$ (on the right), showing the enhanced
  imbalance ($P\sim -\partial_z u$) confined to the edge of the trap.}
\label{fig:trapLOdistort}
\end{figure}

\begin{figure}[bth]
\centering
\setlength{\unitlength}{1mm}
\begin{picture}(50,40)(0,0)
\put(-15,-33){\begin{picture}(50,50)(0,0)
\includegraphics{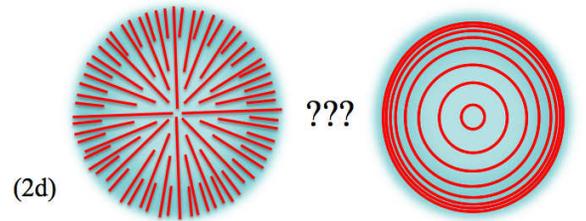}
\end{picture}}
\end{picture}
\caption{Possible alternative forms of the LO state confined in a {\em
    tight} isotropic trap. The form on the right trades off the cost
  of the curvature energy for the dislocation energy.}
\label{fig:onionLOquestion}
\end{figure}

\subsection{Trap-induced elastic distortion}
As discussed earlier, the Larkin-Ovchinnikov state breaks
translational symmetry and as other crystals thereby exhibits a
rigidity to stress. Confinement in a trap induces a variation in the
chemical potential that (as observed experimentally in the BCS
superfluid~\cite{Zwierlein06Science,Partridge06Science,Shin2006prl,
  Navon2009prl}) has the effect of expelling the imbalance
(polarization) to the edge of the cloud. Since in the LO state the BCS
order parameter domain walls ``carry'' the imbalance (the imbalanced
fermions are confined to domain walls in $\Delta_{LO}(z)$, where their
cost is minimum), in the LO state the increased imbalance is
accommodated by the increase in the density of domain walls. Thus, the
trap confinement introduces an effective longitudinal stress that acts
to compress the LO state (increase its local wavevector $q_0(r)$),
where the chemical potential is reduced. This tendency is captured by
introducing a local longitudinal stress, $\sigma(r)$ into the LO
phonon Hamiltonian
\begin{eqnarray}
H_{u}^0&=&\int dz d^2r_\perp\left[
\frac{B}{2}(\partial_z u)^2 + \frac{K}{2}(\nabla_\perp^2 u)^2 
- \sigma(r)\partial_z u\right].\ \ \ \nonumber\\
\label{H0trap}
\end{eqnarray}
Above trap-phonon coupling arises by repeating our derivation in
Secs.\ref{sec:microscopics},\ref{sec:GM} with the trap potential
$V_t(r)$, treating it within the LDA. The effective local chemical
potential $\mu(r)$ enters through many contributions, but the leading
order effect comes from a position-dependent wavevector generalization
of \rfs{eps0Q0J}
\bse
\begin{eqnarray}
q_0(r) &=& 2\alpha\frac{h}{v_F(r)},\\
&=&q_0\left(1-V_t(r)/\mu\right)^{-1/2},\\
&\approx&q_0\left(1+\frac{V_t(r)}{2\mu}\right).
\end{eqnarray}
\ese
In above, consistent with the LDA applicability we utilized the
$V_t(r)/\mu\ll 1$ limit. Since $\delta q_0/q_0\approx 
-\partial_z u\approx -\sigma/B$, we deduce that the local stress is
given by
\bse
\begin{eqnarray}
\sigma(r) &\approx& -\oh \frac{B}{n_0\mu} n(r) V_t(r),\\
&\approx&-\sigma_0 \frac{r^2}{R^2}\left(1-\frac{r^2}{R^2}\right),
\end{eqnarray}
\ese
where $\sigma_0 = B/2$ is the trap induced stress scale,
and we inserted an additional factor of $n(r)/n_0$ to crudely account
for the breakdown of LDA based analysis near the cloud edge, where
atom density, $n(r)$ vanishes at $R$. As expected on physical grounds,
$\sigma(r)$ is a function that vanishes at the trap center and peaked
on its outer shell set by $R$. 

With this, the distortion $u_0(\rv)$ is given by a standard form
\begin{eqnarray}
u(\rv)&=&\int d^3r'G_z(\rv-\rv')\sigma(r'),
\end{eqnarray}
in terms of a Greens function $G_z(\rv)=\partial_z G(\rv)$ that is a
derivative of the smectic Greens function, with
\bse
\begin{eqnarray}
G_z(\rv)&=& \int\frac{d^3q}{(2\pi)^3}\frac{i q_z e^{i\qv\cdot\rv}}
{K q_\perp^4 + B q_z^2},\\
&=&\frac{-1}{8\pi B\lambda|z|}e^{-\frac{r_\perp^2}{4\lambda|z|}}.
\end{eqnarray}
\ese
From this the trap-induced imbalance distortion, $\delta P(\rv)$ is
straightforwardly computed
\bse
\begin{eqnarray}
\hspace{-1cm}  \delta P(\rv)&\approx& -P\partial_z u(\rv),\\
  &=&P\int_{\rv'}\partial_zG_z(\rv-\rv')\sigma(r'),\\
  &\approx& \frac{P}{\mu n_0}\;n(r)V_t(r)\approx P 
\frac{r^2}{R^2}\left(1-\frac{r^2}{R^2}\right),
\end{eqnarray}
\ese
with final unnecessary crude approximation by construction of
$\sigma(r)$ and its coupling to $u$ recovering the expected result.

\subsection{Trap-induced LO 3d-2d dimensional crossover}
\label{sec:3d2dcrossover}
A microscopically accurate account of the trap inside the LO state,
and in particular its coupling to the Goldstone modes with appropriate
boundary conditions (beyond above kludge treatment) is not currently
available. Perhaps a numerical analysis (e.g., numerical solution of
BdG equations or full quantum Monte Carlo
simulation)\cite{Mizushima,Torma,BulgacFFLO} can provide the desired
description.

However, lacking such first principles analysis, we are constrained to
proceed phenomenologically, working directly with the effective
Goldstone mode theory of $u$. Expanding the LO state about the
trap-distorted state $u_0(\rv)$ discussed above, the distortion
$\delta u(\rv)$ is again governed by the smectic Hamiltonian
\rf{H0loSum}. Although, some of the effects of the trap will enter
through the position dependent elastic moduli $B(r), K(r)$, we expect
that the leading effects of the trap are incorporated through the
physically motivated boundary conditions on $u(r)$. Because atom
density vanishes at the edge of the cloud, the LO inner shell is
expected to be surrounded by an outer shell of a fully polarized
normal cloud lacking any positional order. Thus, we expect Neumann
boundary condition 
\begin{eqnarray}
\nh_t\cdot\nabla u = 0,
\end{eqnarray}
($\nh_t$ the unit normal to the cloud's boundary) to be the most
appropriate one to supplement the Euler-Lagrange equation for
$u(\rv)$.  However, the appearance of forth derivative along
$\rv_\perp$, \rfs{H0loSum} requires that this boundary condition be
supplemented by additional ones on the domain-wall normal,
$\nabla_\perp u$.

Given the cylindrical form of smectic elasticity, \rfs{H0loSum},
analytically implementing such spherically symmetric boundary
conditions is quite challenging. However, because our goal here is
more modest, a qualitative understanding of the trap-induced
dimensional crossover can be obtained by simply using cloud size scale
$R$ to cutoff long scales appearing in a bulk analysis. Technically,
our analysis amounts to instead working with periodic boundary
conditions and a cylindrical trap.  The calculational convenience of
such boundary conditions is that they do not modify the form of the
eigenmodes (still Fourier modes), and enter only through a restriction
on the allowed eigenvalues, $q_z=2\pi n_z/L_z$, $q_\perp=2\pi
n_\perp/L_\perp$, with $n_{z,\perp}\in \Z$. Thus, length scales
associated with the trap and finite cloud size, $L_z, L_\perp$,
crudely enter through the minimum allowed momentum eigenvalues
(roughly set by the cloud size $R_{z,\perp}$), and thereby capture the
spatial extent of the lowest phonon eigenmode even for correct
boundary conditions.

\begin{figure}[tbp]
\vskip0.25cm 
\epsfig{file=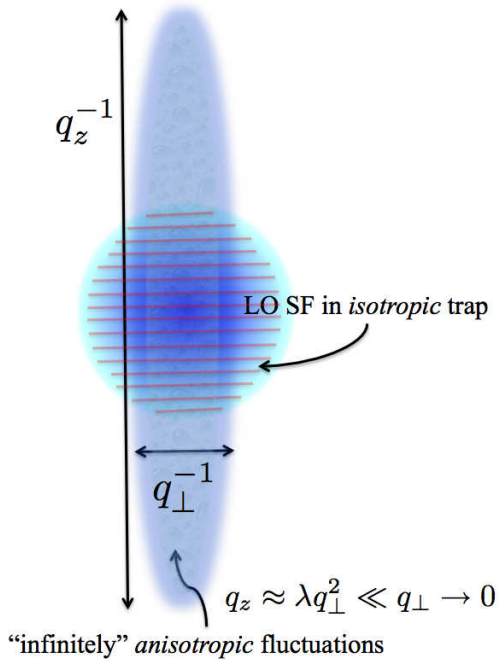,width=7cm,angle=0}
\caption{Illustration of the $q_z=0$ ($z$-independent) phonon modes,
  even in an isotropic trap (indicated by a blue circle) dominating
  over the (scaling-wise) anisotropic, $q_z\sim\lambda q_\perp^2$ bulk
  smectic modes (indicated by an ellipse).}
\label{fig:zeromodeTrap}
\end{figure}

A key observation that emerges from such treatment is that because of
the ``infinite'' anisotropy of the bulk smectic modes in
\rfs{H0loSum}, with $z\sim r_\perp^2/\lambda$ (or equivalently
$q_z\sim\lambda q_\perp^2$), the dimensional crossover is {\em
  qualitatively} different than in systems with more conventional,
scaling-wise isotropic elasticity. Namely, examining smectic bulk
propagator $G_q = 1/(B q_z^2 + K q_\perp^4)$, it is clear that unless
$L_z \approx L_\perp^2/\lambda\gg L_\perp$, the phonon fluctuations
will not be controlled by bulk modes. That is, as illustrated in
Fig.\ref{fig:zeromodeTrap}, for any reasonably shaped trap (even quite
anisotropic one, other than an extremely anisotropic with $L_z \approx
L_\perp^2/\lambda$), fluctuations will be controlled by the $d-1$
dimensional ``zero'' modes, $u_0(\rv_\perp)$, that are uniform along
the $z$ axis, representing compression-free LO undulations. Indeed
this is allowed because of the expected Neumann boundary condition on
$u(z,\rv_\perp)$, that allows a zero-energy cost rigid displacement
along $z$ of the LO domain-walls.

To demonstrate above expectation, we account for these ``zero'' modes
by using a generalized mode expansion of $u(\rv)$ that explicitly
includes the $z$-independent modes
\begin{eqnarray}
  u(\rv) &=& \frac{1}{L_z L_\perp^2}\left[\sum_{q_z,\qv_\perp}
  u(q_z,\qv_\perp) e^{i\qv\cdot\rv} 
  + \sum_{\qv_\perp} u^0(\qv_\perp) 
e^{i\qv_\perp\cdot\rv_\perp}\right].\nonumber\\
\end{eqnarray}
In terms of these the Hamiltonian becomes
\begin{eqnarray}
H_{u}^0&=&\frac{1}{2L_z L_\perp^2}\sum_{q_z,\qv_\perp}
\left[(B q_z^2+K q_\perp^4)|u_\qv|^2 + K\delta_{q_z,0} q_\perp^4|u^0_{\qv_\perp}|^2\right],
\nonumber\\
\label{HuZeroModes}
\end{eqnarray}
and can be used to calculate smectic phonon fluctuations. For example
using $H_u^0$ and equipartition, local root-mean-squared (rms)
fluctuations are given by
\begin{widetext}
\bse
\begin{eqnarray}
\langle u^2(\rv)\rangle&=&\frac{1}{L_z L_\perp^2}\sum_{q_z,\qv_\perp}
\frac{T}{B q_z^2+K q_\perp^4} + 
\frac{1}{L_z L_\perp^2}\sum_{\qv_\perp}\frac{T}{K q_\perp^4} ,\\
&=&\int_{q_z,\qv_\perp}\frac{T}{B q_z^2+K q_\perp^4} + 
\frac{1}{L_z}\int_{\qv_\perp}\frac{T}{K q_\perp^4},\\
&\approx&\frac{T}{4\pi\sqrt{B K}}\ln(\sqrt{L_z\lambda}/a)
+\frac{T}{8\pi^3 K}\frac{L_\perp^2}{L_z},\\
&\approx&\frac{T}{8\pi^3 K}\frac{L_\perp^2}{L_z},\ \ \ \text{for $L_z <
L_\perp^2/\lambda$},
\label{uuZeroModes}
\end{eqnarray}
\ese
\end{widetext}
where in the second line we approximated mode sums by integrals in a
standard way. Indeed as summarized by \rf{uuZeroModes}, by comparing
the two (bulk and zero mode) contributions to $u_{rms}^2$, as
anticipated on general grounds above it is clear that for isotropic
(and even highly anisotropic) traps, the ``zero'' mode second
contribution dominates as long as $L_z < L_\perp^2/\lambda$.

Applying this to a spherically symmetric trap with $L_z\sim
L_\perp\approx R\ll R^2/\lambda$, we conclude that indeed fluctuations
are controlled by the ``zero''-mode LO phonons.  We thus conclude that
all the {\em correlation-function} properties of the LO state will be
even more anomalous, characterized by a $d-1$ dimensional zero-mode
action, that at a harmonic level is given by
\begin{eqnarray}
  S^0_{u} &=&\int_0^\beta d\tau d^{d-1}r_\perp
\left[\frac{\kappa}{2}(\partial_\tau u_0)^2
+ \frac{K}{2}(\nabla_\perp^2 u_0)^2 \right].\ \ \ \ \ \ \ 
\label{S0u}
\end{eqnarray}
More detailed implication of these observations, particularly in
contexts of specific experimental geometries remain to be explored.
We hasten to add, however, that by equipartition, the finite
temperature the LO {\em thermodynamics} will nevertheless be dominated
by bulk modes, simply due to their bulk-to-surface dominance.

\section{Experimental implications}
\label{sec:experiments}

There is a large number of experimentally observable effects, that
emerge from our study. A comprehensive treatment of these requires
further extensive studies, that lie outside the scope of the present
manuscript.  Here we simply sketch out a few of the most important
experimental signatures of our predictions.

\subsection{Larkin-Ovchinnikov order parameter}

As discussed in the Introduction and in Sec.\ref{sec:GMfluctuations},
at finite temperature in the thermodynamic limit (see below for the
discussion in the trap) {\em isotropically}-trapped Larkin-Ovchinnikov
phase is characterized by a {\em vanishing} average LO order
parameter, i.e., $\langle\Delta_{LO}\rangle=0$. Reminiscent of 2d
superfluids and crystals\cite{ChaikinLubensky,KT}, this is a
reflection of its enhanced thermal fluctuations. As with these
well-known examples and other topological phases, this does not
however imply that the state is unstable (at least not in 3d), but
that it requires a finer characterization (e.g., correlation
functions, topological defects, etc.) beyond a simple Landau order
parameter. One of the experimental implications is, that, consequently
the leading nonzero Landau order parameter characterizing the LO state
is the translationally-invariant ``charge''-4 (4-atom pairing)
superconducting order parameter, $\Delta_{sc}$, introduced in
\rfs{Deltasc}. Thus, in the presence of thermal fluctuations the LO
phase corresponds to an exotic state in which the off-diagonal order
is exhibited by pairs of Cooper pairs, i.e., a bound quartet of atoms,
rather than by the conventional 2-atom Cooper
pairs\cite{Berg09nature}. In 2d and 3d this higher order pairing is
driven by arbitrarily low-$T$ fluctuation, rather than by a fine-tuned
attractive interaction between Cooper pairs, and therefore akin to 2d
superfluids and crystals has no mean-field description. While a direct
experimental probe of $\Delta_{sc}$ may be challenging, enhanced
fluctuations in the LO state can be directly observed through
correlation functions to which we turn next.

\subsection{Momentum distribution function}

\begin{figure}[tbp]
\vskip0.25cm 
\epsfig{file=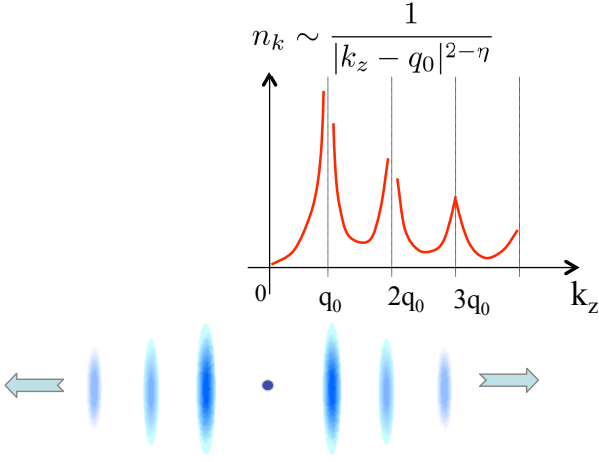,width=6.5cm,angle=-90}
\caption{The finite momentum pairing at $q_0$ and divergent 3d smectic
  phonon fluctuations in the LO state are predicted to be reflected in
  the Cooper-pair center-of-mass momentum distribution function,
  $n_\kv$, displaying power-law Bragg peaks, characteristic of the
  spatial quasi-long-range order. We expect these to be observable
  through the time-of-flight measurements.}
\label{fig:nk}
\end{figure}

Probably the most striking signature of the LO state is the novel form
of the Cooper-pair momentum distribution function, $n_\kv=\langle
\Delta^\dagger_\kv \Delta_\kv\rangle$. As for a conventional
resonantly-paired superfluid, $n_\kv$ should be accessible by a
detuning sweep (controlled by a magnetic field) that projects the
finite-momentum LO Cooper pairs $\Delta_\kv$ on the BCS side onto
tightly bound molecules on the BEC side of the Feshbach
resonance\cite{Regal2004prl}, and then observed through a standard
time-of flight imaging of the resulting molecular condensate. In
contrast to a conventional bosonic and BCS condensates (that in a trap
display a single peak, associated with a condensation into a lowest
trap state, or its interaction-swelled equivalent), the LO condensate
is expected to display a {\em spontaneous} reciprocal lattice of
condensate peaks associated with its periodic structure,
Fig.\ref{fig:nk}.

While in mean-field approximation $n_\kv$ is
predicted\cite{YangFFLOdetect,SRprl,SRaop,RVprl} to resemble a
superfluid in an optical periodic
potential\cite{GreinerOL,BlochReview}, the spontaneous nature of its
translational and orientational symmetry breaking, in the presence of
thermal fluctuations lead to important qualitative distinctions.
Although the full anisotropic form is quite complex and best evaluated
numerically, the asymptotic form of $n_\kv^{LO}$ can be readily
obtained analytically. As calculated in Sec.\ref{sec:GMfluctuations},
$n^{LO}_\kv$ exhibit a power-law (algebraic) peaks around harmonics
$\qv_n$ of the ordering wavevector $\qv_0$, replacing the mean-field
$\delta$-function Bragg peaks of bosons in a periodic
potential\cite{GreinerOL,BlochReview}:
\begin{eqnarray}
  n^{LO}_\kv &\approx& \sum_{q_n\neq0}\frac{n_{q_n}}{|k_z - n
    q_0|^{2-n^2\eta}},\ \ \mbox{for $d=3$},
\label{nkExp}
\end{eqnarray}
where for simplicity we specialized to $\kv = k_z\zh$, the form factor
$n_{q_n}$ is given after \rfs{nkresult} and $\eta = q_0^2
T/(8\pi\sqrt{B K})$. This form is reminiscent of (1+1)d Luttinger
liquids and two-dimensional
crystals\cite{Landau1dsolid,Peierls1dsolid,MerminWagner,KT}, and is a
reflection of the quasi-long-range order of the nonzero-temperature 3d
LO state.  While physically quite distinct, $n^{LO}_\kv$ is
mathematically closely related to the structure function of a
conventional smectic\cite{Caille,deGennesProst}. Another feature of
$n^{LO}_\kv$ is the absence of the $k=0$ condensate peak, that
qualitatively distinguishes the LO state from a
supersolid\cite{Andreev69,Chester70,Leggett70,KimChan}, where
crystalline and superfluid orders merely independently coexist. We
expect these novel features to be the smoking gun for the LO state, in
principle observable in time-of-flight imaging.

\subsection{Structure function}
The density-density correlations, conventionally measured in the
reciprocal space using x-ray or neutron scattering is another
important quantity that can be experimentally probed. The simplest is
the static structure function, that, using $\rho(\rv) \approx
|\Delta_{LO}(\rv)|^2$ is straightforwardly computed (see
Sec.\ref{sec:GMfluctuations}):
\bse
\begin{eqnarray}
  S^{LO}(\qv) &=& \langle \rho_{-\kv}\rho_\kv\rangle,\\
  &\approx& \sum_{q_n}\frac{A_{2q_n}}{|q_z - 2 n q_0|^{2-4n^2\eta}},
\end{eqnarray}
\ese
where for simplicity we evaluated it at $\qv = q_z\zh$ and $A_{2q_n}$
is a form factor. $S^{LO}(\qv)$ contrasts with $n^{LO}_\kv$ by its
insensitivity to the off-diagonal (i.e., superfluid) order. In three
dimensions it also displays quasi-Bragg peaks, but at twice the
reciprocal lattice vectors, $2n q_0$, with the $4n^2\eta$ fluctuation
exponent, and just like conventional
smectics\cite{SqSmExp,deGennesProst,ChaikinLubensky} (unlike
$n^{LO}_\kv$) does exhibit the $q_{n=0}=0$ peak.

After a projection onto the molecular BEC state, a density profile
$\rho(\rv)$ for a large atomic cloud should be measurable in
situ. Computing a Fourier transform of its correlations then allows
one to test $S^{LO}_\kv$, above. This way, in principle the dynamic
structure function $S_{\kv,\omega}$ (a Fourier transform of two-time
density correlation function) should also be accessible.

Alternatively, the dynamic structure function can be directly measured
using Bragg
spectroscopy\cite{KetterleBragg,Steinhauer02prl,Papp08prl}, with its
static limit obtained by integrating over frequencies. Based on
successful measurements of the Bogoliubov mode in bosonic
condensates\cite{Steinhauer02prl,Papp08prl}, we expect that the peaks
in the $S^{LO}_{\kv,\omega}$ (associated with its poles) can be used
to give the dispersion of the collective modes, the superfluid phase
$\phi$ and LO phonon $u$,
\bse
\begin{eqnarray}
\omega_\phi(\kv) &=&\chi_0^{-1}\sqrt{\rho_s^\perp k_\perp^2 + \rho_s^\parallel
k_z^2},\\
\omega_u(\kv) &=&\chi_0^{-1}\sqrt{K k_\perp^4 + B k_z^2},
\end{eqnarray}
\ese
with the corresponding lifetimes remaining to be determined. 

Related to the dynamic structure function is the generalized density
and polarization (spin-imbalance) response function to a periodic
potential. The periodic nature of the LO state makes it possible to
use a perturbation at half the wavelength of the LO state to
resonantly excite long-scale collective modes, such as the
polarization-dipole mode, that can be readily detected in a trapped
gas. This was proposed and carefully investigated for a 1d LO state in
Ref.\onlinecite{CooperModesPRL} and should also apply in the two and
three dimensional cases considered here. We leave the analysis of
these and related experiments to a future more detailed study.

\subsection{Trap effects}
As is clear from the discussion in Sec.\ref{sec:LDA} a detailed study
of the effects of the trap is essential for an explicit contact with
experiments. By introducing system boundaries, a trap modifies the
structure of and fluctuations in the LO state, that, away from
$h_{c1}$ (where LO period can potentially
diverge\cite{PokrovskyTalapov,BurkhardtRainerLO}) can be treated
within LDA.

\subsubsection{Fluctuations}
The detailed phenomenology of the LO state in a trap depends on the
nature of the boundary conditions, that still remain to be
understood. However, for large atomic clouds with small surface to
volume ratio, the boundary effects are expected to be a weak
perturbation to the bulk. As discussed in Sec.\ref{sec:LDA}, in this
case its effects can be taken into account phenomenologically in the
spirit of the finite size scaling of critical
phenomena\cite{ChaikinLubensky,WilsonKogut} by introducing a long
scale cutoff (set by the cloud size $R$) into our bulk quantities. For
observables related to the soft LO phonon $u$, that is strongly
affected by fluctuations this introduces an experimentally observable
sensitivity to the cloud size, $R$ and by extension to the atom number
$N$, trapping frequency $\omega_t$ and the explicit trap anisotropy
$\omega^\perp_t/\omega^\parallel_t$.

Referring to the analysis of
Secs.~\ref{sec:GMfluctuations},\ref{sec:LDA}, that give phonon rms
fluctuations for a gas in a ``box'' of size $L_\perp\times L_z$, we
conclude that in a isotropic trap with cloud size $R\sim L_\perp=L_z$,
the phonon fluctuations are large but finite, in 3d given by
\begin{eqnarray}
u^2_{rms}
&\approx&\frac{T}{8\pi^3K} R + \frac{T}{4\pi\sqrt{B K}}\ln R/a,
\label{exp:uuT}
\end{eqnarray}
where we approximated the phonons by Gaussian fluctuations, neglecting
weak nonlinear elastic effects of Sec.\ref{sec:RG} (that can become
important on scales $R > \xi^{NL}_{\perp,z}$, \rf{xiNL}), and assumed
Neumann boundary conditions that allow for the $q_z=0$ ``zero''
modes. The crossover from the bulk to the ``zero'' modes dominated
regimes takes place for a cloud size $R\gtrsim R_* =
\sqrt{K/B}\ln\sqrt{K/Ba^2}$.

One consequence of this result is that in a trap the thermally
averaged LO order parameter, $\langle\Delta_{LO}(\rv)\rangle^T
=2\tilde{\Delta}_{q_0}(R)\cos\big(\qv_0\cdot\rv)$ no longer vanishes (as it
does in an infinite bulk system), but strongly depends on the cloud
size. For Neumann boundary conditions, its thermally suppressed
amplitude in 3d is given by
\begin{eqnarray}
\hspace{-1cm}
\tilde\Delta_{q_0}(R)&\approx&
\Delta_{q_0}e^{-\oh T/(\xi_0\sqrt{\rho_s^
\perp\rho_s^\parallel})}
e^{-R/R_0}\left(\frac{a}{R}\right)^{\eta/2},
\label{exp:DeltaR}
\end{eqnarray}
with $R_0\approx q_0^2 T/K$ the scale beyond which the ``zero'' modes
become particularly important. Thus through $R(N,\omega_t)$ the order
parameter amplitude is also exponentially sensitive to the atom number
$N$ and trap frequency $\omega_t$. We note that the exponential
dependence on $R$ is a consequence of the dominance of the $q_z=0$
``zero''-modes (under Neumann boundary conditions on $u$) over the
anisotropic smectic bulk modes in an isotropic trap. In contrast,
under Dirichlet boundary conditions that exclude these ``zero''-modes,
the exponential factor is suppressed, and the LO amplitude is a weaker
power-law function of the cloud size, reminiscent of 2d xy-systems,
such as a superfluid confined to two-dimensions. We note, that
Kosterlitz-Thouless phase fluctuation physics has been reported in 2d
trapped atomic superfluids\cite{Hadzibabic,ChengChinKT}, despite the
finite trap size. We therefore expect our predictions for strong
fluctuations effects in the LO phase to also be experimentally
accessible.

Finally we note that given our prediction that generically the LO
state is sandwiched by its descendents phases ($SF_N$, $FL_{Sm}$,
$FL_N$, $FL^*$, etc, rather than a simple vacuum) the analysis and
implementation of boundary conditions for a trap is further
significantly complicated, particularly near $h_{c1}$.

\subsubsection{Phase diagram through shell structure}
As discussed and analyzed in Sec.\ref{sec:LDA}, an even stronger
effect of the trap is that it leads to an effective locally varying
chemical potential $\mu(r)=\mu(1-r^2/R^2)$, and therefore gives slices
through the chemical potential phase diagram as a function of radius
$r$. We thus expect that the phase diagram in
Fig.\ref{fig:phasediagramLO3d} can be ``imaged'' in the spatial cloud
profile, with critical phase boundaries $\mu_c^{(i)}$ (to phase $i$)
translating to critical radii of shells defined by $\mu_c^{(i)} =
\mu(r_c^{(i)},h)$. Given the past success of such phase detection with
bosons in optical potentials (exhibiting ``wedding cake''
profiles)\cite{BlochReview}, and the uniformly paired superfluids and
Fermi liquids of an imbalanced Fermi
gas~\cite{KetterleZwierleinReview,Zwierlein06Science,Partridge06Science,Navon2009prl},
we expect that similar identification will be possible for some of the
LO liquid crystal phases.

\subsection{Response to rotation}
Consistent with its neutral superfluid order, above a critical rate of
rotation, $\Omega_{c1}$ a FFLO state responds to an imposed rotation
by nucleating quantized vortices. Because of its crystalline
superfluid form, even within a mean-field description its vortices are
predicted to form a rich variety of vortex
lattices\cite{ShimaharaRainerVL,KleinVL,YangMacDonaldVL}, that depend
on the nature of the ``host'' FFLO state.

We predict a number of additional interesting vortex features that are
special to strongly fluctuating LO state. One distinguishing feature
of the smectic LO state studied here is its uniaxial anisotropy, that
is strongly tunable with species imbalance. It manifests itself in the
(quantitative) anisotropy of the superfluid densities (stiffnesses),
$\rho_s^{\perp}\ll \rho_s^\parallel$, \rfs{derive:ratio}, and the {\em
  qualitatively} anisotropic dispersion of the LO phonon,
\rfs{omegasPhiU}. The former leads to vortices with a spontaneously
elliptical vortex cores, \rf{anisotropicCore}, that we expect to form
a centered rectangular lattice for vortices (rotation axis) oriented
transversely to $\qv_0$.  This striking feature should be easily
identifiable in the rotation experiments of the type previously used
to identify phase separation in imbalanced Fermi
gases\cite{Zwierlein06Science,KetterleZwierleinReview}. In this
transverse geometry we also expect vortices to be pinned to LO phase
fronts, localized in zeros of domain-walls in $\Delta_{LO}(\rv)$, akin
to vortices in layered superconductors e.g., BISCCO\cite{RMPvortices}.
In contrast for rotation about an axis along $\qv_0$, a conventional
hexagonal vortex lattice is expected.

Another striking feature of the LO state found in
Sec.\ref{sec:defects} is the $\pi$-vortex bound to a half-
$a/2$-dislocation, with a $1/4$ energy cost (in thermodynamic limit)
of a conventional $2\pi$ vortex. We therefore predict that it is this
fractional vortex that will be preferentially induced by the rotation.
One experimental consequence of this is the reduction in the
lower-critical frequency $\Omega_{c1}$ down to a $1/2$ of the
conventional value $\frac{\hbar}{2m R^2}\ln R/\xi$ in a large cloud of
radius $R$. Concomitantly, in addition to its imbalance-tunable
anisotropy (that diverges near the upper, $h_{c2}$ phase boundary),
the vortex lattice is characterized by a vortex density $n_{v} =
\frac{4m\Omega}{\pi\hbar}$, that is increased by a factor of $2$ from
the conventional value of a rotated paired superfluid. Perhaps the
most striking feature that we predict is that a rotation-induced
lattice of half-dislocations must accompany and be locked to this
$\pi$-vortex lattice. We expect above novel features to be most
pronounced for a rotation axis transverse to $\qv_0$, with a more
conventional rotational response for $\Omega\parallel\qv_0$. Finally,
the LO state may exhibit more than one critical rotational velocity
$\Omega_{c i}$ corresponding to distinct onsets of the penetration of
half-integer and integer vortices.

\subsection{Fermionic excitations}

As we have seen above, in addition to the low-energy Goldstone modes,
$\phi, u$, the LO state is characterized by gapless fermionic
excitations associated with the imbalanced atom, that are localized on
zeros of domain-walls in $\Delta_{LO}(\rv)$. The most direct probe of
these mid-gap fermionic states is through the Feshbach
resonance\cite{GreinerFBR} and the RF
spectroscopies\cite{RegalJinRF03,ChinRF,KetterleRF,TormaZollerRF,VeilletteRF},
with the momentum-resolved extension\cite{StewartRFk,GaeblerRFk},
allowing one to measure the dispersion and Fermi surfaces pockets of
the Andreev states calculated in Sec.\ref{sec:fermions}. An
observation of coexisting gapped and (Andreev) gapless features in
these spectroscopies, along with the superfluid phase coherence in
$n_\kv$ would provide strong evidence for realization of a FFLO state.

Such measurements can be complemented with shot-noise correlation
spectroscopy\cite{AltmanNoise}, that has been successfully used to
probe bosonic Mott insulators\cite{BlochReview} and fermionic paired
condensates\cite{GreinerNoise05}. In the LO state, we expect pairing
shot-noise correlations to be peaked for $-\kv_F+\qv_0/2,\downarrow
\longleftrightarrow \kv_F+\qv_0/2,\uparrow$ and
$\kv_F-\qv_0/2,\downarrow \longleftrightarrow -\kv_F-\qv_0/2,\uparrow$
atom pairs. In contrast to a conventionally paired BCS state,
shot-noise correlations are furthermore anisotropically distributed in
the center-of-mass pair momentum, with peaks around $\pm \qv_0$,
reflecting the spontaneous nematic anisotropy of the LO state.

The low-energy excitations in the LO state can also be probed less
directly through thermodynamics, modifying the power-law in $T$
behavior of, for example, the heat-capacity. They will also manifest
themselves in thermal transport, though its experimental
implementation in the context of trapped atomic gases remains an open
problem.

Probably the most direct way to detect the existence of the LO state
is to simply image the population species imbalance. Because
imbalanced fermions are confined to Andreev states localized on the LO
domain-walls, the polarization density is periodic, with its phase
locked to that of $\Delta_{LO}(\rv)$. As illustrated for
the thermally-averaged LO order parameter, in a trap the 
amplitude of this periodic component of polarization will be nonzero
and strongly $R$ dependent.

Quite clearly, significant detailed theoretical analysis is necessary
beyond above qualitative discussion, but is left for a future
research.

\subsection{Phase transitions and novel phases}

In addition to our findings of the rich fluctuation-driven
phenomenology of the LO (superfluid smectic) state, in
Sec.\ref{sec:transitions} we predicted a variety of putative
descendent phases, that emerge by disordering the LO state through a
set of continuous phase transitions. If indeed realized as stable
phases (something that our phenomenological approach is unable to
determine for any specific system) organized into phase diagram in
Fig.\ref{fig:phasediagramLO3d}, then many of the predicted features
should be readily detectable in experiments on imbalanced resonant
Fermi gases. Each of these states ($SF_{N}$ $FL_N$, $FL_{Sm}$, etc.)
exhibits its own qualitatively distinct phenomenology, discussed in
Sec.\ref{sec:transitions}, where they were defined. For example, Bragg
peaks in the time-of-flight images can distinguish the periodic
$SF_{Sm}$ (LO superfluid smectic) state from the homogeneous $SF^4_N$
(superfluid nematic), which are in turn distinguished from the
$FL^{2q}_{Sm}$ and $FL_N$ (normal smectic and nematic) by their
superfluid properties, broken spatial symmetries (periodicity and
anisotropy), collective modes, quantized vortices, and condensate
peaks. Standard thermodynamic signatures (e.g., heat capacity) will
identify the corresponding phase transitions, though some associated
with strongly non-meanfield topological type between two gapped
(disordered) phases (e.g., FL$^*$ and FL) maybe more difficult to
detect. In a trap (see above), the most vivid manifestation of these
states is the appearance of a shell structure, corresponding to slices
through the chemical grand-canonical phase diagram of an imbalanced
resonant Fermi gas, Fig.\ref{fig:phasediagramLO3d}. Again, we leave
the detailed analyses of all these features to extended future
studies.

%

\section{Open questions}
\label{sec:open}

While this manuscript addresses a broad range of phenomena associated
with the Larkin-Ovchinnikov state, it leaves many interesting
questions open. Certainly the most important of these is the
long-standing question of the range of energetic stability of the
crystalline superconductor discussed in the Introduction. If the state
is indeed stable over a sufficiently broad range of detuning and
imbalance to be experimentally accessible, is its lowest energy form
indeed the striped collinear LO type, assumed throughout this
manuscript? While for large atomic clouds and shallow traps (such that
LDA remains valid) we expect only a small deformation of the LO state
near the boundaries of the phase, for tighter traps a more detailed
treatment of the trap is necessary, and may lead to a distinct global
form of the LO state, such as the ``onion'' and ``radial'' structures
illustrated in Fig.\rf{fig:onionLOquestion}. To address such questions
undoubtedly requires numerical solutions in experiment-specific
geometries.

Furthermore, the nature of the (2d and 3d) transition into the LO
state at the lower-critical Zeeman field
$h_{c1}$,\cite{BurkhardtRainerLO}, and the extent to which it
resembles a commensurate-incommensurate transition (as in
1d\cite{MachidaNakanishiLO,YangLL}) remains an open question. More
broadly, in Sec.\ref{sec:transitions} we predicted a number of novel
LO descendent phases adjacent to the LO smectic superfluid state, but
have left open the detailed nature of their phenomenology, stability
to quantum and thermal fluctuations, as well the nature of the
associated phase transitions. Similarly to the LO state, these phases
are expected to exhibit gapless fermionic excitations coupled to their
Goldstone modes. Understanding the effects of these fermionic modes on
the properties of the phases and the associated transitions remain
wide open and extremely interesting problems.

Finally, as is clear from the discussion of the previous section, much
detailed theoretical analysis remains to be done to make contact of
our general predictions with specific experiments. We leave these and
many other interesting questions to future studies.

\section{Summary and Conclusions}
\label{sec:summary}

To summarize, we studied a wide range of fluctuation phenomena in a LO
state, expected to be realizable in an imbalanced resonant Fermi gas.
Starting with a microscopic description of a resonant Fermi gas,
supported by robust model-independent and very general symmetry
arguments we have demonstrated that in an isotropic trap the LO state
is a gapless superfluid smectic liquid crystal, whose elastic moduli
and superfluid stiffness we derived near $h_{c2}$. Consequently, the
state is extremely sensitive to thermal fluctuations that destroy its
long-range positional order even in three dimensions, replacing it by
a quasi-longer range order of the resulting algebraic quantum smectic
state, characterized by power-law correlations akin to a system tuned
to a critical point or two-dimensional xy-model systems. We showed
that this exotic state also exhibits vortex fractionalization, where
the basic superfluid vortex is half the strength of a vortex in a
regular paired condensate, and is accompanied by half-dislocations in
the LO smectic (layered) structure. 

Studying the fluctuation-driven disordering of the LO smectic, we
predicted a rich variety of descendant phases such as the superfluid
($SF_N$) and Fermi liquid nematics ($FL_N$) and the fractionalized
nonsuperfluid states ($FL^*$), that generically intervene between the
LO state and the conventional BCS superfluid (at low population
imbalance) and a conventional Fermi liquid (at high population
imbalance). We outlined a large variety of experimental implications
of our findings, but leave their detailed analysis to future studies.
\vspace{0.5cm}
\section{Acknowledgments}

I thank A. Vishwanath for stimulating discussions and a collaboration
on the early stages of this work\cite{RVprl}, as well as for his and
the Berkeley Physics Department's hospitality during a sabbatical
stay. I acknowledge fruitful discussions with V. Gurarie, M. Hermele,
D. Huse, and M. Levin, and thank S. Choi for proofreading of the
manuscript. This work was supported by the National Science Foundation
through a grant No.\ DMR-1001240.

\appendix

\vspace{0.5cm}
\section{Ginzburg-Landau expansion}
\label{app:GLexpansion}

In this appendix we provide some of the technical details necessary to
derive the quartic interaction in $\Delta_\qv$ of the Ginzburg-Landau
expansion appearing in Eqs.\rf{HDelta_qhc2},\rf{H_GL} and in
particular the current-current contribution \rf{H4jj}, that determines
the transverse superfluid stiffness $\rho_s^\perp$. As outlined in the
main text, we use the coherent-state imaginary-time path-integral
formulation of the BCS partition function. We decouple the quartic
fermion interaction $\fermiint$ by introducing the Cooper-pair
Hubbard-Stratonovich field $\Delta(\xv)$, and formally integrate out
the fermionic atoms, obtaining 
\begin{eqnarray} 
Z&=&\int[d\Delta^*
  d\Delta]e^{-S_{eff}[\Delta^*,\Delta]}, 
\end{eqnarray}
where the effective Ginzburg-Landau action is given by
\begin{eqnarray}
\hspace{-1cm}
 S_{eff}[\Delta^*,\Delta]&=&-\ln\left[\int[d\psi^*_\sigma d\psi_\sigma]
    e^{-S_\tau[\psi^*_\sigma,\psi_\sigma,\Delta^*,\Delta]}\right],
\end{eqnarray}
where $S_\tau$ is the microscopic action defined in \rfs{Stau}. 

Taylor-expanding $S_\tau$ in powers of $S_{int}$, \rfs{Sint} we obtain
\begin{widetext}
\bse
\begin{eqnarray}
S_{eff}[\Deltab,\Delta]&=&-\ln Z_0 
-\ln\left[1-\langle S_{int}\rangle_0 
+\frac{1}{2!}\langle S_{int}^2\rangle_0 
-\frac{1}{3!}\langle S_{int}^3\rangle_0 
+\frac{1}{4!}\langle S_{int}^4\rangle_0 +\ldots\right],\\
&=&-\ln Z_0 - \frac{1}{2!}\langle S_{int}^2\rangle_0 
-\frac{1}{4!}\langle S_{int}^4\rangle_0 
+\frac{1}{8}\langle S_{int}^2\rangle_0^2 +\ldots,
\end{eqnarray}
\ese
\end{widetext}
where $Z_0=\int[d\psi^*_\sigma d\psi_\sigma]e^{-S_0}$, $\langle\ldots
\rangle_0 =\int[d\psi^*_\sigma d\psi_\sigma]\ldots e^{-S_0}/Z_0$.
Above, all odd-power in $\Delta$ terms clearly vanish, and the
quadratic $\Delta^*\hat\eps\Delta$ term has already been analyzed in
the main text\cite{LO,SRprl,SRaop}.

Thus, we focus on the contribution quartic in $\Delta$, that is a connected
forth cumulant. Using the definition of $S_{int}$, \rfs{Sint} we find
\bse
\begin{eqnarray} 
\hspace{-1.5cm}
S_4&=&-\frac{1}{4!}\left[\langle S_{int}^4\rangle_0
    -3\langle S_{int}^2\rangle_0^2\right]
=-\frac{1}{4!}\langle S_{int}^4\rangle_0^c,\\
&=&\frac{12}{4!}\int_{\xv_1\xv_2\xv_3\xv_4} V(\xv_1,\xv_2,\xv_3,\xv_4)
\Delta^*_{\xv_1}\Delta_{\xv_2}\Delta^*_{\xv_3}\Delta_{\xv_4},\nonumber\\
\end{eqnarray}
\ese
where
\begin{eqnarray}
&&\hspace{-0.7cm}V(\xv_1,\xv_2,\xv_3,\xv_4)=\\
&&G^0_{\uparrow}(\xv_2-\xv_1)G^0_{\downarrow}(\xv_2-\xv_3)
G^0_{\uparrow}(\xv_4-\xv_3)G^0_{\downarrow}(\xv_4-\xv_1),\nonumber
\end{eqnarray}
with the noninteracting fermionic Green's function (in Fourier space)
as usual given by
\bse
\begin{eqnarray}
G^0_{\sigma}(\omega_n,q)&=&-\langle\psi_\sigma\psi^*_\sigma\rangle_0,\\
&=&\frac{1}{i\omega_n-\eps_{q\sigma}}.
\end{eqnarray}
\ese
Because we are interested in $\tau$-independent $\Delta(\rv)$, the
$\tau$ integrals can be taken, giving $S_4=\int d\tau
H_4$, with
\begin{widetext}
\begin{eqnarray}
H_4&=&\frac{1}{2}\int_{\qtv_i}
(2\pi)^d\delta^d(\qtv_1-\qtv_2+\qtv_3-\qtv_4)
\bigg[\sum_{\qv_1=\pm\qv}
\Vt(\qv_1+\qtv_1,\qv_1+\qtv_2,\qv_1+\qtv_3,\qv_1+\qtv_4)
\Delta^*_{\qv_1}(\qtv_1)\Delta_{\qv_1}(\qtv_2)
\Delta^*_{\qv_1}(\qtv_3)\Delta_{\qv_1}(\qtv_4)\nonumber\\
&&+4\Vt(\qv+\qtv_1,\qv+\qtv_2,-\qv+\qtv_3,-\qv+\qtv_4)
\Delta^*_{\qv}(\qtv_1)\Delta_{\qv}(\qtv_2)
\Delta^*_{-\qv}(\qtv_3)\Delta_{-\qv}(\qtv_4)\bigg].
\label{app:qH4VV}
\end{eqnarray}
\end{widetext}
A Taylor-expansion of
$\Vt(\qv_{n_1}+\qtv_1,\qv_{n_2}+\qtv_2,\qv_{n_3}+\qtv_3,
\qv_{n_4}+\qtv_4)$ in $\qtv_i$ gives $H_4=H_4^{(0)}+H_4^{(2)}$, where
$H_4^{(0)}$ has already been computed by LO in their mean-field
approximation\cite{LO} and in Ref.~\onlinecite{SRaop} from the
expansion of the BdG ground state energy of the FF state discussed in
Sec.\ref{sec:BdGff}.

We focus on the second $H_4^{(2)}$ term, and Taylor
expand it to second order in $\qtv_i$
\bse
\begin{eqnarray}
&&\Vt(\Qv+\qtv_1,\Qv+\qtv_2,-\Qv+\qtv_3,-\Qv+\qtv_4)\approx
v^{(0)}_{+-}
\hspace{1.5cm}\\
&&+
\frac{v_{ij}^{(2)}(\qv)}
{4m^2}\left(\qt_{1i}\qt_{4j}+\qt_{1i}\qt_{3j}+\qt_{2i}\qt_{4j}
+\qt_{2i}\qt_{3j}\right),
\nonumber\\
\end{eqnarray}
\ese
that gives the key current-current interaction vertex,
$v_{ij}^{(2)}j_i j_j$, with
\begin{widetext}
\bse
\begin{eqnarray}
v_{ij}^{(2)}&=&m^2\int\frac{d\omega d^d k}{(2\pi)^{d+1}}
\Gt^0_{\uparrow}(\kv,\omega)^2
\partial_i\Gt^0_{\downarrow}(\qv-\kv,-\omega)
\partial_j\Gt^0_{\downarrow}(-\qv-\kv,-\omega),\\
&=&\int_{\omega,\kv}\frac{(\Qv-\kv)_i(\Qv+\kv)_j}
{(i\omega -\frac{k^2}{2m} + \mu + h)^2
(-i\omega - \frac{(\kv-\Qv)^2}{2m} + \mu - h)^2
(-i\omega - \frac{(\kv+\Qv)^2}{2m} + \mu - h)^2},\\
&\approx&\int_{\omega,\eps,\kh}\frac{(\Qv-k_F\kh)_i(\Qv+k_F\kh)_j}
{(i\omega -\eps + h)^2
(i\omega + \eps - v_F\kh\cdot\Qv + \frac{q^2}{2m} + h)^2
(i\omega + \eps + v_F\kh\cdot\Qv + \frac{q^2}{2m} +h)^2},
\end{eqnarray}
\ese

We first carry out the $\omega$ integral using simple identity,
\begin{eqnarray}
I_\omega&=&\int\frac{d\omega}{2\pi}\frac{1}{(i\omega - a)^2
(i\omega - b)^2(i\omega - c)^2}
\nonumber\\
&=&\frac{\partial^3}{\partial a\partial b\partial c}
\int\frac{d\omega}{2\pi}\frac{1}{(i\omega - a)
(i\omega - b)(i\omega - c)}
\nonumber\\
&=&\frac{\partial^3}{\partial a\partial b\partial c}
\left[\frac{\Theta(-a)}{(a-b)(a-c)}
+\frac{\Theta(-b)}{(b-c)(b-a)}
+\frac{\Theta(-c)}{(c-a)(c-b)}\right],
\nonumber\\
&=&-\frac{\delta(a)}{b^2c^2}
-\frac{\delta(b)}{c^2a^2}
-\frac{\delta(c)}{a^2b^2}\nonumber\\
&&-2\Theta(-a)\frac{2a-b-c}{(a-b)^3(a-c)^3}
-2\Theta(-b)\frac{2b-c-a}{(b-c)^3(b-a)^3}
-2\Theta(-c)\frac{2c-a-b}{(c-a)^3(c-b)^3},
\end{eqnarray}
which, when used in the expression above, gives
\begin{eqnarray}
v_{ij}^{(2)}&\approx&\int_{\eps,\kh}(k_F\kh-\Qv)_i(k_F\kh+\Qv)_j\bigg[
\frac{4(2\eps + \frac{q^2}{2m})\Theta(h-\eps)}
{\big[(2\eps+\frac{q^2}{2m})^2-(v_F\kh\cdot\Qv)^2\big]^3}
-\frac{(2\eps + \frac{q^2}{2m}+3v_F\kh\cdot\Qv)
\Theta(\eps + \frac{q^2}{2m}+h+v_F\kh\cdot\Qv)}
{4(v_F\kh\cdot\Qv)^3(2\eps + \frac{q^2}{2m}+v_F\kh\cdot\Qv)^3}\nonumber\\
&&+\frac{(2\eps + \frac{q^2}{2m}-3v_F\kh\cdot\Qv)
\Theta(\eps + \frac{q^2}{2m}+h-v_F\kh\cdot\Qv)}
{4(v_F\kh\cdot\Qv)^3(2\eps + \frac{q^2}{2m}-v_F\kh\cdot\Qv)^3}
+\frac{\delta(\eps-h)}
{\big[(2h+\frac{q^2}{2m})^2-(v_F\kh\cdot\Qv)^2\big]^2}\nonumber\\
&&+\frac{\delta(\eps + \frac{q^2}{2m}+h+v_F\kh\cdot\Qv)}
{4(v_F\kh\cdot\Qv)^2(2h + \frac{q^2}{2m}+v_F\kh\cdot\Qv)^2}
+\frac{\delta(\eps + \frac{q^2}{2m}+h-v_F\kh\cdot\Qv)}
{4(v_F\kh\cdot\Qv)^2(2h + \frac{q^2}{2m}-v_F\kh\cdot\Qv)^2}\bigg],\nonumber\\
&\approx&N(\epsilon_F)\frac{1}{4\pi}\int d\Omega_\kh
\frac{(k_F\kh-\Qv)_i(k_F\kh+\Qv)_j}
{\big[(2h+\frac{q^2}{2m})^2-(v_F\kh\cdot\Qv)^2\big]^2}.
\label{app:vij}
\end{eqnarray}
\end{widetext}

In above we split the $\kv$ integration into an integration over
its orientations $\kh$ and magnitude $k$, and approximated the latter
using a nearly constant density of states at the Fermi energy, 
equivalently, ignoring small quadratic 
contributions in $\delta k\equiv k-k_F$, valid in the BCS
limit. However, in contrast to a standard isotropic calculation, here 
integral over $\kh$ is nontrivial because of the anisotropy introduced
by $\Qv$. That is, we used
\begin{eqnarray}
\hspace{-1cm}\int\frac{d^3k}{(2\pi)^3}\ldots
&=&\int_0^\infty dk k^2\frac{1}{(2\pi)^3}\int d\Omega_\kh\ldots,\nonumber\\
&=&\int_0^\infty d\epsilon N(\epsilon)\frac{1}{4\pi}\int
d\Omega_\kh\ldots,\nonumber\\
&\approx&N(\epsilon_F)\int_{-\infty}^\infty 
d\eps\frac{1}{4\pi}\int d\Omega_\kh\ldots,
\end{eqnarray}
where the density of states per spin is
\bse
\begin{eqnarray}
N(\epsilon)&=&\frac{m^{3/2}}{2^{1/2}\pi^2\hbar^3}\epsilon^{1/2}
\equiv c\epsilon^{1/2},\\
&=& N(\eps+\epsilon_F)\nonumber\\
&\approx&N(\epsilon_F)=\frac{3}{4}\frac{n}{\epsilon_F}.
\end{eqnarray}
\ese

By symmetry $v_{ij}^{(2)}$ is clearly uniaxial along $\Qv$. Thus it
can be written as
\begin{eqnarray}
v_{ij}^{(2)} = g_1\delta_{ij}+g_2\Qh_i\Qh_j
\label{app:vij2}
\end{eqnarray}
where
\bse
\begin{eqnarray}
g_1&=&\frac{1}{2}v_{ii}^{(2)}-\frac{1}{2}v_{ij}^{(2)}\Qh_i\Qh_j,\\
g_2&=&-\frac{1}{2}v_{ii}^{(2)}+\frac{3}{2}v_{ij}^{(2)}\Qh_i\Qh_j,
\end{eqnarray}
\ese
with the sums over repeated indices implied. Using this inside
\rfs{app:vij2} we find:
\bse
\begin{eqnarray}
g_1&=&\oh N(\epsilon_F)k_F^2\frac{1}{4\pi}\int d\Omega_\kh
\frac{1-(\kh\cdot\Qh)^2}
{\big[(2h+\frac{q_0^2}{2m})^2-(v_F\kh\cdot\Qv)^2\big]^2},\nonumber\\
&\approx&\frac{N(\epsilon_F)k_F^2}{2v_F^4q_0^4}\int_0^1 d\sigma
\frac{1-\sigma^2}{\big[\sigma^2-\frac{4h^2}{v_F^2q_0^2}\big]^2},\nonumber\\
&\equiv&\frac{N(\epsilon_F)k_F^2}{2v_F^4q_0^4}
\alpha_1\left(\frac{2h}{v_Fq_0}\right),\label{app:g1}\\
g_2&=&\oh N(\epsilon_F)k_F^2\frac{1}{4\pi}\int d\Omega_\kh
\frac{3(\kh\cdot\Qh)^2-1-q_0^2/k_F^2}
{\big[(2h+\frac{q_0^2}{2m})^2-(v_F\kh\cdot\Qv)^2\big]^2},\nonumber\\
&\approx&\frac{N(\epsilon_F)k_F^2}{2v_F^4q_0^4}\int_0^1 d\sigma
\frac{3\sigma^2-1}{\big[\sigma^2-\frac{4h^2}{v_F^2q_0^2}\big]^2}\nonumber\\
&\equiv&\frac{N(\epsilon_F)k_F^2}{2v_F^4q_0^4}
\alpha_2\left(\frac{2h}{v_Fq_0}\right),\label{app:g2}
\end{eqnarray}
\ese
where in above we evaluated $q$ at the dominant dispersion minimum
value of $q_0$ and ignored subdominant $q_0^2/k_F^2$ terms.

The dimensionless functions $\alpha_1(x)$ and $\alpha_2(x)$ defined
by the above polar angle ($\sigma$) integrals are given by
\bse
\begin{eqnarray}
\alpha_1(x)&=&\int_0^1 d\sigma
\frac{1-\sigma^2}{\big[\sigma^2-x^2\big]^2},\nonumber\\
&=&-\frac{1}{2x^2} + \frac{1+x^2}{4x^3}\ln\frac{1+x}{1-x},\\
&\approx&2.07, \ \text{at $h_{c2}$ ($x_{c2}=1/\alpha_{c2}=5/6$)
  transition},\nonumber\\
\alpha_2(x)&=&\int_0^1 d\sigma
\frac{3\sigma^2-1}{\big[\sigma^2-x^2\big]^2},\nonumber\\
&=&\frac{x-3x^3+\oh(3x^4-2x^2-1)\ln\frac{1+x}{1-x}}
{2x^3(1-x^2)},\\
&\approx&-5.75, \ \text{at $h_{c2}$ ($x_{c2}=1/\alpha_{c2}=5/6$)
  transition},\nonumber
\end{eqnarray}
\ese
with the final values obtained by their evaluation at the
upper-critical field $h_{c2}$. Although naively the expressions appear
to be linearly divergent, a careful analysis of the integrals
regularized with an infinitesimal imaginary part (associated with the
analytic continuation to real frequencies, $i\omega\rightarrow\omega +
i 0^+$) give well-defined values given above.

\section{Fluctuation in a finite ``box''}

In the main body of the paper, Sec.\ref{sec:3d2dcrossover} we
demonstrated that due to the smectic phonon dispersion anisotropy, for
Neumann boundary conditions the zero $q_z=0$ mode dominates
fluctuations over the bulk $q_z\approx \lambda q_\perp^2$ modes and
must be explicitly taken into account. This leads to an effective
dimensional reduction for a system confined in any reasonably
isotropic trap. Here, we rederive this result within a complementary,
fully continuum description and contrast the behavior in the smectic
state with that of the more familiar isotropic xy-model.

\subsection{Isotropic models}

For an xy-model (or really any model with isotropic spatial
dispersion) described by 
\begin{equation}
H=\oh\int d^d x (\nabla u)^2,
\end{equation}
the fluctuations inside a ``box'' of aspect ratio $L_z\times L_\perp$,
with $L_z < L_\perp$, for free (Neumann) boundary conditions are given
by
\bse
\begin{eqnarray}
\hspace{-1cm}\langle u^2(\xv)\rangle 
&=& \int\frac{d q_z d^{d-1}q_\perp}{(2\pi)^d}
\frac{1}{q_z^2 + q_\perp^2},\nonumber\\
&\sim& \frac{1}{L_z} L_\perp^{3-d}  + L_z^{2-d}\\
&\sim&\frac{1}{L_z} L_\perp^{3-d} \gg L_z^{2-d},
\ \mbox{for $d < 2, L_z < L_\perp$},
\end{eqnarray}
\ese
where the first term is due to the $q_z=0$ zero-mode and is larger
than the second bulk modes contribution when $L_z < L_\perp$. It
accounts for fluctuations in the reduced dimensionality of a $d-1$
dimensional "film" of thickness $L_z$. We note that for an isotropic
dispersion the crossover to a lower-dimensional scaling takes place
only when the actual geometrical aspect ratio of the system is
``film''-like, i.e., anisotropic.

\subsection{Smectic models}
We contrast the above standard finite-size scaling with that of a
model with a smectic dispersion, described by
\begin{equation}
H = \oh\int d^d x \left((\partial_z  u)^2 +
  (\nabla_\perp^2 u)^2\right).
\end{equation}
In a finite ``box'', its fluctuations are instead given by
\bse
\begin{eqnarray}
\langle u^2(\xv)\rangle 
&=& \int\frac{d q_z d^{d-1}}{(2\pi)^d}
\frac{1}{q_z^2 + q_\perp^4},\nonumber\\
&\sim& \frac{1}{L_z} L_\perp^{5-d}  + L_z^{(3-d)/2},\\
&\sim& 
\frac{1}{L_z} L_\perp^{5-d} \gg L_z^{(3-d)/2},\ 
\mbox{for $d < 3, L_z < L^2_\perp$}.\nonumber\\
&&
\end{eqnarray}
\ese
Now, clearly the first zero-mode ($q_z=0$) term dominates over the
bulk contribution (second term) for $L_z < L_\perp^2 \gg L_\perp$, and
does so even in an isotropic box with all dimensions $L$. Thus, a
smectic confined in a geometrically isotropic environment is
effectively deep in the lower, $(d-1)$-dimensional "film" regime with
phonon fluctuations scaling like $u_{rms}^2\sim L^{4-d}\gg L^{(3-d)/2}$.


\begin{thebibliography}{99}
\bibitem{commentClassFFLO} Throughout we will use FFLO to denote a
  generic class of fermionic superfluid states paired at a finite
  center of mass momentum.
\bibitem{commentNotSS} We emphasize that while they show some
  resemblance to a conventional supersolid, a LO state is
  qualitatively distinguished from it by the absence of a zero
  momentum condensate. More explicitly, while a conventional
  supersolid also breaks translational and gauge symmetry, this
  symmetry breaking is controlled by two independent order
  parameters. In contrast a LO state is characterized by a single
  product order parameters, thereby putting it in a distinct
  univesality class, that is more appropriately referred to as a
  ``pair-density wave'' (PDW)\cite{ZhangPDW}.
\bibitem{commentIsotropic} All our results are quoted for FF and LO
  states that form in an otherwise isotropic environment, i.e., in
  the absence of a crystal lattice as appropriate for a resonant Fermi
  gas confined in an isotropic trap. The case of an anisotropic
  background (e.g., a lattice) can be easily incorporated through an
  explicit orientational symmetry breaking.
\bibitem{commentSq} At longest scales (smaller $q$) this result, based
  on Gaussian fluctuations is modified by 3d logarithmic corrections
  (power-law in 2d) associated with nonlinear elastic effects of the
  smectic phonon $u$\cite{GP,GW}. However, it is unlikely that lengths
  at which this is manifest can be accessible in the trapped atomic
  gases.
\bibitem{commentCharge} We use the term ``charge'' in the generalized
  sense of a conserved quantum number. Here it is associated with the
  number of fermionic atoms, that are in fact electromagnetically
  neutral for the case of an atomic gas that is a focus of our work.
\bibitem{commentMFTtrans} The FFLO$\to$N transition is only continuous
  within mean-field theory, and can be argued to be either generically
  driven first-order by
  fluctuations.~\cite{Landau1dsolid,Brazovskii} or split into
  intermediate phases.
\bibitem{comment2Deltas} Actually in principle for an arbitrarily oriented
  $\qv$ there is no need to introduce both $\Delta_+$ and
  $\Delta_-$. However, in practice, because calculations are done
  perturbatively about a particular spontaneously chosen $\qv$, both
  $\Delta_\pm$ are needed.
\bibitem{comment_irrel} Because near $h_{c2}$ the $\rho_s^\perp$
  vanishes faster than $\rho_s^\parallel$, these may need to be
  included to account for an expected long crossover before these
  irrelevant terms actually vanish.
\bibitem{commentOtherDOF} As we discuss in
  Secs.\ref{sec:defects},\ref{sec:fermions}, FFLO thermodynamics receives
  contribution from topological defects (vortices and dislocations) as
  well as the gapless fermionic (atom) excitations, in addition to
  these bosonic Goldstone modes.
\bibitem{commentBerry} Additional Berry phase effects may arise and
  qualitatively modify the dynamics, as argued in e.g.,
  Ref.\onlinecite{FujimotoFFLOberry}.
\bibitem{QMfluct} Quantum fluctuations have a negligible effect on the
  domain-proliferation (SF-LO) transition at $h_{c1}$, as they lead to
  a correction to domain-wall interaction that is Gaussian in $z$ and
  is therefore even shorter range than the microscopic (exponential)
  interaction.
\bibitem{1dLLcomment} This $d>1$ long-range ordered ground state
  contrasts sharply with the well-explored $d=1$ system that exhibits
  quasi-long-range order even at $T=0$, consistent with the expected
  Luttinger liquid phenomenology\cite{YangLL} in both charge and
  spin sectors, and with the exact Bethe-Ansatz
  solution\cite{OrsoBA,HuBA}.
\bibitem{commentWI} The rotational invariance gaurantees that the form
  of the nonlinear strain tensor is preserved by graphical
  corrections, thereby imposing the relation between $\eta_{B}$ and
  $\eta_K$, \rf{WI}. This relation can also be obtained from the
  fixed-point condition $d g/d\ell = 0 = (3-d -\oh\eta_B -
  \frac{3}{2}\eta_K)g$ at the nontrivial fixed point $g_*$.
\bibitem{commentZ2} In terms of the ``rotated'' Goldstone modes
  $\phi=\oh(\phi_++\phi_-)$ and $\theta=\oh(\phi_+-\phi_-)$, the
  low-energy LO manifold is described as a half-twisted torus,
  $\big(U(1)\otimes U(1)\big)/\Z_2$. Because of the structure of the
  Hamiltonian \rf{H0loSum} the energetics is best analyzed in terms of
  $\phi,\theta$, while topological constraints on defects are most
  transparently implemented in terms $\phi_\pm$.
\bibitem{BergDual} An analogous dual sine-Gordon model for an xy-PDW
  (non-rotationally invariant) was recently reported in
  Ref.\onlinecite{Berg09nature}.
\bibitem{comment_morephases} In addition to the conventional phases
  classifiable by a Landau order parameter and symmetry breaking, we
  also find states that are symmetry-wise identical, but are
  distinguished by a more subtle topological order (see
  Fig.\ref{fig:phasediagramLO3d},\ref{fig:flowchart} and below).
\bibitem{comment1dPockets} In contrast, in 1d the effective chemical
  potential generically lies in the gap between Andreev
  bands\cite{SamokhinFFLO}. Thus, a 1d LO state exhibits only
  collective low-energy excitations $\phi,\theta$, consistent with its
  Luttinger-liquid description\cite{YangLL,ZhaoLiuLL}.
\end{thebibliography}
\end{document}